\documentclass[fleqn,10pt]{wlscirep}
\usepackage[utf8]{inputenc}
\usepackage[T1]{fontenc}
\usepackage[page,title]{appendix}
\usepackage{url}
\usepackage{graphicx}
\usepackage{amssymb}
\usepackage{bm}
\usepackage{mathrsfs}
\usepackage{amsmath}
\usepackage{amsfonts}
\usepackage{color}
\usepackage{xspace}
\usepackage{tikz}
\usepackage{slashed}
\usepackage{cancel}
\usepackage{hyperref}
\usepackage[utf8]{inputenc}  
\usepackage{xcolor}
\usepackage{slashed}
\usepackage{fancyhdr}
\usepackage{rotating}
\usepackage{multirow}
\usepackage{xcolor,colortbl}
\usepackage[sort, numbers]{natbib}
\usepackage{graphicx}
\usepackage[freestanding]{hepunits}
\usepackage{cleveref}
\usepackage{tocloft}

\newcommand{\cf}{\textit{cf.}\xspace}
\newcommand{\ie}{\textit{i.e.}\xspace}
\newcommand{\eg}{\textit{e.g.}\xspace}
\newcommand{\etc}{\textit{etc.}\xspace}
\newcommand{\rfr}[1]{ref.~\cite{#1}\xspace}
\newcommand{\rfrs}[1]{refs.~\cite{#1}\xspace}
\newcommand{\Rfr}[1]{Ref.~\cite{#1}\xspace}

\begin{document}

\title{ 
{\normalsize \textit{Solicited White Paper: Topic area -- RF6} }
\\Experiments and Facilities for Accelerator-Based Dark Sector Searches
}

\author[1]{Philip~Ilten (ed.)}
\author[2]{Nhan~Tran (ed.)}
\author[3]{Patrick~Achenbach}
\author[4,5]{Akitaka~Ariga}
\author[6]{Tomoko~Ariga}
\author[7]{Marco~Battaglieri}
\author[8]{Jianming~Bian}
\author[7,9]{Pietro~Bisio}
\author[7]{Andrea~Celentano}
\author[10]{Matthew~Citron}
\author[11]{Paolo~Crivelli}
\author[12,13]{Giovanni~de Lellis}
\author[12,13,14]{Antonia~Di Crescenzo}
\author[15]{Milind~Diwan}
\author[8]{Jonathan L.~Feng}
\author[16]{Corrado~Gatto}
\author[17]{Stefania~Gori}
\author[18]{Felix~Kling}
\author[7]{Luca~Marsicano}
\author[17]{Simone M.~Mazza}
\author[19]{Josh~McFayden}
\author[20]{Laura~Molina-Bueno}
\author[7,9]{Marco~Spreafico}
\author[21]{Natalia~Toro}
\author[2]{Matthew~Toups}
\author[22,23]{Sebastian~Trojanowski}
\author[8]{Yu-Dai~Tsai}
\author[24]{Mike~Williams}
\author[2]{Jacob~Zettlemoyer}
\author[25]{Yiming~Zhong}
\affil[1]{University of Cincinnati, Cincinnati, Ohio 45221, USA}
\affil[2]{Fermi National Accelerator Laboratory, Batavia, IL 60510, USA}
\affil[3]{Institute for Nuclear Physics, Johannes Gutenberg University Mainz, 55099 Mainz, Germany}
\affil[4]{Albert Einstein Center for Fundamental Physics, Laboratory for High Energy Physics, University of Bern,Sidlerstrasse 5, CH-3012 Bern, Switzerland}
\affil[5]{Department of Physics, Chiba University, 1-33 Yayoi-cho Inage-ku, Chiba, 263-8522, Japan}
\affil[6]{Kyushu University, Nishi-ku, 819-0395 Fukuoka, Japan}
\affil[7]{Istituto Nazionale di Fisica Nucleare, Sezione di Genova, Genova, Italy}
\affil[8]{Department of Physics and Astronomy, University of California, Irvine, CA 92697, USA}
\affil[9]{Universit\'a degli Studi di Genova, Dipartimento di Fisica, 16146 Genova, Italy}
\affil[10]{University of California, Santa Barbara,  CA 93106, USA}
\affil[11]{ETH Zurich, Institute for Particle Physics and Astrophysics, CH-8093 Zurich, Switzerland}
\affil[12]{Dipartimento di Fisica, Universit\`a ``Federico II'', Napoli, Italy}
\affil[13]{Istituto Nazionale di Fisica Nucleare, Napoli, Italy}
\affil[14]{CERN, Geneva, Switzerland}
\affil[15]{Department of Physics, Brookhaven National Laboratory, Upton, NY 11973, USA
}
\affil[16]{Sezione di Lecce, Istituto Nazionale di Fisica Nucleare}
\affil[17]{University of California, Santa Cruz, CA 95064, USA}
\affil[18]{Deutsches Elektronen-Synchrotron DESY, Notkestr. 85, 22607 Hamburg, Germany}
\affil[19]{University of Sussex, Falmer, UK}
\affil[20]{Instituto de Fisica Corpuscular (CSIC/UV), 46980 Paterna, Valencia, Spain}
\affil[21]{SLAC National Accelerator Laboratory, Menlo Park, CA 94025, USA}
\affil[22]{Astrocent, Nicolaus Copernicus Astronomical Center Polish Academy of Sciences, ul. Rektorska 4, 00-614, Warsaw, Poland}
\affil[23]{National Centre for Nuclear Research, Pasteura 7, 02-093 Warsaw, Poland}
\affil[24]{Massachusetts Institute of Technology, Cambridge, MA 02139, USA}
\affil[25]{Kavli Institute for Cosmological Physics, University of Chicago, Chicago, IL 60637 USA}

\begin{abstract}
This paper provides an overview of experiments and facilities for accelerator-based dark matter searches as part of the US Community Study on the Future of Particle Physics (Snowmass 2021). Companion white papers to this paper present the physics drivers: thermal dark matter, visible dark portals, and new flavors and rich dark sectors.
\end{abstract}

\maketitle

\vspace{5mm}
\noindent\makebox[\linewidth]{\rule{0.5\paperwidth}{0.4pt}}
\begin{center}
\normalsize \textit{Submitted to the Proceedings of the US Community Study on the Future of Particle Physics (Snowmass 2021)} 
\end{center}
\noindent\makebox[\linewidth]{\rule{0.5\paperwidth}{0.4pt}}

\clearpage
\tableofcontents

\clearpage
\section{Executive Summary}
\begin{figure}[h]
  \centering  
  \includegraphics[width=\textwidth]{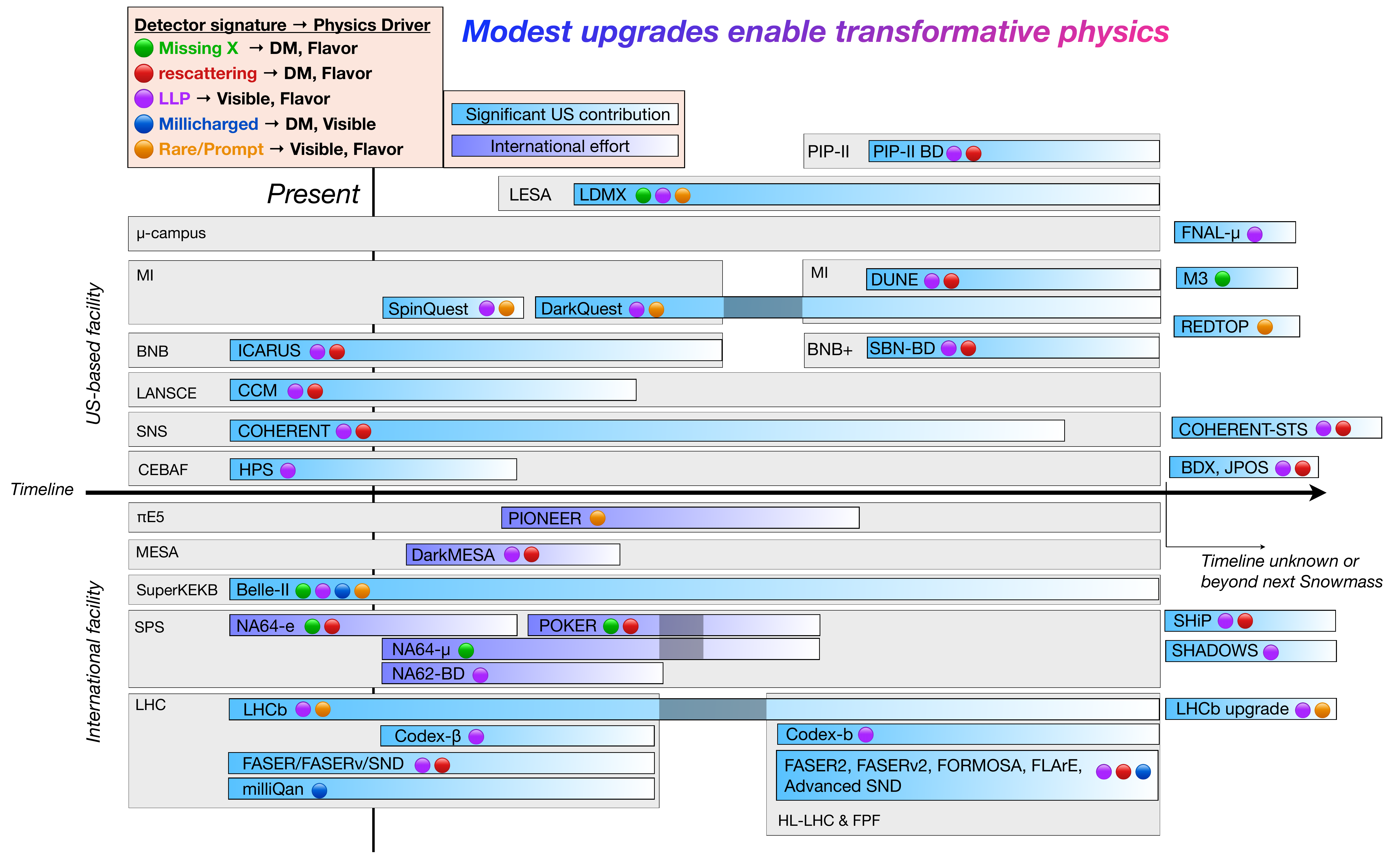}
  \caption{Summary of accelerator facilities and detector signatures of experiments.\label{fig:exec}}
\end{figure}

Searches for dark sector particles in the GeV mass range and below at relativistic particle accelerators are a highly-motivated physics opportunity in the next decade. This paper summarizes and characterizes experiments and facilities for accelerator-based dark sector searches. The physics drivers are characterized into three main thrusts and those are described in companion reports: thermal dark matter~\cite{bi1}, visible dark portals~\cite{bi2}, and new flavors and rich dark sectors~\cite{bi3}. Motivated by these physics drivers, we enumerate a number of experimental initiatives, describing them and characterizing them by their types of experimental signatures: long-lived particles (LLP), DM rescattering, millicharged particles, missing $X$, and rare prompt decays. Given the large interest in this physics and the number of current and proposed experiments, this paper provides a summary, in one central place and including brief descriptions, of these experiments and facilities. We also provide references to more detailed studies where the reader can find more information. This information is summarized compactly in \cref{fig:exec} based on the beam facility producing the dark sector particles and the types of detector signatures proposed at those facilities along a schematic timeline. 

\clearpage
\section{Summary of Experimental Approaches}
We summarize various experimental initiatives and accelerator facilities that can be used to search for dark sectors. In \cref{sec:experiments} are brief descriptions of the experimental initiatives by the proponents which include details on unique features, experiment status and feasibility studies, and timeline. In \cref{sec:facilities} are brief descriptions of the accelerator facilities by the proponents which include details on the capabilities of each of the accelerator particle beams. In this section, we summarize all the experimental initiatives. We characterize the experiments based on certain features that can help to highlight their capabilities:
\begin{itemize}
    \item experiment and facility;
    \item detector signature -- missing $X$, rescattering, long-lived particles (LLP), prompt, and millicharged (more on this below);
    \item particle beam, energy, interaction;
    \item sensitivity to the RF6 physics drivers;
    \item status -- in terms of if they are US-based initiatives and are currently operational.
\end{itemize}
The physics drivers are described in complementary white papers in \rfrs{bi1,bi2,bi3}. We summarize the motivating scenarios briefly here as:
\begin{itemize}
    \item \textbf{thermal dark matter}: We denote this as ``DM'' below. This physics driver focuses on searches for dark matter produced in accelerator based-experiments. These types of searches have sharp milestones to reach in mass-coupling space under the assumption that dark matter was in thermal equilibrium with the SM in the early universe. 
    \item \textbf{visible dark portals}: We denote this as ``Visible'' below. This physics driver focuses on searches for dark mediator particles that decay back to SM particles. In certain mass hierarchies of the dark sector, we are most sensitive to such scenarios by looking for visible SM signatures. 
    \item \textbf{new flavors and rich dark sectors}: We denote this as ``Flavor'' below. This physics driver has focuses on two classes of models: dark sectors with rich structure, as we have in the SM, that may result in both invisible and visible particle final states; and dark sectors motivated by current anomalies from existing experimental results such as $g-2$~\cite{Muong-2:2001kxu,Muong-2:2015xgu}, the MiniBooNE excess~\cite{MiniBooNE:2018esg}, and flavor anomalies~\cite{Belle:2009zue,BaBar:2012mrf,BaBar:2013mob,LHCb:2015gmp,LHCb:2017avl,Belle:2017oht,LHCb:2017vlu,LHCb:2019hip,LHCb:2020lmf}.
\end{itemize}
To characterize the dark sector experiments and their strategies to search for the physics drivers presented above, we classify the experiments into the following categories: 
\begin{itemize}
    \item \textbf{LLP}: Here LLP stands for long-lived particles where the primary detector signature is a portal particle with a significant lifetime that it decays visibly to SM particles on macroscopic scales relative to detector capabilities.
    \item \textbf{DM rescattering}: This denotes experiments where the primary signature is a dark matter particle that is rescattered off of a dense detector of SM particles after being produced at an accelerator.
    \item \textbf{millicharged}: this denotes experiments which have specialized detection capabilities targeting particles that have fractional charge at the $10^{-3}e$ to $10^{-6}e$ level. 
    \item \textbf{missing X}: This denotes experiments where the primary signature is an invisible particle and detection is based on missing momentum, energy, or mass. 
    \item \textbf{rare prompt}: This denotes experiments where the primary detector signature is a portal particle that decays visibly to SM particles promptly (near the interaction point) on macroscopic scales relative to detector capabilities.
\end{itemize}

This information is summarized in \cref{tab:summary}.
\begin{sidewaystable}
    \centering
    \caption{Summary of experimental initiatives, facilities, and key features. \label{tab:summary}}
    \resizebox{\textwidth}{!}{
    \begin{tabular}{|c|c|c|c|c|c|c|}
    \hline
    \multirow{2}{*}{Experiment} & \multirow{2}{*}{Facility} & \multirow{2}{*}{Beam Config} & \multirow{2}{*}{Beam Energy} & \multirow{2}{*}{Det Signature} & \multirow{2}{*}{Timeline} & \multirow{2}{*}{Refs.} \\
    & & & & & & \\    
    \hline
    \multicolumn{7}{|c|}{\cellcolor{gray}\textbf{US-based}} \\
    \hline
    HPS & CEBAF $@$ JLab & electron FT & 1-6\,GeV & LLP & running & \cref{sec:hps}, \cite{HPS:2018xkw} \\
    \hline
    COHERENT & SNS $@$ ORNL & proton FT & 1\,GeV & rescattering & running & \cref{sec:pip2}, \cite{COHERENT:2020iec} \\
    \hline
    CCM & LANSE $@$ LANL & proton FT & 0.8\,GeV & rescattering & running & \cite{CCM:2021leg} \\
    \hline
    SpinQuest/DarkQuest & MI $@$ FNAL & proton FT & 120\,GeV & LLP & construction, proposed upgrade & \cref{sec:darkQuest}, \cite{Batell:2020vqn} \\
    \hline 
    LDMX & LESA $@$ SLAC & electron FT & 4-8\,GeV & Missing X & R\&D funding, 2024 & \cref{sec:ldmx}, \cite{LDMX:2019gvz}\\
    \hline 
    BDX & CEBAF $@$ JLab & electron BD & 11\,GeV & rescattering, Millicharged  & proposed & \cref{sec:bdx}, \cite{Battaglieri:2019ciw} \\
    \hline 
    JPOS & CEBAF $@$ JLab & positron FT & 11\,GeV & Missing X & proposed & \cref{sec:jpos}, \cite{Marsicano:2018oqf} \\
    \hline 
    PIP-II BD & PIP-II $@$ FNAL & proton FT & 1\,GeV & rescattering, LLP & proposed (2029) & \cref{sec:pip2bd}, \cite{Toups:2022yxs} \\
    \hline 
    SBN-BD & Booster $@$ FNAL & proton BD & 8\,GeV & rescattering & proposed (2029) & \cite{Toups:2022knq} \\
    \hline 
    REDTOP & TBD & proton FT & 1-5\,GeV & Missing X, LLP, Prompt & proposed & \cref{sec:redtop}, \cite{REDTOP:2022slw} \\
    \hline 
    M$^3$ & MI $@$ FNAL & muon FT & 15\,GeV muons & Missing X & proposed & \cite{Kahn:2018cqs} \\
    \hline 
    FNAL-$\mu$ & muon campus $@$ FNAL & muon FT & 3\,GeV & LLP & proposed & \cref{sec:fnalMu}, \cite{Chen:2017awl} \\
    \hline   
    \multicolumn{7}{|c|}{\cellcolor{gray}\textbf{International}} \\
    \hline 
    Belle-II & SuperKEKB $@$ KEK & e+e- collider & 150\,MeV & Missing X, LLP, Prompt & running & \cref{sec:belleII}, \cite{Belle-II:2010dht} \\
    \hline
    CODEX-$\beta$ & LHC $@$ CERN &  pp collider & 6.5-7\,TeV & LLP  & construction (2023) & \cref{sec:codex-beta}, \cite{Aielli:2022awh} \\
    \hline
    CODEX-b & LHC $@$ CERN &  pp collider & 6.5-7\,TeV & LLP  & proposed (2026) & \cref{sec:codex-b}, \cite{Gligorov:2017nwh} \\
    \hline
    LHCb & LHC $@$ CERN &  pp collider & 6.5-7\,TeV & LLP, Prompt  & running, future upgrade planned & \cref{sec:lhcb}, \cite{LHCb:2014set} \\
    \hline
    NA62 & SPS-H4 $@$ CERN &  proton BD & 400\,GeV & LLP  & dedicated running planned & \cite{NA62:2017rwk} \\
    \hline
    FASERnu & LHC $@$ CERN & pp collider & 6.5-7\,TeV & rescattering  & running & \cref{sec:faserNu}, \cite{FASER:2020gpr} \\
    \hline 
    milliQAN & LHC $@$ CERN & pp collider & 6.5-7\,TeV & Millicharged  & running & \cref{sec:milliqan}, \cite{milliQan:2021lne} \\    
    \hline
    DarkMESA & MESA $@$ Mainz & Electron FT & 150\,MeV & rescattering, LLP & construction (2023) & \cref{sec:darkMesa} \\
    \hline
    NA64-e & SPS-H4 $@$ CERN & electron FT & 100-150\,GeV & Missing X, Prompt  & running & \cref{sec:na64}, \cite{Banerjee:2019pds} \\
    \hline
    NA64-mu & SPS-M2 $@$ CERN & muon FT & 100-160\,GeV & Missing X & commissioning & \cref{sec:na64Mu} \\
    \hline
    NA64/POKER & SPS-H4 $@$ CERN & positron FT & 100\,GeV & Missing X & planned (2024) & \cref{sec:poker}, \cite{Banerjee:2019pds} \\
    \hline
    PIONEER & $\pi$E5 $@$ PSI & proton FT & 10-20\,MeV pions & Prompt & planned (2028) & \cref{sec:pioneer}, \cite{PIONEER:2022yag} \\
    \hline
    FASER2 & FPF $@$ CERN & pp collider & 6.5-7\,TeV & LLP & proposed (2029) & \cref{sec:faser2} \cite{Ariga:2018zuc} \\
    \hline
    FORMOSA & FPF $@$ CERN & pp collider & 6.5-7\,TeV & Millicharged & proposed (2029) & \cref{sec:formosa}, \cite{Foroughi-Abari:2020qar} \\
    \hline
    FASERnu2 & FPF $@$ CERN & pp collider & 6.5-7\,TeV & rescattering & proposed (2029) & \cref{sec:faserNu2}, \cite{FASER:2020gpr} \\
    \hline
    FLArE & FPF $@$ CERN & pp collider & 6.5-7\,TeV & rescattering & proposed (2029) & \cref{sec:flare}, \cite{Feng:2022inv} \\
    \hline
    SND$@$LHC & LHC $@$ CERN & pp collider & 6.5-7\,TeV & rescattering & running & \cref{sec:snd}, \cite{Collaboration:2729015} \\
    \hline
    Advanced SND$@$LHC & FPF & pp collider & 6.5-7\,TeV & rescattering & proposed (2029) & \cref{sec:snd}, \cite{Collaboration:2729015} \\
    \hline
    \end{tabular}
    }
\end{sidewaystable}

\clearpage
\section{Experimental Initiatives}
\label{sec:experiments}
\subsection{BDX}
\label{sec:bdx}

In an electron beam-dump experiment, light dark matter (LDM) particles are pair-produced in the thick target and traverse unimpeded through sufficient shielding that eliminates all standard model particles aside from neutrinos. BDX will detect LDM through scattering on atomic electrons in a state-of-the-art electromagnetic calorimeter with excellent forward geometric acceptance located $\sim20~\meter$ downstream of the Jefferson Lab Hall-A beam dump. The experiment will run parasitically, making use of the high-energy (up to $11~\GeV$), high-current (up to $65~\si{\micro\ampere}$) electron beam delivered to Hall A for its scheduled hadron physics program. Unlike other efforts, BDX is the only proposed experiment that features both dark matter production and detection utilizing only its coupling to electrons. It can therefore test viable models which do not require any couplings to baryons (which SBND, COHERENT, MiniBooNE, and SHiP all require) and it can directly observe the dark matter scatter (or invisible dark matter decay) in the downstream detector (which LDMX, NA64, and Belle~II cannot). Furthermore, BDX is unique even among other electron-based approaches in its ability to observe the particles produced in the fixed-target, whereas other electron-beam experiments rely on missing-energy signatures, which indirectly infer the production of signal events, thus being exposed to unknown backgrounds.

\begin{figure}[h]
  \centering
  \includegraphics[width=.56\textwidth]{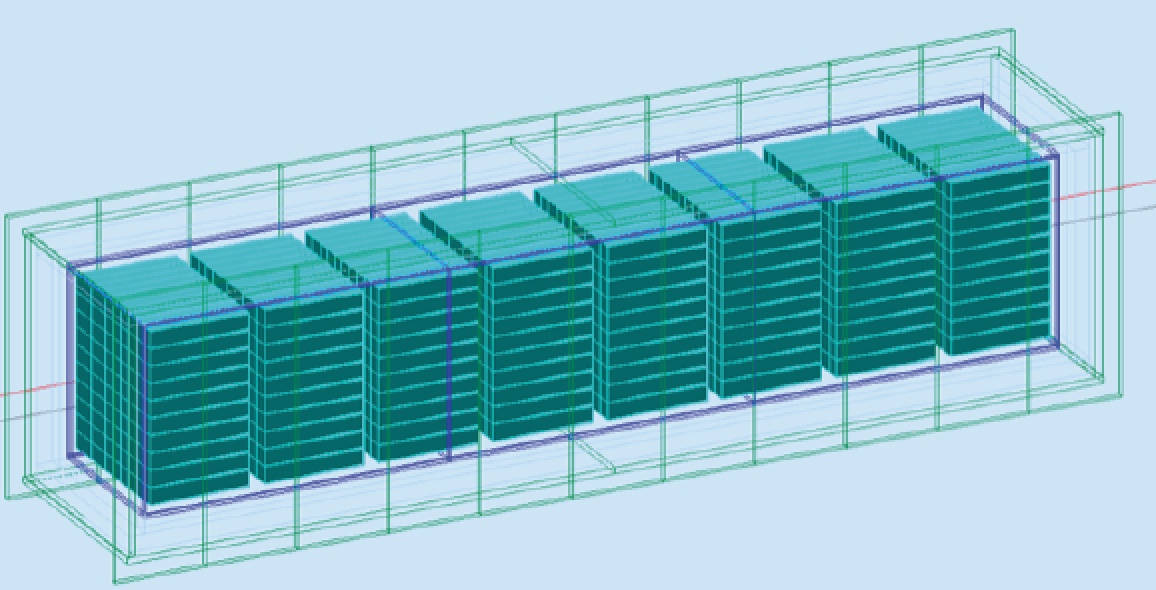}
  \includegraphics[width=.38\textwidth]{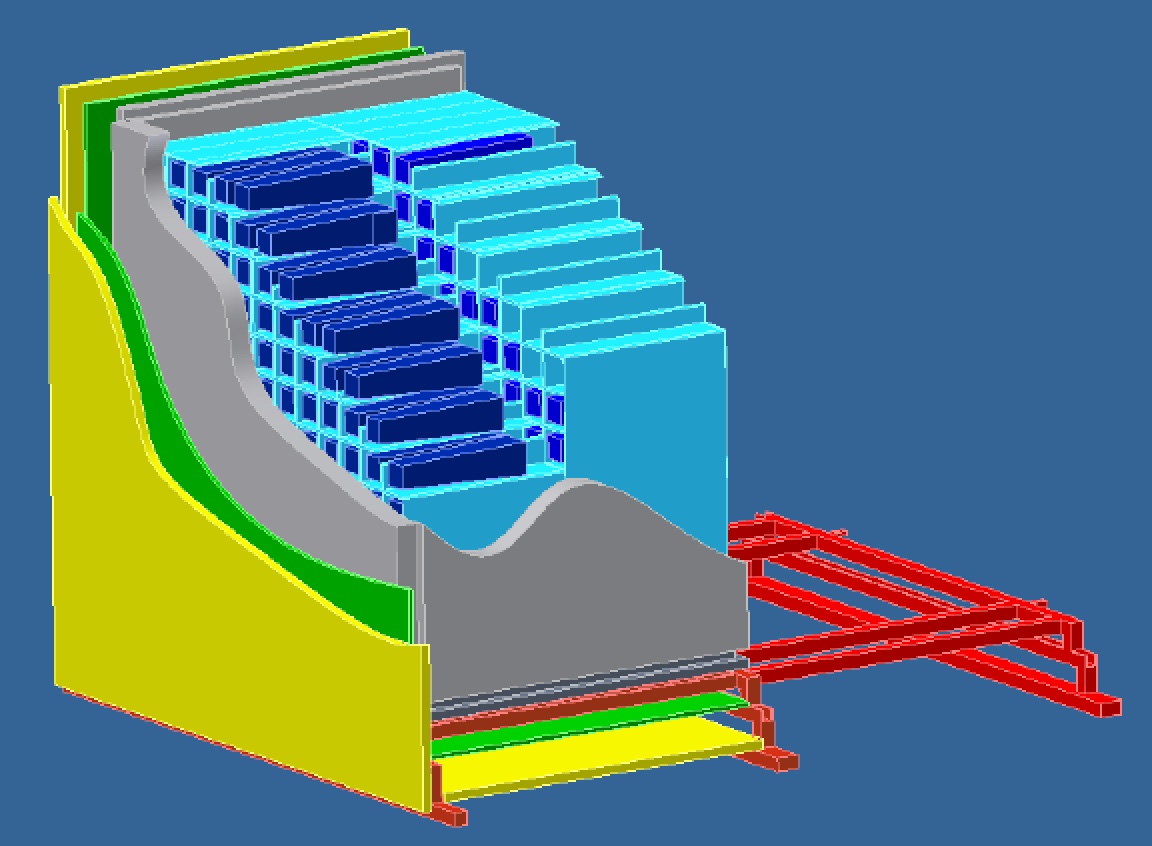}
  \caption{A sketch of the BDX detector. (left) The full reference design composed of $8$ modules of 10x10 CsI(Tl) crystals is shown. On the right the single module drawing showing the (dark blue) crystals, (light blue) aluminum alveoli, (grey) lead vault, (green) inner veto, (yellow) and outer veto.\label{fig:bdx}}
\end{figure}

The BDX detector consists of two main components: a homogeneous electromagnetic calorimeter used to detect signals produced by the interacting dark matter, and a veto system used to reject the cosmic background. The detector has a modular construction to allow for future flexibility in rearranging or increasing the interaction volume. A sketch of the detector is shown in \cref{fig:bdx}. 

The electromagnetic calorimeter is the main component of BDX. The detector uses CsI(Tl) crystals, which are available from the decommissioned BaBar EM calorimeter. It is compact and achieves the large target thickness required for the experiment. BaBar crystals have different shapes and tapering: they will be inserted in new regular-shaped parallelepiped aluminum alveoli for ease of assembly. Since the LDM beam is focused in the forward direction, a rectangular detector provides a long target length along the LDM beam. The reference setup foresees 8 modules of $10\times10$ crystals each, arranged with the long dimension along the beam direction. This arrangement has a cross section of  $50\times55~\si{\centi\meter}^2$ for a total length of $295~\si{\centi\meter}$, see \cref{fig:bdx}. Crystals will be read out by SiPM for fast time coincidence with the plastic veto counters. In order to reject cosmic rays, the EM calorimeter is operated inside two hermetic layers of plastic scintillator veto and a $5~\si{\centi\meter}$ thick lead vault. The design of the active and passive vetoes is driven to achieve the highest hermeticity  in order to shield and/or track and reject standard model particles entering in the EM calorimeter detection volume from outside. The inner and outer vetoes are based on plastic scintillators, which are read out with SiPMs.

The concept of the BDX detector has been tested by means of a prototype (BDX-PROTO) consisting of a matrix of 16 BaBar CsI(Tl) crystals, a lead vault, and a veto system (inner and outer). The detector was deployed in 2017 at LNS-CT (Italy) to assess the cosmic background under similar conditions (overburden and acquisition time) expected for the full BDX experiment. The results showed that with a sufficiently high threshold on the energy deposited in the crystals ($\sim100~\MeV$) the background counts can be reduced to zero. Another prototype detector (BDX-HODO), based on the same technology (a single BaBar CsI(Tl) crystal surrounded by plastic scintillator counters), was deployed in 2018 in two wells dug $\sim20~\meter$ downstream of the Hall-A beam-dump at the location of the planned BDX facility. The test measured the  muon flux produced by the interaction of  the $10.6~\GeV$ electron beam with the dump and propagating to the detector through the present concrete and dirt shielding~\cite{Battaglieri:2019ciw}. The good agreement with predictions from the BDX simulation framework (based on \textsc{Geant4}~\cite{GEANT4:2002zbu} and \textsc{Fluka}~\cite{Ferrari:2005zk}) provides confidence in the ability to design a shielding configuration to eliminate all beam-related backgrounds except for neutrinos.

A small, compact version of the BDX detector (BDX-MINI) consisting of 44 high-density PbWO$_4$ crystals\footnote{The compact PbWO$_4$ crystals are used as a substitute for CsI(Tl) in order to fit into the well.} enclosed in a tungsten vault and  two concentric cylindrical veto systems, has been constructed and installed in one of the wells, and is currently taking data. During the low-energy experiments in Hall-A (E$_{beam}\simeq 2~\GeV$), the existing dirt between the dump and detector becomes an effective shield to all penetrating radiation. This allows for a beam-induced background-free measurement at low energy. BDX-MINI collected $2\times10^{21}$ electrons on target. This test run validated the capabilities of BDX in a realistic experimental setup over an extended period.

In parallel to the leading calorimetry option, an alternative option for the BDX experiment was developed, based on a low-pressure, negative-ion, time-projection-chamber detector. The Directional Recoil Identification From Tracks (DRIFT) technology grew out of the requirements to detect a directional signal from halo dark matter recoils~\cite{Snowden-Ifft:1999reu}. It is ideally suited to search for light dark matter recoils behind a beam-dump. Despite its low mass the BDX-DRIFT detector has sensitivity rivaling and complimentary to the BDX experiment due to its low threshold and low background, and provides information on the directionality of light dark-matter-recoil candidates~\cite{Snowden-Ifft:2018bde}.

\subsection{Belle II}
\label{sec:belleII}

\begin{figure}[tbh]
  \centering
  \includegraphics[width=0.5\textwidth]{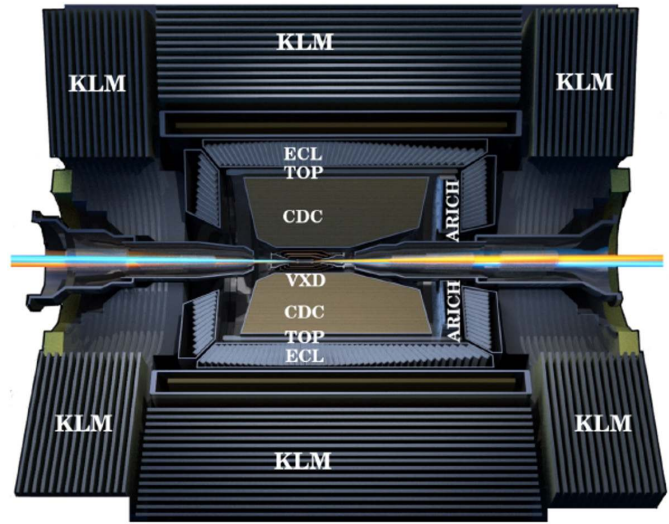}
  \caption{The Belle~II detector, located at the SuperKEKB $e^+ e^-$ collider.\label{fig:BelleII}}
\end{figure}

Belle~II is a multipurpose detector located at the SuperKEKB $e^+ e^-$ collider (\cref{fig:BelleII})~\cite{Belle-II:2010dht}. It is an extensive upgrade of the original Belle detector. It uses silicon pixels for the innermost layer of the vertex detector, achieving a factor of two improvement over Belle and BaBar. It includes a new large-volume drift chamber; powerful particle identification detectors based on Cherenkov light radiated in fused silica bars or in aerogel; a rebuilt muon detector; and next-generation trigger and data acquisition systems.

SuperKEKB operates at or near $10.58~\GeV$ center-of-mass energies, corresponding to the $\Upsilon(4S)$ resonance. It collides $7~\GeV$ electrons with $4~\GeV$ positrons to provide a boost in the lab frame of the $B$ mesons produced in the $\Upsilon(4S)$ decay. SuperKEKB is the world's highest instantaneous luminosity particle collider, and has the goal to deliver a dataset of $50~\invab$ over the next decade, $100\times$ that recorded by BaBar. 

Belle~II will undertake analyses sensitive to a wide range of dark sector models. These include dark photons decaying to invisible final states (the so called ``single photon'' search)~\cite{Kou:2018nap}, or to prompt standard model particles (typically a pair of leptons)~\cite{Kou:2018nap}, or to a displaced vertex~\cite{Ferber:2022ewf}. More complex models, such as $L_{\mu} - L_{\tau}$~\cite{Adachi:2019otg} or inelastic dark matter~\cite{Duerr:2019dmv}, also produce distinctive signatures. A search for axion-like particles has already been published with a small dataset~\cite{Belle-II:2020jti}. Other production mechanisms are also possible, including two-photon fusion~\cite{Dolan:2017osp}, or $B$ decays~\cite{Izaguirre:2016dfi, Ferber:2022rsf}. A dark Higgs could be produced in a similar fashion in $B$ decays, with either a prompt or displaced decay~\cite{Filimonova:2019tuy}, or it could be produced directly in $e^+e^-$ annihilation in conjunction with a dark photon~\cite{Duerr:2020muu}. Belle~II will have an enormous sample of $\tau$ leptons, which provides sensitivity to a sterile neutrino that mixes predominately with the $\tau$~\cite{Dib:2019tuj}. The projections in these cases are a combination of studies done by the collaboration, extrapolations of Belle~II papers on smaller data sets, and phenomenological studies. 

Belle~II started collecting data with a complete detector in March 2019, and expects to collect $500~\invfb$ -- approximately the size of the BaBar data set -- by summer 2022. The experiment will then take an extended shutdown to install a new two-layer pixel vertex detector, with data taking resuming in October 2023. Projections show an accumulated dataset of approximately $2-3~\invab$ by April 2026. A second extended shutdown beyond this date is planned to enable the SuperKEKB collider to achieve the design instantaneous luminosity of $6\times 10^{35}~\invcmsqpersec$. 

There are approximately 1100 Belle~II collaborators, including 450 PhD physicists and 390 graduate students.

\subsection{CODEX-b}
\label{sec:codex-b}

\begin{figure}[h]
  \centering
  \includegraphics[width=0.45\columnwidth]{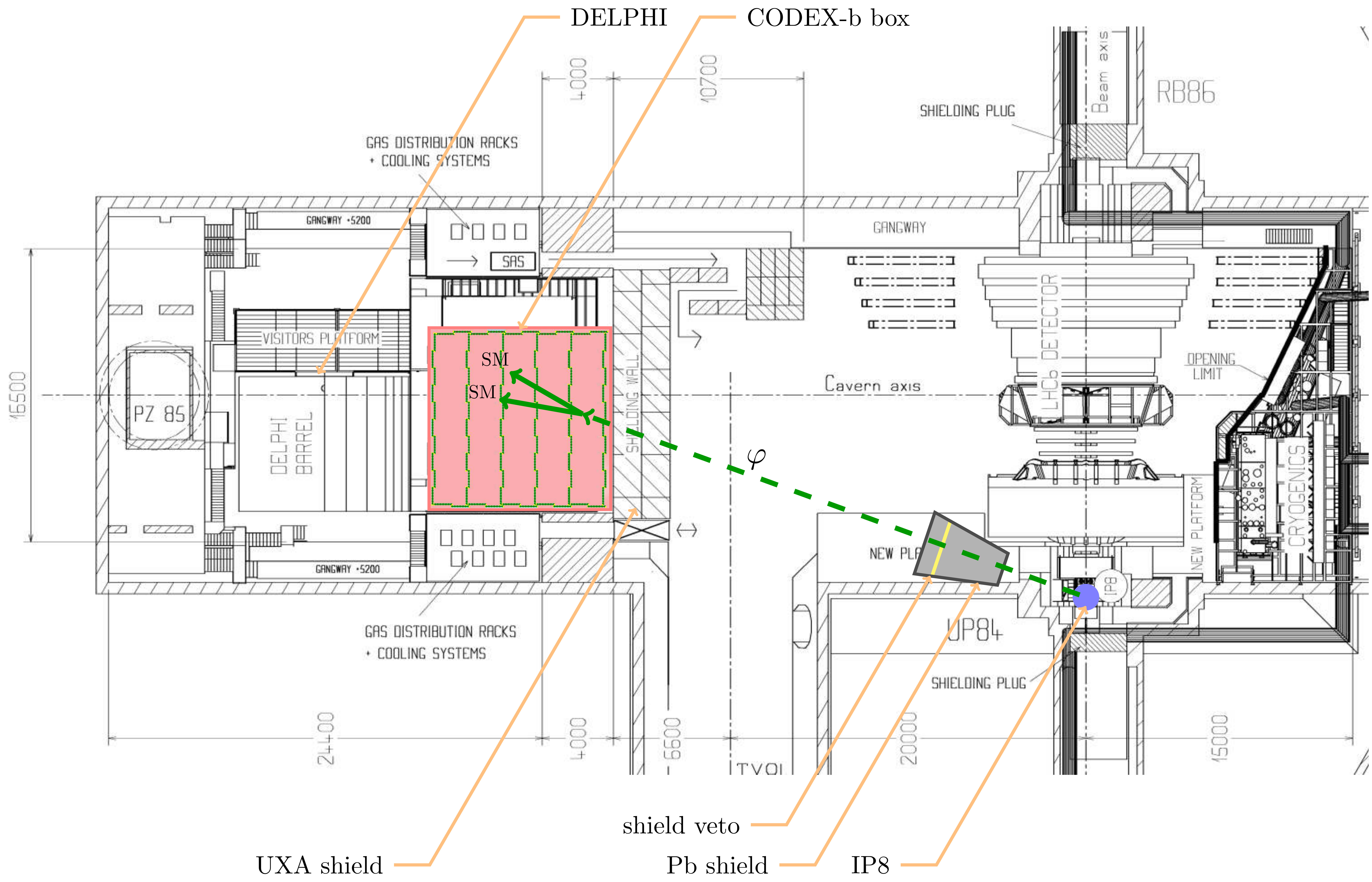}\hfill
  \includegraphics[width=0.4\columnwidth]{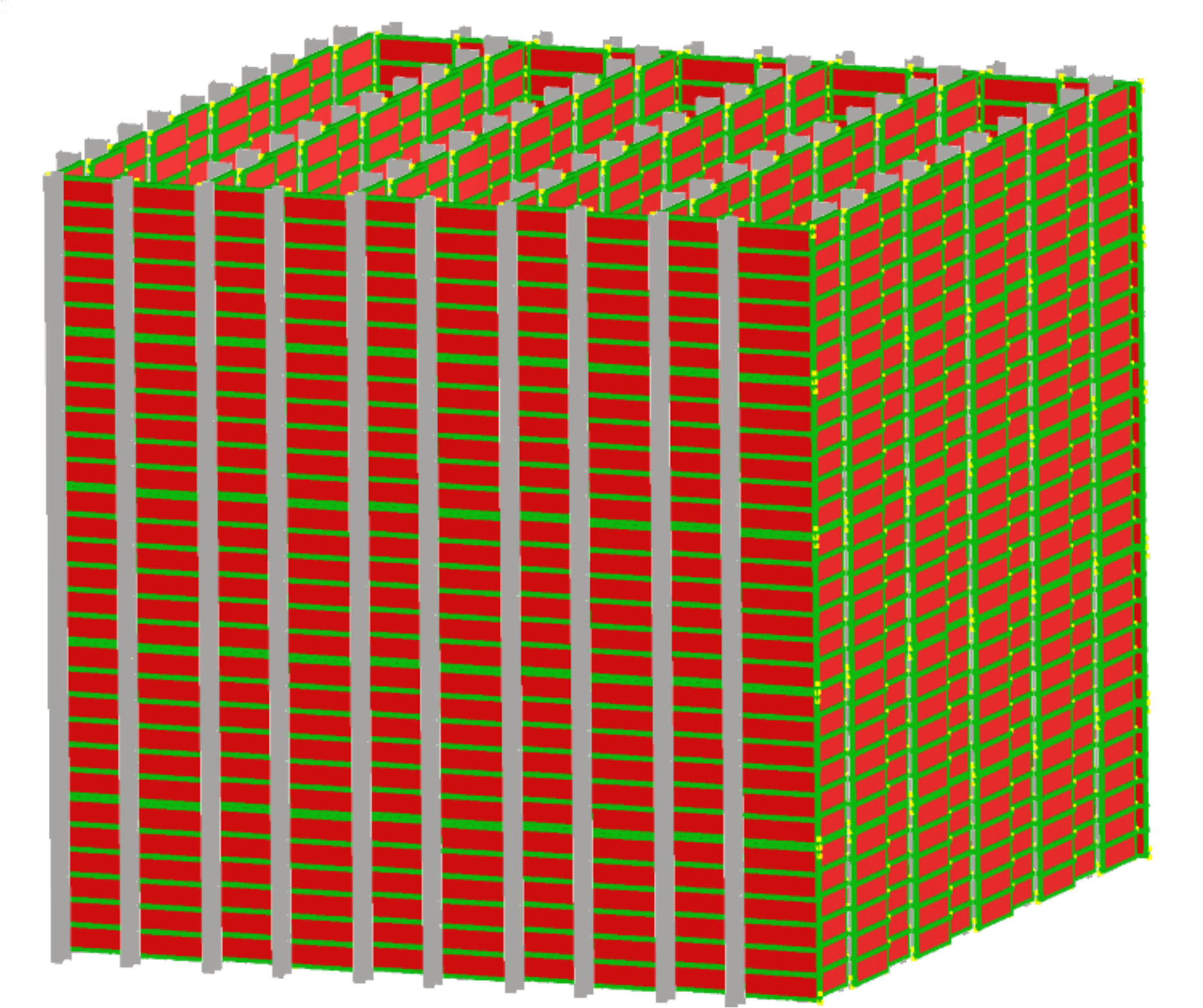}
  \caption{(left) Schematic of the LHCb cavern, with the proposed location of CODEX-b outlined in red. (right) Perspective drawing of the proposed CODEX-b detector design, with no optimization in place.\label{fig:codex-b}}
\end{figure}

The proposed CODEX-b detector~\cite{Gligorov:2017nwh} is a hermetic detector to be installed $25~\meter$ transverse to the LHCb (see \cref{sec:lhcb}) interaction point (IP8) on the LHC (see the left diagram of \cref{fig:codex-b}). CODEX-b is sensitive to long-lived particles (LLPs) over a wide range of production and decay mechanisms~\cite{Aielli:2022awh} at low LLP masses, long LLP mass lifetimes, and high center-of-mass energies. This parameter space has never been explored, and cannot be probed by existing experiments, since such low LLP masses and long LLP lifetimes are only accessible via a transverse detector. The CODEX-b proposal utilizes already existing infrastructure to provide a cost effective version of such a transverse detector.

In addition to its excellent sensitivity and low cost, CODEX-b also will interact with the LHCb data acquisition system, and store full LHCb events when the CODEX-b detector records an event of interest (see \cref{sec:lhcb} for details on LHCb). While the CODEX-b detector will interact closely with LHCb systems, it will remain separate, allowing participation from researchers across the LHC collaborations. The proposed detector design is shown in the right diagram of \cref{fig:codex-b}, and will consist of a $10\times10\times10~\meter^3$ cube instrumented with resistive plate chambers (RPCs). This detector technology has been chosen for its resolution, reliability, and relatively low cost, as a large area will need to be instrumented, including four internal planes. The mass of LLPs will be determined using the LLP decay vertex and opening angle~\cite{Curtin:2017izq}, eliminating the need for any magnetic field.

Detailed sensitivity of the studies have been performed and excellent reach well beyond existing bounds can be achieved for a wide range of models, including, but not limited to, dark photons via a Higgs portal, dark Higgses, axion-like particles, and heavy neutral leptons~\cite{Aielli:2019ivi}. Detailed simulation studies have been performed on expected background rates, including a dedicated measurement campaign at the proposed location~\cite{Dey:2019vyo}. Optimization studies of the detector design are underway, with current results indicating possible configurations with only $40\%$ of the instrumentation, while maintaining signal efficiency at the level of $50$ to $90\%$~\cite{Aielli:2022awh}.

The CODEX-b collaboration currently consists of approximately 50 members from 20 institutes, with a full organizational structure in place. The proposed timeline for CODEX-b begins with the installation and running of a smaller $2\times2\times2~\meter^3$ detector, CODEX-$\beta$ (see \cref{sec:codex-beta}), from 2022 to 2026 during Run 3 of the LHC. This will be used to validate the full designs and installation plans of the CODEX-b detector, with production targeted for 2025 and partial installation in 2026.

\subsection{CODEX-$\beta$}
\label{sec:codex-beta}

The CODEX-$\beta$ detector~\cite{Aielli:2022awh} is a proof-of-concept for the proposed CODEX-b detector (see \cref{sec:codex-b}), a transverse detector that will be located near the LHCb interaction point on the LHC. The purpose of CODEX-$\beta$ is to: validate background estimates from both simulation and a dedicated measurement campaign~\cite{Dey:2019vyo}; demonstrate integration with the LHCb data acquisition system; test the resistive plate chambers (RPC) to be used in CODEX-b, including spatial granularity and timing resolution; reconstruct and determine the rate for known standard model backgrounds; and prototype the mechanical structure for both the RPCs and the detector frame.

\begin{figure}[h]
  \centering
  \includegraphics[width=0.7\columnwidth]{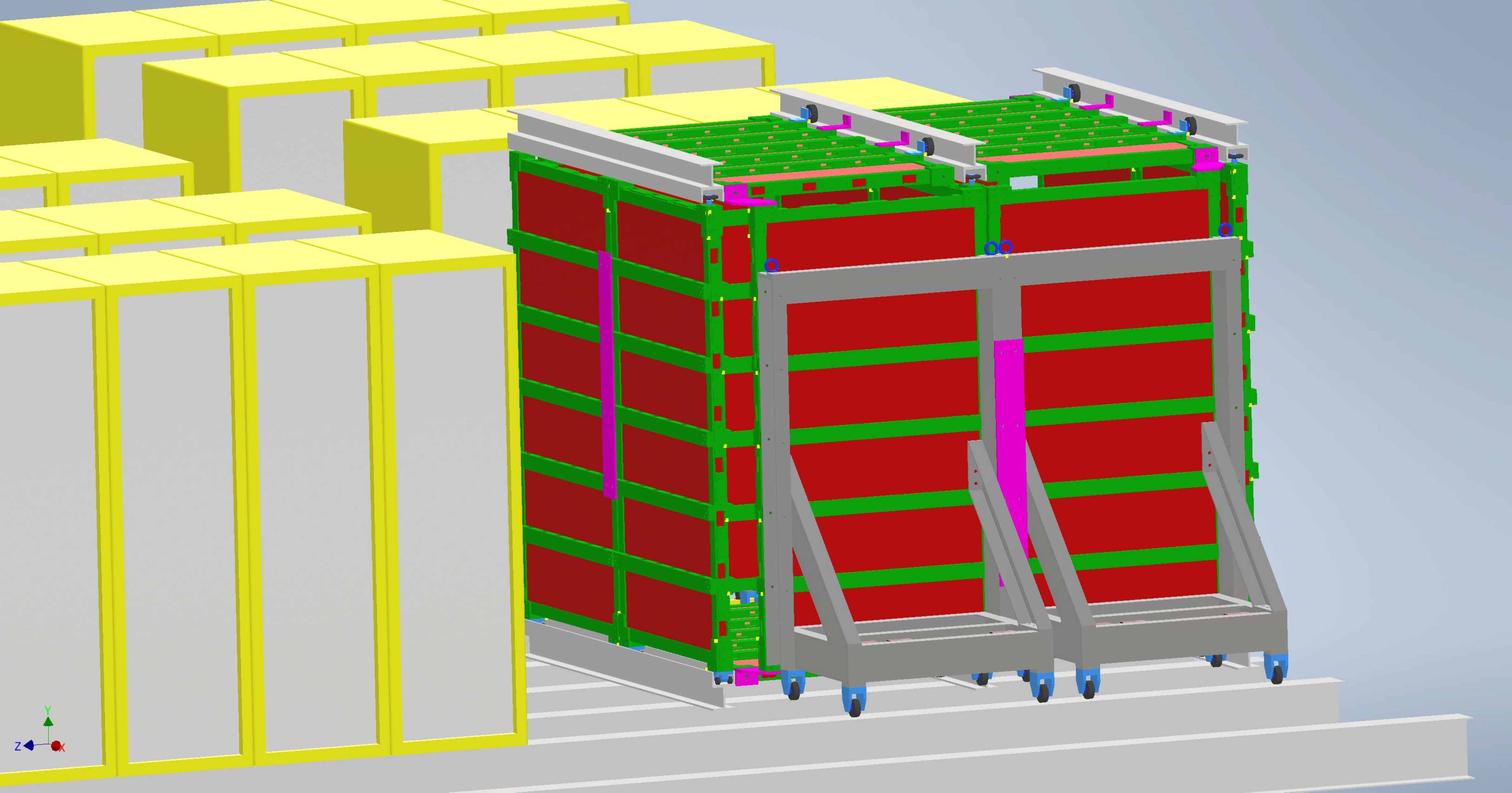}
  \caption{Diagram of the CODEX-$\beta$ detector installed in the IP8 counting barracks: (yellow) server racks, (gray) detector support structure, (red) RPC modules, and (green) RPC frames.\label{fig:codex-beta}}
\end{figure}

A part of the counting barracks located at IP8 which previously housed the LHCb high level trigger, will be used for CODEX-$\beta$ with a minimal impact on LHCb operation. Because this location is behind concrete shielding, the installation of CODEX-$\beta$ can occur while the beam is running, and not interfere with LHCb installation plans. The CODEX-$\beta$ detector will be a $2\times2\times2~\meter^3$ cube consisting of 14 RPCs, 2 RPCs per face and a central plan of 2 RPCs. The design of the detector is shown in \cref{fig:codex-beta}, where the yellow frames indicate the location of LHCb server racks.

The CODEX-$\beta$ detector is being built by the CODEX-b collaboration (for details see \cref{sec:codex-b}). A detailed schedule for the production and installation of the detector can be found in \rfr{Aielli:2022awh}. Production of RPC modules is expected to begin in late 2022, with installation of the detector in the first half of 2023 and commissioning during the second half of 2023. This schedule may shift to ensure no conflicts with LHCb installation or operation.

\subsection{DarkQuest}
\label{sec:darkQuest}

The DarkQuest experimental concept is a high sensitivity, near-term, modest-cost opportunity to explore new parameter space in dark-sector physics scenarios. DarkQuest is a proton fixed-target beam-dump spectrometer experiment on the neutrino-muon beamline of the Fermilab accelerator complex, where it would receive a high-intensity beam of $120~\GeV$ protons from the main injector. It takes advantage of the long history of investment by the Department of Energy (DOE) nuclear physics (NP) and high energy physics (HEP) programs in the existing E906/E1039 SeaQuest/SpinQuest spectrometer experiments at Fermilab, which focus on proton parton-distribution-function measurements in Drell-Yan events.

The DarkQuest detector concept proposes to add an electromagnetic calorimeter (EMCal) detector to the SpinQuest spectrometer that will open up two additional orders of magnitude in mass parameter space. An additional tracking layer is also proposed to extend the acceptance of the experiment and enable it to withstand higher instantaneous luminosity. This will allow DarkQuest to explore dark sector signatures from $\mathcal{O}(\MeV)$ to $\mathcal{O}(\GeV)$ in a variety of new final states, thus enabling DarkQuest to be a high impact dark sector experiment on the world stage.

DarkQuest can reach interesting parameter space in several dark sector scenarios. The purely standard model signals are studied in the context of dark photon, sterile neutrinos, and axion-like particle models (ALPs), while the dark matter (DM) and rich dark sectors are captured by models of inelastic dark matter  and  strongly interacting massive particles (SIMPs). The latter models offer the possibility of explaining the dark matter of the universe in a predictive framework with concrete experimental targets. Many more details on the physics sensitivity of the experiment are given in \rfrs{Berlin:2018pwi,Batell:2020vqn,Blinov:2021say}. DarkQuest can also play an important part of a potential program at Fermilab to explore how light new physics could contribute to $g-2$~\cite{g2workshop}.

The DarkQuest proponents have worked over the past few years to establish strong connections with the existing SpinQuest collaboration, perform detailed \textsc{Geant}-based~\cite{GEANT4:2002zbu} simulations to understand detector performance for dark sector signatures, and build a strong proto-collaboration to ensure the success of this experimental initiative. This is described in more detail in \rfr{Apyan:2022tsd}.

Further upgrades beyond DarkQuest are also envisioned. In the LongQuest proposal~\cite{Tsai:2019mtm}, three potential types of installations were proposed. They are: installing long-baseline detectors behind the iron block in the backroom to search for long-lived and millicharged particles; installing a ring-imaging Cherenkov detector or a hadron blind detector to improve particle identification and background reduction; and adding a new front-dump and fast-tracking detector to search for promptly decaying particles. More details are discussed in~\rfrs{Tsai:2019mtm,Apyan:2022tsd}. 

\begin{figure}[h]
  \centering
  \includegraphics[width=0.8\textwidth]{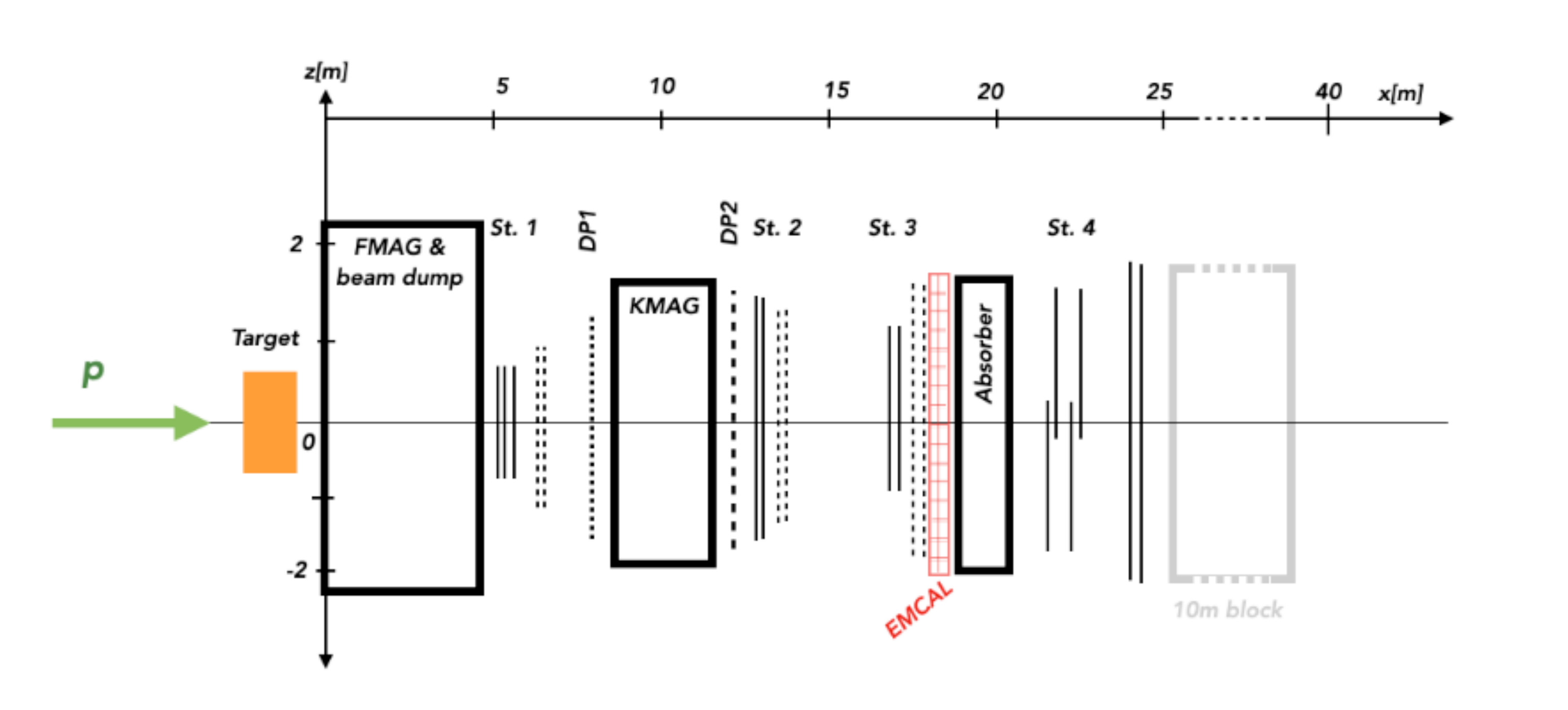}
  \caption{The proposed DarkQuest detector upgrade (in red) to the SpinQuest experiment.\label{fig:darkQuest}}
\end{figure}

\subsection{DarkMESA}
\label{sec:darkMesa}

\begin{figure}[h]
    \centering
    \includegraphics[width=0.8\textwidth]{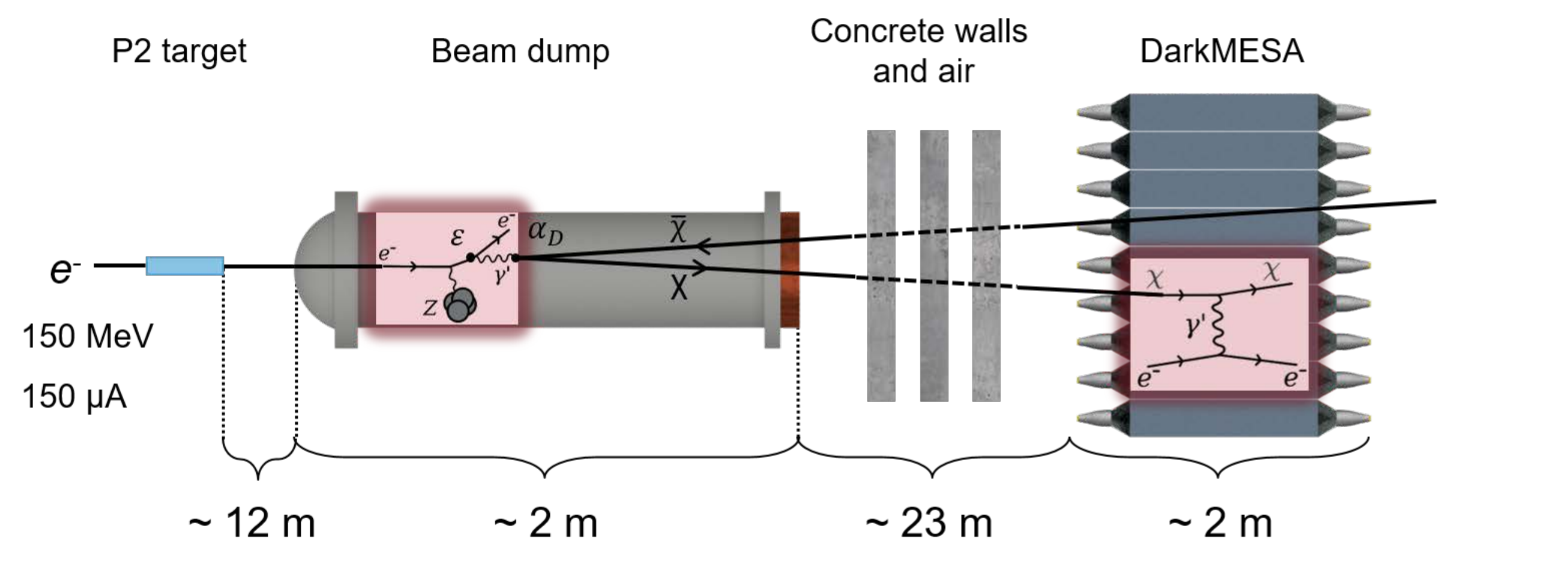}
    \caption{The main elements, from left to right, of the DarkMESA beam dump experiment at MESA, Germany: the electron beam penetrates a target and impinges upon the dump; electrons and positrons can undergo dark bremsstrahlung, which can then decay into $\chi\bar{\chi}$ pairs; and shielding will be located in front of the calorimetric DM scattering detectors.\label{fig:darkmesa}}
\end{figure}

DarkMESA is an electron beam dump experiment under development at the new MESA accelerator of the Johannes Gutenberg University in Mainz, Germany, to detect dark matter (DM) production through subsequent DM scattering. It focuses on exploring thermal DM interaction strengths in the mass range of \MeV to few tens of \MeV, see \cref{fig:darkmesa} for a schematic.

Using the funding made available to the Cluster of Excellence ``Precision Physics, Fundamental Interactions and Structure of Matter'' (PRISMA+), the civil construction of the  accelerator is expected to be completed in 2022 with first beam operation and data taking foreseen to start in 2023. The accelerator and key experiments will include a new experimental hall provided by the ``Center for Fundamental Physics'' and will benefit from the unique local research infrastructure of the University campus; for more than 50 years the Institute for Nuclear Physics has constructed and operated high-intensity electron accelerators, most notably the MAinz MIkrotron (MAMI).

The DM scattering detectors will be placed $23~\meter$ behind the beam dump, shielded from neutrons and other standard model particles. DarkMESA has the unique advantage that neutrino production is kinematically forbidden at the MESA beam energy of $150~\MeV$. The detectors will consist of Cherenkov light radiators complemented by a plastic scintillator veto system for rejecting cosmogenic backgrounds. The advantage of Cherenkov crystals is their speed and relatively low sensitivity to background neutrons. In the near future the experiment will demonstrate its feasibility with a 25-crystal prototype as Phase A. In Phase B, a 1000-crystal PbF$_2$ calorimeter and a 1000-block lead glass calorimeter will be operated.

DarkMESA will run parasitically to the scheduled program for parity-violation experiments covering $20,000$ hours of operation time at a $150~\si{\micro\ampere}$ beam current. Simulations have shown that already in these phases it will be possible to reach the thermal relic targets in an unexplored parameter space of DM masses. Details on the detector technology and the sensitivity of the experiment within the dark photon model are given in \rfrs{Doria:2019sux,Christmann:2020qav}. The DarkMESA sensitivity to axion-like particle models is under study. Currently, alternative and more innovative detector concepts are being investigated, among them a low-pressure, negative-ion, time-projection-chamber and a calorimeter based on radiation shielding window glasses.  The DarkMESA group has worked in the past few years to establish strong connections with the existing BDX collaboration at Jefferson Lab.

\subsection{FASER}
\label{sec:faser}

FASER, the Forward Search Experiment~\cite{Feng:2017uoz}, is an LHC experiment that is composed of two detectors: the main FASER detector, designed to search for new long-lived particles, and FASER$\nu$ (see \cref{sec:faserNu}), designed to study $\TeV$-energy collider neutrinos for the first time. Both FASER and FASER$\nu$ are constructed and installed, and will begin collecting data in LHC Run 3 from mid-2022 to 2025.  

\begin{figure}[h]
  \centering
  \includegraphics[width=0.57\textwidth]{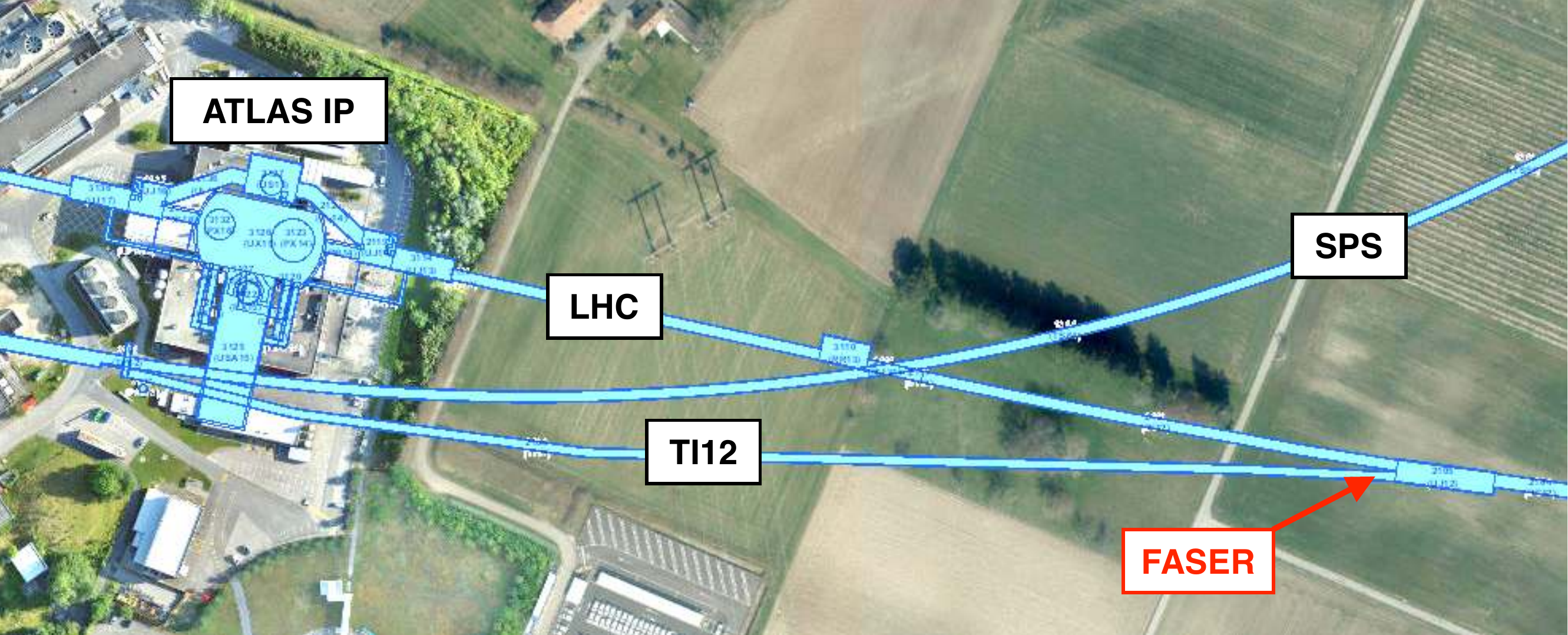}\hfill
  \includegraphics[width=0.41\textwidth]{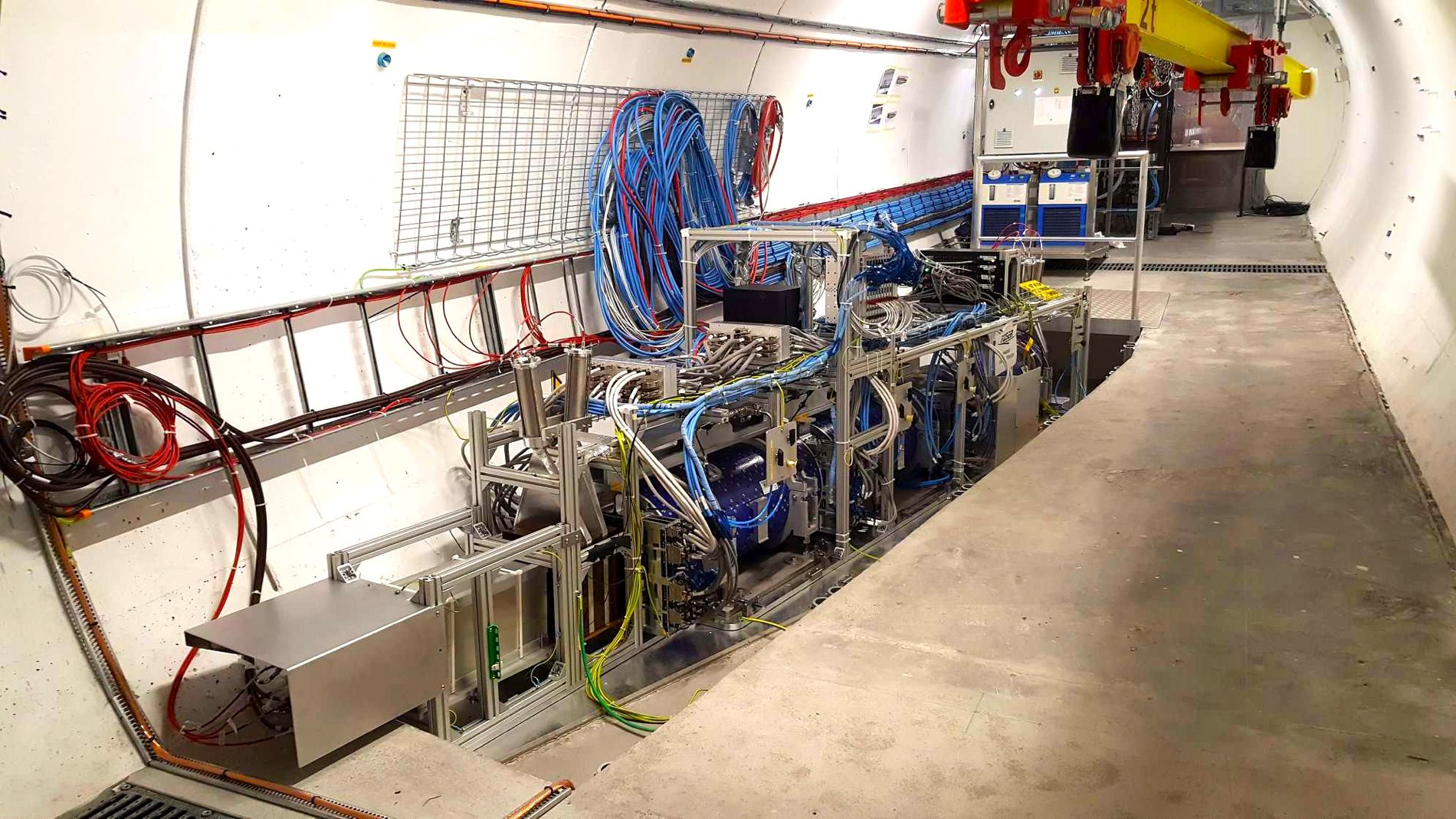}
  \caption{(left) The arrow points to FASER's location in service tunnel TI12, roughly $480~\meter$ east of the ATLAS interaction point. (right) FASER as currently installed in TI12.\label{fig:FASERinfrastructure}} 
\end{figure}

The main FASER detector is unique in being the only detector dedicated to new particle searches in the far-forward direction at the LHC. FASER is located in the TI12 side tunnel, along the beam collision axis, $480~\meter$ to the east of the ATLAS interaction point, see \cref{fig:FASERinfrastructure}. It covers pseudorapidities $\eta > 9.2$, and so is sensitive to particles that are produced along the beam-pipe and escape all existing large LHC detectors. Starting with the first \invfb of integrated luminosity, FASER will begin probing new regions of parameter for many well known models, in which light and weakly-interacting particles are produced at the ATLAS IP, pass through roughly $100~\meter$ of concrete and rock, and then decay to visible particles, such as electrons, muons, photons, and hadrons, in the FASER detector volume. FASER has discovery potential for a wide range of proposed new particles, including dark photons, light gauge bosons, and axion-like particles~\cite{Feng:2017uoz, Feng:2017vli, Kling:2018wct, Feng:2018noy, Berlin:2018jbm, FASER:2018eoc, Kling:2021fwx}.

The FASER collaboration currently consists of approximately 75 collaborators from 21 institutions in 9 countries. FASER was proposed in 2017~\cite{Feng:2017uoz}, and the collaboration was formed in 2018. The FASER letter of intent~\cite{FASER:2018ceo} and technical proposal~\cite{FASER:2018bac} were submitted later that year, and FASER was constructed from 2019--2021.  The detector has been commissioned with cosmic rays and test beam data. More details describing the detector as built may be found in \rfrs{FASER:2021cpr,FASER:2021ljd,FASER:inprep}.

\subsection{FASER2}
\label{sec:faser2}

The existing FASER experiment (see \cref{sec:faser}) is already set to probe new phase space in the search for beyond the standard model (BSM) physics. However, the overall size of FASER, and therefore its possible decay volume, has been heavily constrained since the initial stages of planning, by the available space underground. This directly affects the sensitivity and reach obtainable by FASER, as for many representative BSM models the sensitivity is directly related to the length and radius of the decay volume. This strongly motivates the case for an enlarged detector, FASER2, which was already explored in the FASER letter of intent~\cite{Ariga:2018zuc}, technical proposal~\cite{Abreu:2020ddv}, and physics reach~\cite{Ariga:2018uku} documents. 

\begin{figure}[h]
  \centering
  \includegraphics[width=0.75\textwidth]{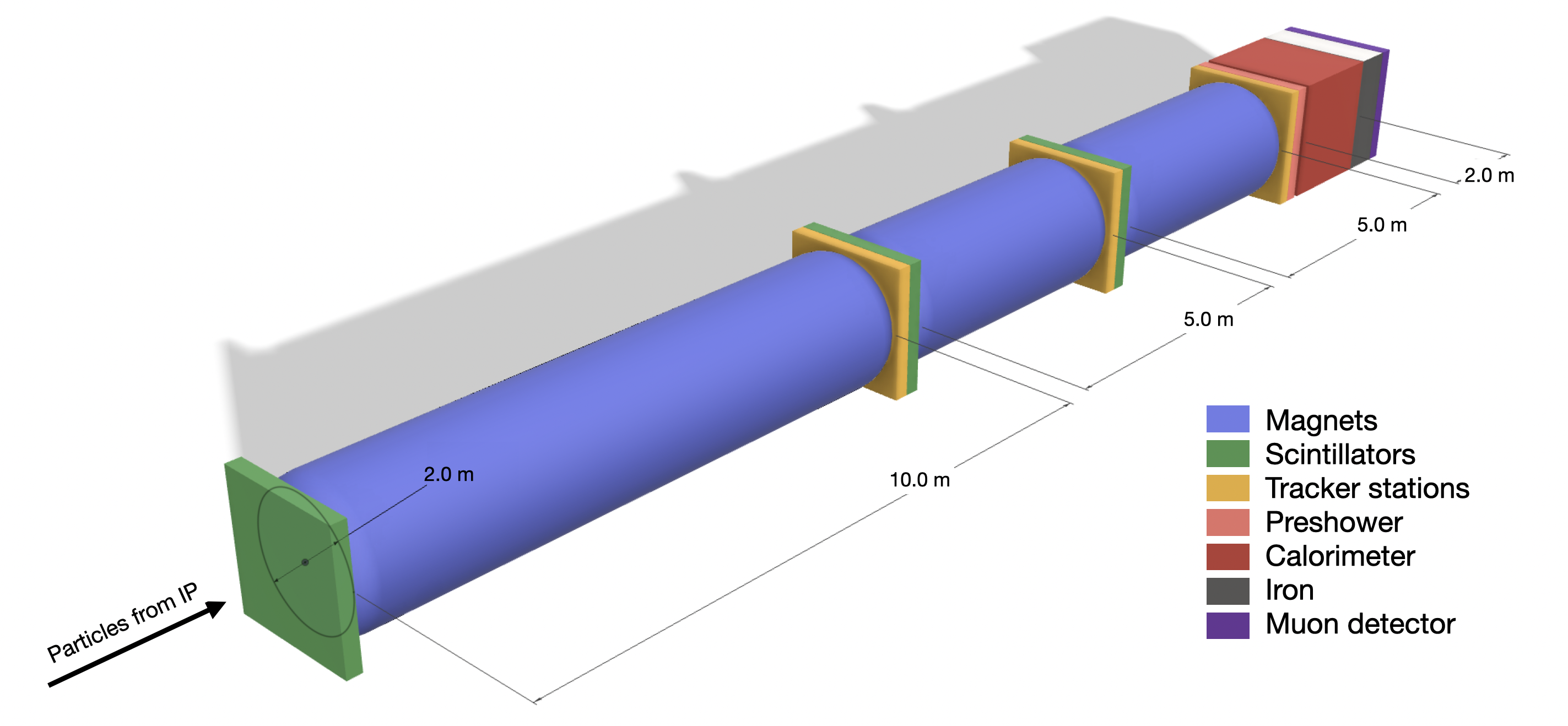}
  \caption{Schematic diagram of the proposed FASER2 detector.\label{fig:FASER2-Design}}
\end{figure}

The FASER2 design can be optimized for either Alcove or Cavern Forward Physics Facility (FPF) scenarios~\cite{Feng:2022inv} (see \cref{sec:fpf}). The design is not yet strictly defined but it will be similar to FASER in its general philosophy, modulo changes needed to ameliorate some of the additional challenges already described. A schematic layout of the FASER2 detector, assuming the Cavern FPF scenario, is given in \cref{fig:FASER2-Design}. The veto system will be scintillator-based, similar to FASER. The significantly increased area of the active volume makes it impractical to use silicon tracker technology. A SiPM and scintillating fiber tracker technology, such as LHCb's SciFi detector~\cite{Hopchev:2017tee} (see \cref{sec:lhcb}), is a strong candidate to replace the ATLAS SCT modules used in FASER. In addition, Monitored Drift Tube (MDT) technology, similar to that used in the ATLAS New Small Wheel~\cite{Kawamoto:1552862}, is also being considered, although this option requires the use of gases in the LHC tunnel that could be problematic for the Alcove scenario. Superconducting magnet technology would be required to maintain sufficient field strength across the much larger aperture; suitable technology for this already exists and can be built for FASER2. There are several possibilities for the cooling of such magnets, the use of cryocoolers and the possibility to share a single cryostat across several magnets are being considered. The calorimeter must have: sufficient spatial resolution to be able to identify particles at $\sim1-10~\si{\milli\meter}$ separation; good energy resolution; improved longitudinal separation with respect to FASER; and the capability to perform particle identification, separating \eg electrons and pions. Dual readout calorimetry~\cite{Lee:2017xss,Antonello:2018sna} is a good candidate to satisfy all these requirements. Finally, the ability to identify separately electrons and muons would be very important for signal characterization, background suppression, and for the interface with FASER$\nu$2 (see \cref{sec:faserNu2}. To achieve this, a mass of iron will be placed after the calorimeter, with sufficient depth to absorb pions and other hadrons, followed by a detector for muon identification.

As already discussed  the sensitivity to dark photons of the default FASER2 configuration has a significant increase with respect to FASER. However, the Alcove FPF scenario does not allow for such a large detector, while an FPF with a dedicated cavern is able to recover and even improve upon the default FASER2 sensitivity, making it strongly the preferred scenario. The only downside to the Cavern scenario is the slight reduction in sensitivity due to increased distance from the interaction point, but this is a rather small effect. Similar conclusions can be drawn for dark Higgs sensitivity, where the effect of the increased radius is even stronger due to the enhancement in acceptance to $B$-meson decays already discussed.

A key metric to define the requirements on the tracking detector technology is the spatial resolution required to efficiently separate the long-lived particle decay (LLP) products. For a requirement of $10~\si{\milli\meter}$ separation there is a sufficiently small reduction in sensitivity such that the tracker technologies under consideration would be more than sufficient. The particle separations are also large enough at the calorimeter that even a relatively coarse granularity could be sufficient given the size field strength of the proposed magnets.

To conclude, the physics potential of a larger scale successor to FASER is clear. Possible scenarios for this larger detector are being explored and initial studies strongly indicate a preference for a FPF with a dedicated new cavern. Much progress has been made on defining the possible FASER detector designs and identifying detector technologies. Several studies are still required to finalize the design boundaries of FASER2, such as understanding the physics needs and possible detector performance capabilities for LLP mass and pointing reconstruction and particle identification.

\subsection{FASER$\nu$}
\label{sec:faserNu}

FASER$\nu$~\cite{FASER:2019dxq,FASER:2020gpr} is designed to directly detect collider neutrinos of all three flavors, and provide new measurements of their cross-sections at $\TeV$ energies, higher than those seen from any previous artificial sources.

FASER$\nu$ can provide measurements of neutrino cross-sections at the $\TeV$ energy scale, where they are currently poorly constrained, and tests of lepton universality in high-energy neutrino scattering. In addition to the measurements of charged-current interactions, neutral-current interactions can be measured. Such measurements can probe new physics that change the neutrino interaction rate, such as non-standard interactions~\cite{Ismail:2020yqc}. FASER$\nu$ can also study other beyond the standard model physics scenarios, including new light particles that can decay into $\tau$ neutrinos~\cite{Kling:2020iar} and sterile neutrino oscillations~\cite{FASER:2019dxq, Bai:2020ukz}. Furthermore, FASER$\nu$ measurements of electron neutrinos at high energies above $\sim500~\GeV$, mainly originating from charm decays~\cite{Kling:2021gos}, can provide the first data on forward charm production as valuable input for QCD.

\begin{figure}[h]
  \centering
  \includegraphics[width=0.8\textwidth]{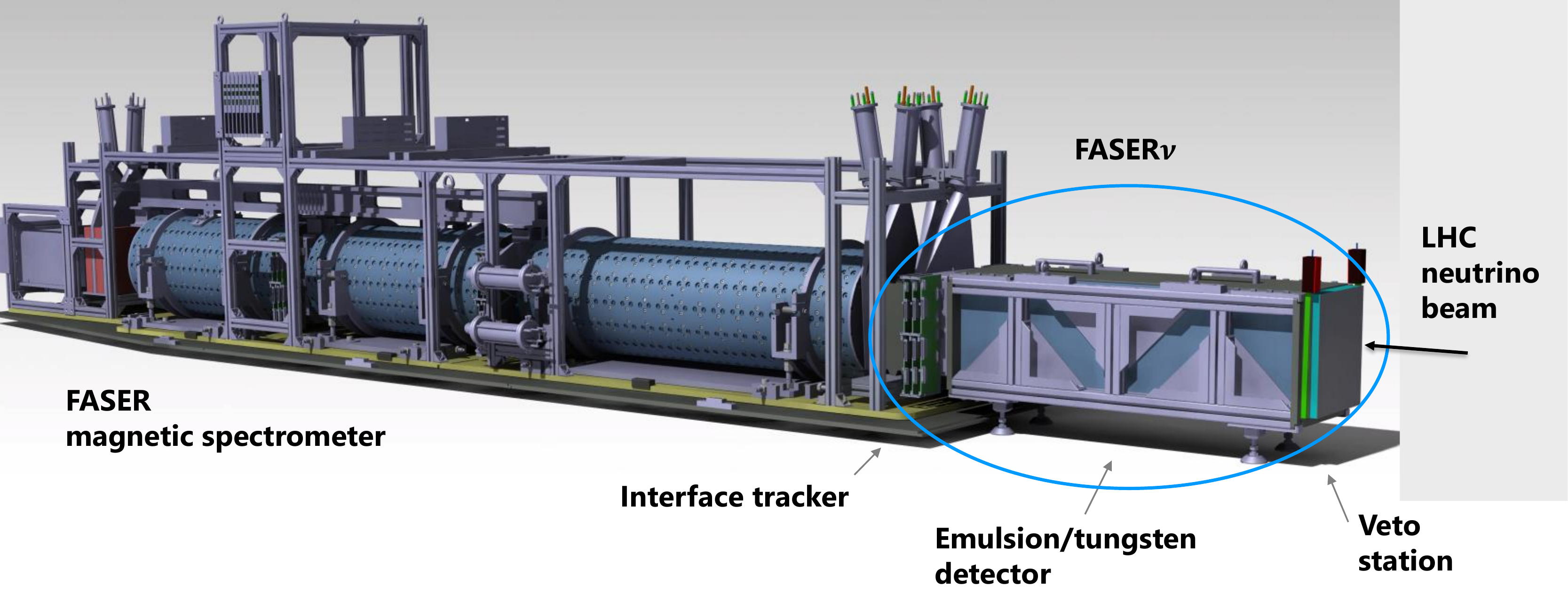}
  \caption{Sketch of the FASER detector, highlighting the FASER$\nu$ detector.\label{fig:fasernu_detector}}
\end{figure}

In Run 3 of the LHC, proton-proton collisions at a center-of-mass energy of $14~\TeV$, and with an expected integrated luminosity of $150~\invfb$, will produce a high-intensity beam of $\mathcal{O}(10^{12})$ neutrinos in the far-forward direction with a mean interaction energy of $\sim 1~\TeV$. The FASER$\nu$ detector is located in front of the main FASER detector (see \cref{sec:faser}) along the beam-collision axis. A sketch of the FASER$\nu$ detector is shown in \cref{fig:fasernu_detector}. The FASER$\nu$ detector includes a veto station, an emulsion/tungsten detector, and an interface tracker coupled to the FASER magnetic spectrometer. The FASER$\nu$ emulsion detector has: the ability to identify different lepton flavors; sufficient target material to identify muons; finely sampled detection layers to identify electrons and to distinguish them from gamma rays; and good position and angular resolutions to detect $\tau$ and charm decays. The detector can also measure the momenta of muons and hadrons, the energy of electromagnetic showers, and estimate neutrino energy.

In 2018 during LHC Run 2, a pilot run was performed in TI18 of the LHC to demonstrate neutrino detection at the LHC for the first time. The depth of the pilot detector was only $0.6~\lambda_{int}$, which is much shorter than the $8~\lambda_{int}$ of the FASER$\nu$ detector being prepared for LHC Run 3. The data from the pilot detector are used to demonstrate the feasibility of high-energy neutrino measurements in this experimental environment. Neutrino interactions were searched by analyzing the data corresponding to $11~\si{\kilo\gram}$ of the target mass. The first neutrino interaction candidate events at the LHC were observed~\cite{FASER:2021mtu}. These results pave the way for current and future collider neutrino experiments. 

In 2022--2025 during LHC Run 3, FASER$\nu$ is expected to collect $\sim 2,000$ $\nu_e$, $\sim 6,000$ $\nu_{\mu}$, and $\sim 40$ $\nu_{\tau}$ charged-current interactions in FASER$\nu$~\cite{Kling:2021gos}, along with neutral-current interactions. In March 2022, the first physics run module was installed into the tunnel, and data taking will commence soon.

\subsection{FASER$\nu$2}
\label{sec:faserNu2}

FASER$\nu$2 is a proposed experiment that will utilize the large flux of $\TeV$-energy neutrinos produced in LHC collisions to perform neutrino measurements and search for signs of new physics. It is designed as a successor to the FASER$\nu$ detector~\cite{FASER:2019dxq, FASER:2020gpr} (see \cref{sec:faserNu}) and uses an emulsion-based technology.

\begin{figure}[h]
  \centering
  \includegraphics[width=0.8\textwidth]{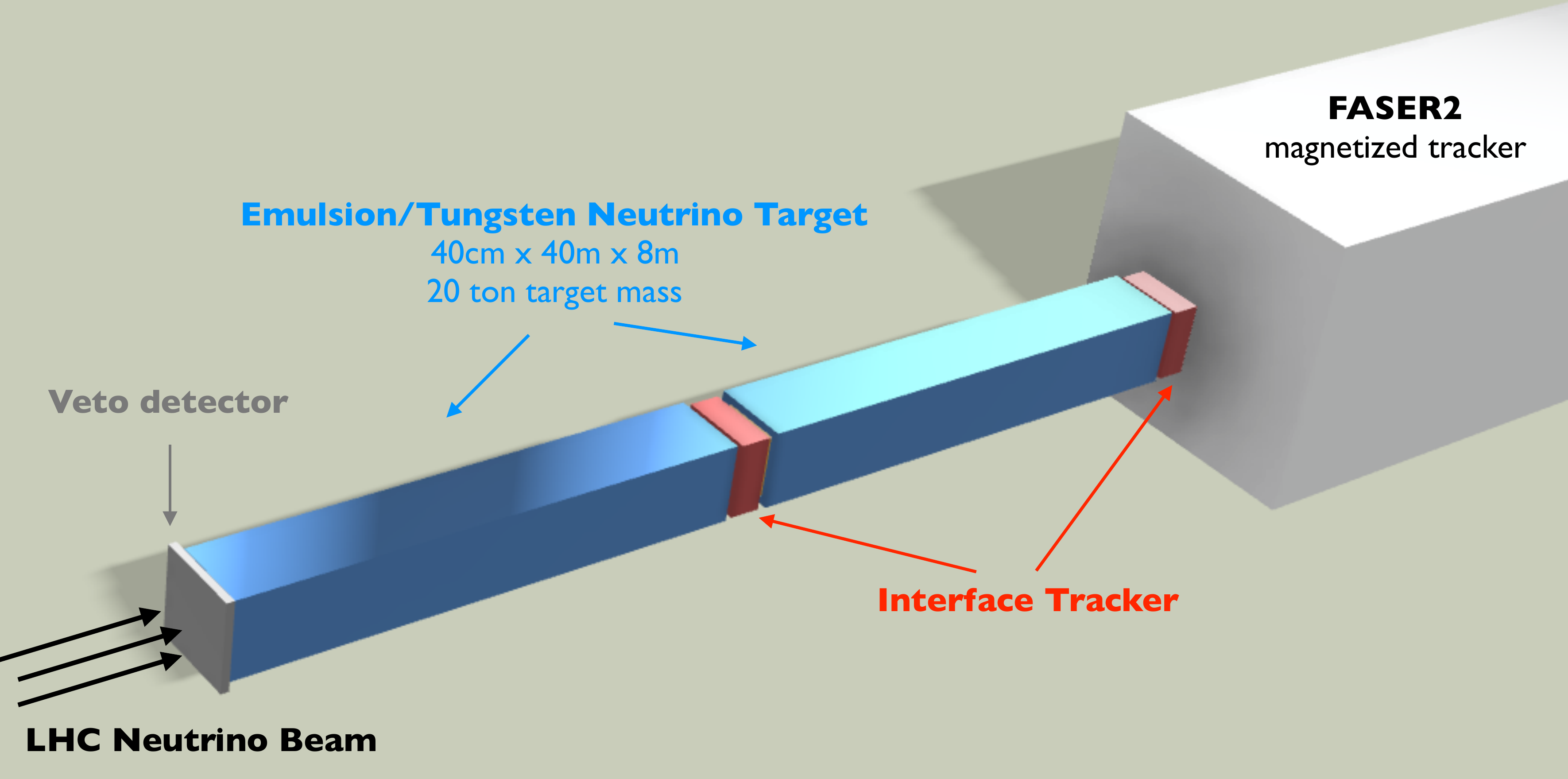}
  \caption{Conceptual design of the FASER$\nu$2 detector.\label{fig:fasernu2}}
\end{figure}

During the high luminosity era, the LHC will collide protons with protons at a center-of-mass energy of $14~\TeV$. These collisions will produce an intense and strongly-collimated neutrino beam of all three flavors with \TeV energies in the forward direction. FASER$\nu$2 will be located about $620~\meter$ downstream from ATLAS in a purpose build Forward Physics Facility (FPF)~\cite{MammenAbraham:2020hex, Anchordoqui:2021ghd, Feng:2022inv} (see \cref{sec:fpf}). It will be placed in the center of the neutrino beam where the event rate of neutrino interactions for all three flavors is maximized. A schematic view of the detector is presented in \cref{fig:fasernu2}. The main component of the FASER$\nu$2 detector is $\sim 3300$ emulsion layers that are interleaved with $2~\si{\milli\meter}$ thick tungsten plates corresponding to a target mass of 20 tonnes. During the HL-LHC, with a nominal integrated luminosity of $3~\invab$, more than $10^5$ electron neutrinos, $10^6$ muon neutrinos, and $10^3$ $\tau$ neutrinos are expected to interact in the detector~\cite{Anchordoqui:2021ghd,Kling:2021gos}. Additionally, the detector design includes a front veto and two interface tracking stations. This would permit a combined analysis with the FASER2 detector (see \cref{sec:faser2}) which is located behind FASER$\nu$2, and would: reject possible backgrounds associated with incoming muons; identify the muon change, and provide an additional measurement of the muon momentum. An important issue is the sizable flux of LHC muons traversing the detector, which limits the possible exposure time. To address this issue, the installation of an upstream sweeper magnet to deflect these muons is considered. 

FASER$\nu$2 will provide extraordinary opportunities for a broad range of neutrino studies: measurements of neutrino fluxes as probes of forward particle production, which provide valuable input for QCD and astroparticle physics~\cite{Bai:2021ira,Anchordoqui:2022fpn}; the study of neutrino interactions at \TeV energies, which includes the measurement of poorly-constrained neutrino cross-sections at the \TeV energy scale, tests of lepton universality in neutrino scattering, and measurements of hadronic and nuclear effects in the initial and final state (probing similar physics as the EIC, but with neutrino beams)~\cite{Mosel:2022tqc}; and precision measurements of $\tau$ neutrino properties.  In addition, FASER$\nu$2 will be able to probe a variety of beyond the standard model physics scenarios, including: new physics that changes the neutrino interaction rate, such as non-standard interactions or light new mediators~\cite{Ismail:2020yqc,Falkowski:2021bkq}; neutrino dipole moments~\cite{Ismail:2021dyp}; new light particles that can decay into $\tau$ neutrinos and modify neutrino fluxes~\cite{Bahraminasr:2020ssz,Kling:2020iar,Ansarifard:2021dju}; new neutrino-philic mediators that are produced in neutrino interactions~\cite{Kelly:2021mcd,Berryman:2022hds}; sterile neutrino oscillations~\cite{FASER:2019dxq,Bai:2020ukz}; scattering of light dark matter~\cite{Batell:2021blf,Batell:2021aja,Batell:2021snh,Kling:2022ykt}; and the decay of long-lived particles~\cite{Jodlowski:2019ycu,Jodlowski:2020vhr}. Generically, in these models FASER$\nu$2 probes the range where the new particles have masses between an \MeV and a few \GeV. 

The FPF~\cite{Feng:2022inv} is envisioned to be constructed during Long Shutdown~3, while the detector would be installed at the beginning of LHC Run~4. Data taking would start soon after. The collaboration would form out of the existing FASER collaboration. For more information on the detector and physics case see \rfrs{Anchordoqui:2021ghd,Feng:2022inv}.

\subsection{FerMINI}
\label{sec:fermini}

\begin{figure}[h]
  \centering
  \includegraphics[width=0.9\textwidth]{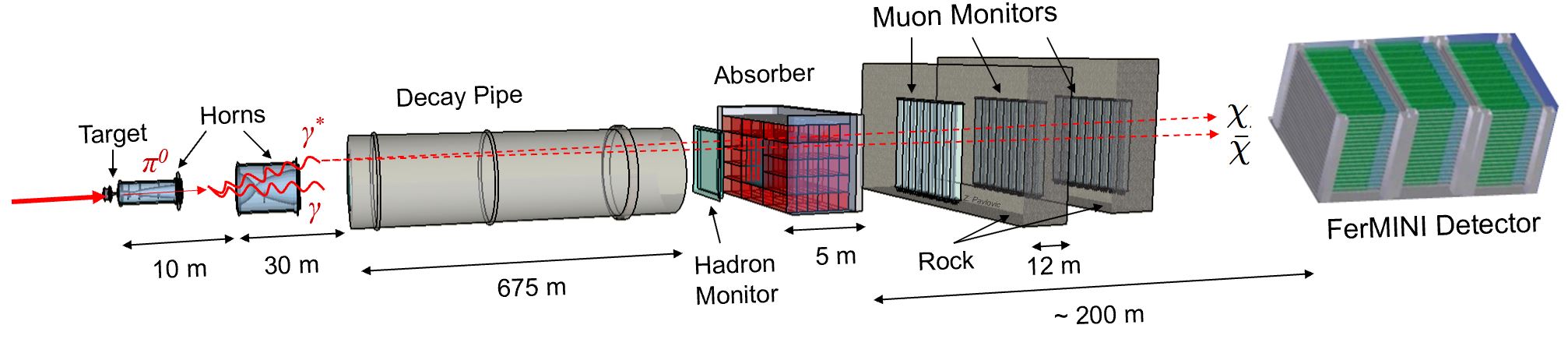}
  \caption{A figure of the FerMINI experiment placed downstream of the NuMI beam to search for millicharged particles.~\cite{Kelly:2018brz}}
  \label{fig:FerMINI}
\end{figure}

High-intensity beam facilities can be used to host symbiotic detectors to conduct dark sector searches. One of such proposal is FerMINI~\cite{Kelly:2018brz}, see \cref{fig:FerMINI}, placing a milliQan-like detector in any high-intensity proton beam-dump/fixed-target facilities, to search for millicharged particles. In \rfr{Kelly:2018brz}, two specific sites are considered, one is at the Long Baseline Neutrino Facility (LBNF)~\cite{Papadimitriou:2016ksv}, one is in the NuMI beam facility~\cite{Adamson:2015dkw}. LBNF has the advantage of both high energy ($120~\GeV$) and high intensity ($\sim$ $10^{22}$ total protons on target (POT)), while the NuMI beam absorber provides additional production near the detector site, which enhances the mCP flux. One can also use the main injector beam of SpinQuest~\cite{SeaQuest:2019hsx} and the booster beam~\cite{Machado:2019oxb}.

Beyond the scope of \rfr{Kelly:2018brz}, a new strategy were developed in J-PARC to enhance the sensitivity to mCPs for milliQan-like detectors. The proposal, SUBMET at J-PARC~\cite{Choi:2020mbk}, developed the technique to further enhance the sensitivity at the low-mass mCP region, by using the excellent timing information of a fixed-target beam to reduce the number of layers of bay detectors down to two. This strategy can be applied to FerMINI as well to enhance its low-mass mCP sensitivity.

\subsection{FLArE}
\label{sec:flare}

The Forward Liquid Argon Experiment (FLArE) has been proposed to perform high-energy neutrino measurements and beyond the standard model (BSM) searches, including for light sub-\GeV dark matter (DM) particles, in the far-forward region of the LHC. The detector will be placed in the proposed Forward Physics Facility (FPF, see \cref{sec:fpf}) new cavern~\cite{Anchordoqui:2021ghd,Feng:2022inv} and will operate during the HL-LHC era. It will employ liquid argon (LAr) time projection chamber (TPC) technology to identify products of scatterings of SM and BSM species which can be abundantly produced in proton-proton collisions at the LHC. Such light and high-energy particles are preferentially produced with small transverse momenta and travel along the beam collision axis. In their scatterings, they can deposit energies ranging from tens of \MeV to about one \TeV, therefore a flexible detector technology is required to study such interactions and encompass a broad physics program of the FPF. 

To answer these needs, the FLArE detector will take advantage of the excellent spatial resolution (\si{\milli\meter}-scale) and electromagnetic calorimetry of LAr TPCs. In addition, event triggering will be achieved with an additional light collection system, while high-energy signal containment could be improved with the use of a downstream hadronic calorimeter and a muon range detector. The use of liquid krypton (LKr) has also been envisioned which could offer improved high-energy event calorimetry due to a lower radiation length in LKr than in LAr. During the entire HL-LHC phase, FLArE is expected to measure as many as a million charged-current neutrino interactions with an average energy of several hundred \GeV, including $\sim 10^3$ scatterings of $\tau$ neutrinos. This will significantly extend the neutrino physics program at the LHC to be initiated during Run 3 with the FASER$\nu$ (see \cref{sec:faserNu}) and SND@LHC (see \cref{sec:snd}) experiments. The thermal relic targets of popular light DM models will also be studied with up to $\mathcal{O}(100)$ of expected events that can be differentiated from neutrino-induced backgrounds based on much smaller expected energy depositions and with the use of a combination of various scattering signatures~\cite{Batell:2021blf,Batell:2021aja,Batell:2021snh,Kling:2022ehv,Kling:2022ykt}.

\begin{figure}[h]
  \centering
  \includegraphics[width=0.9\textwidth]{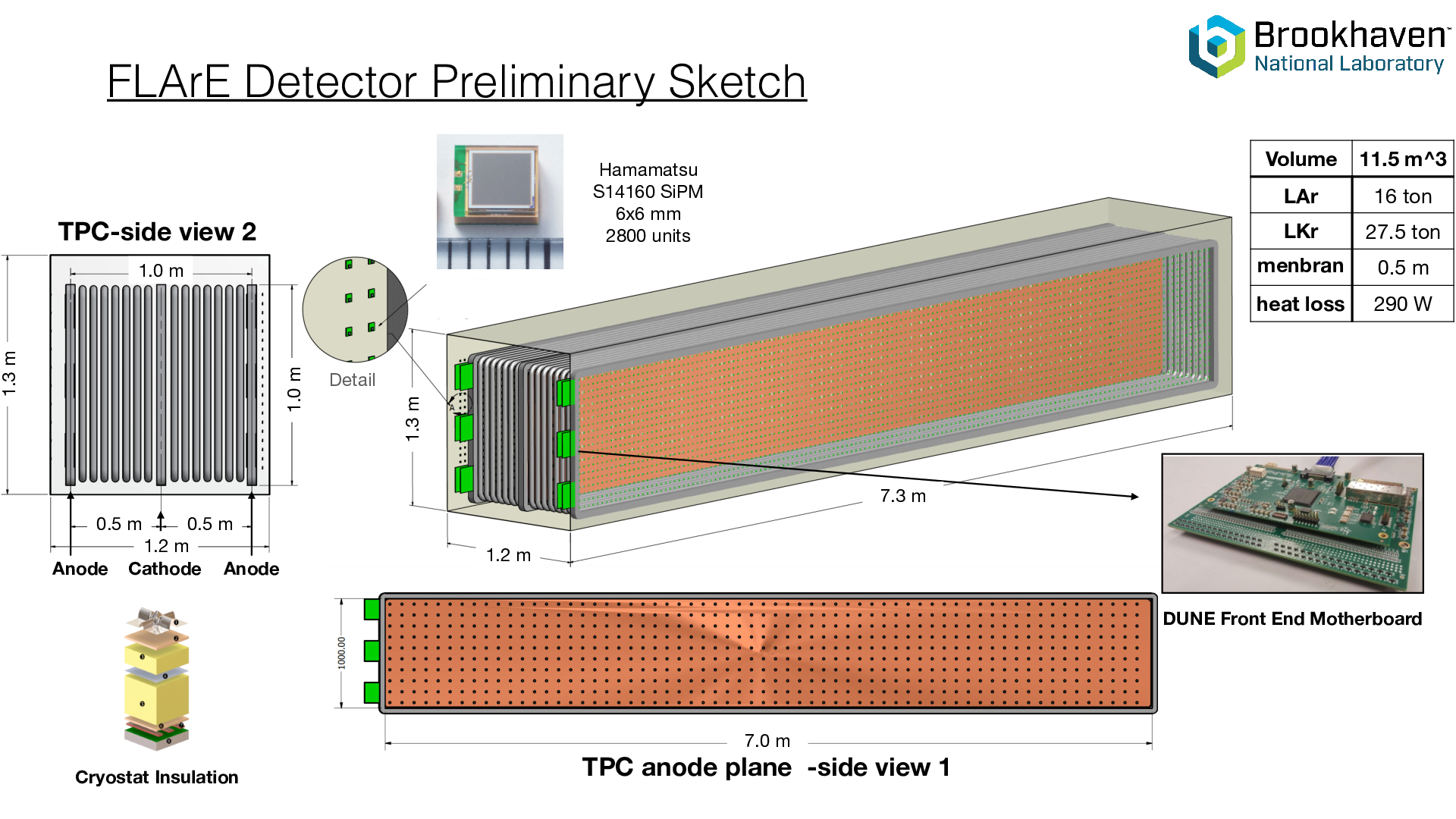}
  \caption{Preliminary conceptual design of the FLArE detector. Taken from Ref.~\cite{Feng:2022inv}.\label{fig:flare}}
\end{figure}

The preliminary conceptual design of FLArE is given in \cref{fig:flare}. The experiment will employ a $7~\meter$-long detector with a $1~\meter^2$ transverse size and about $10$~ton fiducial mass scale for LAr target material (about $24$~ton for LKr). The LAr TPC design will include a central cathode and two anode planes on two sides of the detector parallel to the LHC beam axis. The envisioned electric field to provide a drift field for ionization electrons will be at the level of $\sim500~\volt/\si{\centi\meter}$. This results in the drift time of about $0.3~\si{\milli\second}$ assuming a $0.5~\meter$-long drift. For the TPC, a readout using wires or pixels is currently being considered, including the use of several planes of wires to improve efficient 3D event reconstruction~\cite{Qian:2018qbv}. The light collection and detection system is expected to be able to identify events originating from inside the detector fiducial volume and simultaneously keep a low trigger rate of order $1~\Hz$. The use of photomultiplier tubes or silicon-photomultiplier (SiPM) sensors is considered, with the latter following the DUNE detector design~\cite{DUNE:2018hrq}. The main LArTPC will be installed in a membrane cryostat with inner dimensions of $2.1\times2.1\times8.2~\meter^3$ and with passive insulation to allow high voltage safety. The design and installation of the cryosystem in the FPF will benefit from the experience of running the protoDUNE experiment at CERN~\cite{DUNE:2021hwx}. Further efforts towards the FLArE detector design and research and development studies are currently ongoing, including detector response simulations with a dedicated \textsc{Geant4}~\cite{GEANT4:2002zbu} model.

During its operation, FLArE is expected to collect data from more than 500 high-energy neutrino interactions per day. These large event statistics will allow for measuring high-energy neutrino interaction cross-sections with very high precision, as well as, indirectly, it will provide unique information to constrain predicted forward meson spectra at the LHC. The latter has important implications for QCD studies and cosmic-ray physics, \cf \rfr{Feng:2022inv} for further discussion. It will also constrain the parameter space of popular light DM models, including scenarios that remain challenging to probe in DM direct detection experiments due to their suppressed non-relativistic scattering rates. The first such bounds going beyond current constraints are expected with the initial $300~\invfb$ of data, \ie with about a tenth of the total duration of the HL-LHC phase.

\subsection{FNAL-$\mu$}
\label{sec:fnalMu}

\begin{figure}[h]
  \centering
  \includegraphics[width=0.5\textwidth]{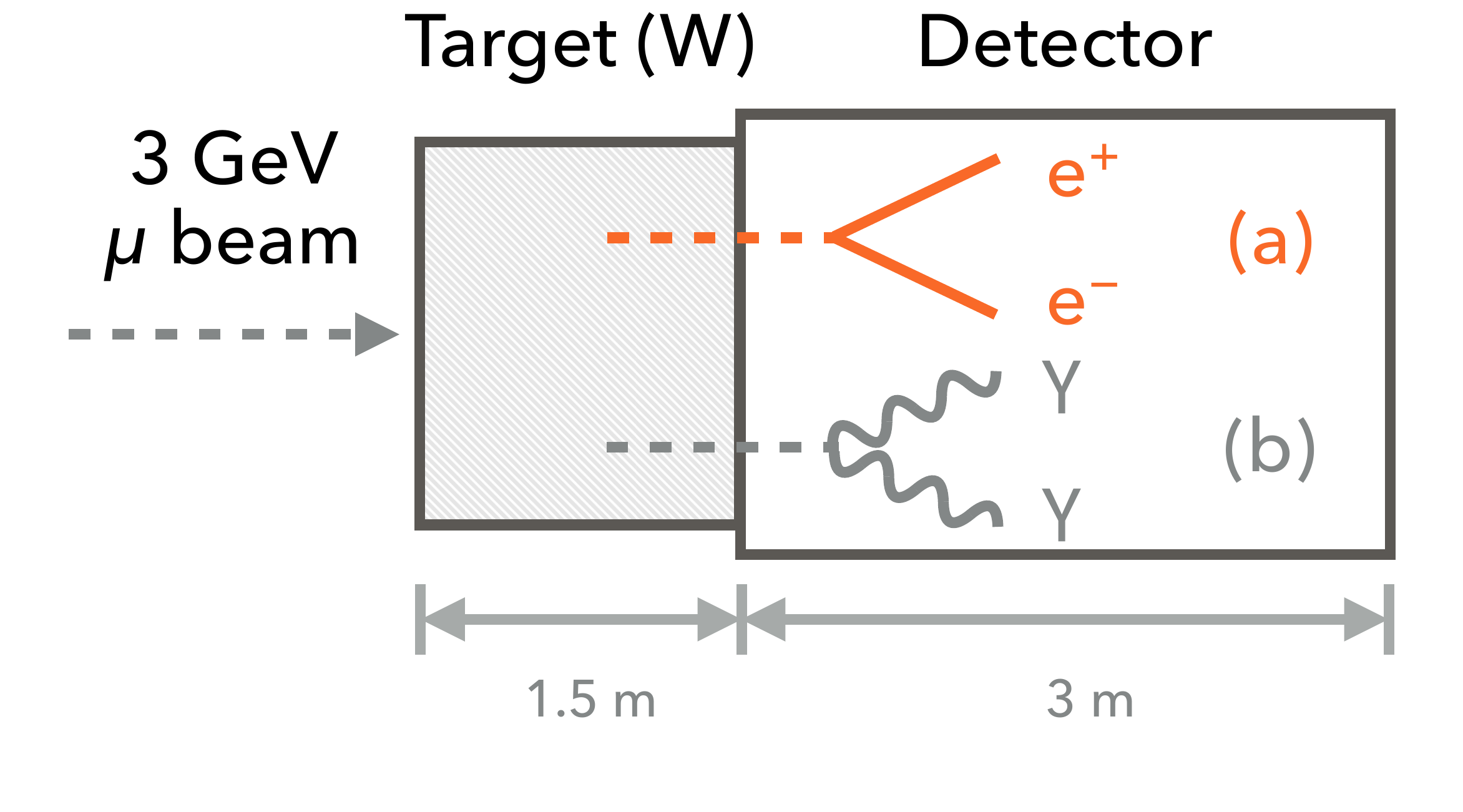}
  \caption{Setup of the proposed FNAL-$\mu$ muon beam dump experiment placed downstream of the muon beam used for the Muon $g-2$ experiment at Fermilab~\cite{Chen:2017awl}.\label{fig:FNALmu}}
\end{figure}

FNAL-$\mu$ has been proposed to search for light dark-sector particles that dominantly couple to muons that can explain the observed muon $g-2$ anomaly~\cite{Chen:2017awl}. FNAL-$\mu$ is a muon beam-dump experiment at the muon campus of Fermilab using the existing Fermilab muon beam source with the anomalous energy deposition downstream from the dump. The proposed incident muon beam energy is around $3~\GeV$, as the accelerator complex is already tuned to this energy for the Muon $g-2$ experiment. Such a beam will be completely stopped within a $1.5~\m$-thick tungsten target. Dark sector particles that are produced through muon-nucleon bremsstrahlung interactions can then visibly decay inside a $3-\meter$ detector equipped with an electron or photon tracker/calorimeter. See \cref{fig:FNALmu} for a schematic.

FNAL-$\mu$ has a simple and compact design that could be facilitated at the $g-2$ hall of Fermilab. It could run in parallel with the on-going Muon $g-2$ experiment. With a beam intensity of $10^7$ muons per second, a one-month run corresponding to $2.5\times 10^{13}$ muons on target is expected to reach a sensitivity of $3\times10^{-4}$ for the muonic dark scalar couplings. Such a sensitivity completely explores the parameter space for light muonic dark scalars ($m_S< 2 m_\mu$) that explains the muon $g-2$ anomaly. A one-year run or $3\times 10^{14}$ muons on target could reach a sensitivity of $\mathcal{O}(10^{-5})$ for the muonic dark scalar coupling. Combined with the E137 experiment that probed the muonic dark sector through the secondary muons~\cite{Marsicano:2018vin}, such a sensitivity could completely explored the parameter space for light muonic dark scalars scalars ($m_S< 2 m_\mu$) above the supernova 1987A cooling limit.

\subsection{FORMOSA}
\label{sec:formosa}

The FORMOSA detector is a successor to the milliQan experiment (see \cref{sec:milliqan}) to search for millicharged particles at the HL-LHC. This detector will be placed within the Forward Physics Facility (FPF, see \cref{sec:fpf}), which provides an ideal location for a next generation experiment to search for millicharged particles.

\begin{figure}[h]
  \centering
  \includegraphics[width=0.7\textwidth]{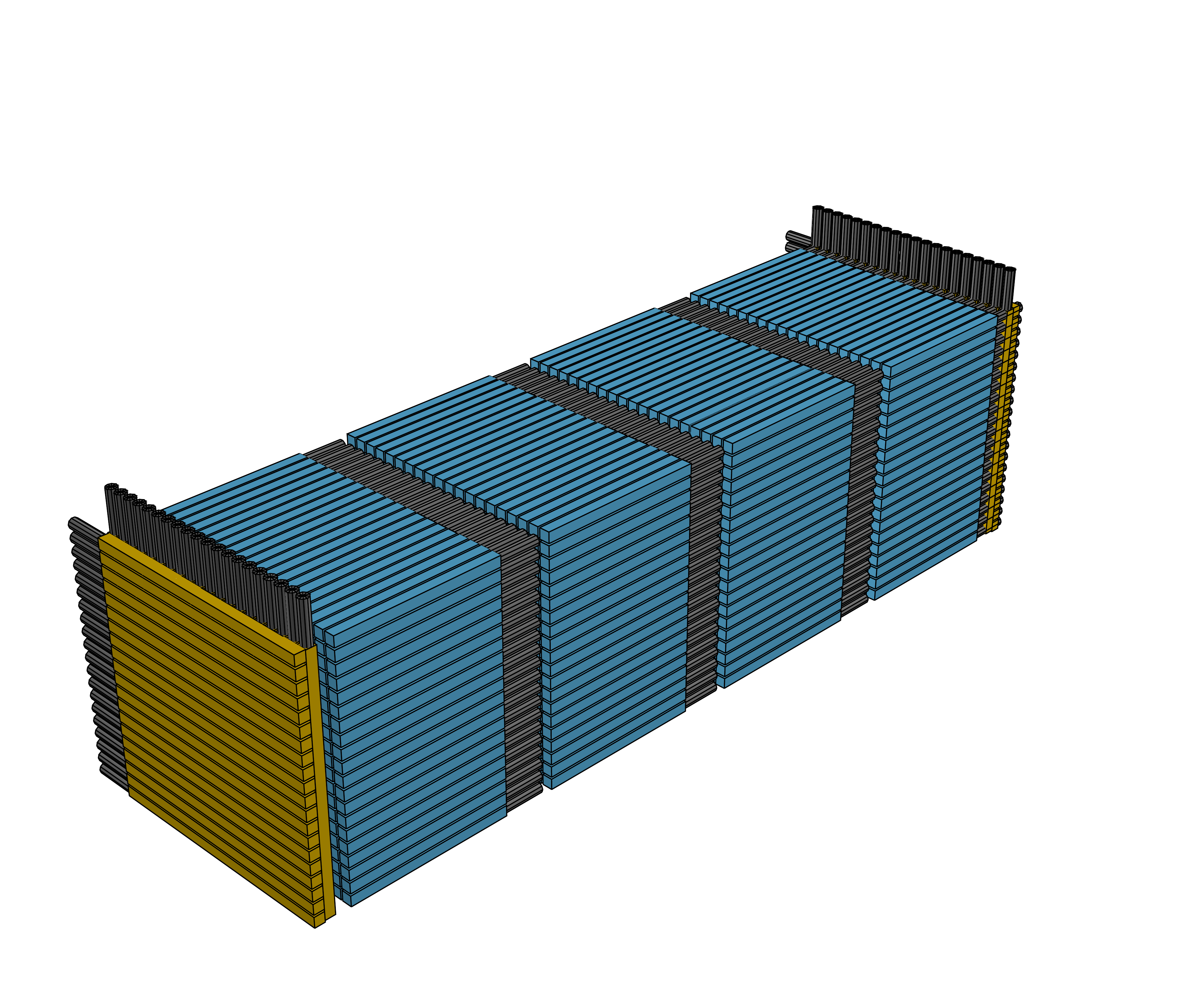}
  \caption{A diagram of the FORMOSA-II detector components. The scintillator bars are shown in blue connected to PMTs in black.\label{fig:formosa}}  
\end{figure}

In order to be sensitive to the small $\mathrm{d}E/\mathrm{d}x$ of a particle with $Q \lesssim 0.1e$, an mCP detector must contain a sufficient amount of sensitive material in the longitudinal direction pointing to the interaction point (IP). FORMOSA is planned to be a $1\times1\times5~\meter^3$ array of suitable plastic scintillator. The array will be oriented such that the long axis points at the ATLAS IP1 and is located on the beam axis. The array contains four longitudinal ``layers'' arranged to facilitate a four-fold coincident signal for feebly interacting particles originating from the ATLAS IP. Each layer in turn contains four hundred $5\times5\times100~\si{\centi\meter}^3$ scintillator ``bars'' in a $20\times20$ array. To maximize sensitivity to the smallest charges, each scintillator bar is coupled to a high-gain photomultiplier tube (PMT) capable of efficiently reconstructing the waveform produced by a single photoelectron (PE). In order to reduce random backgrounds, mCP signal candidates will be required to have a quadruple coincidence of hits with $\bar{N}_\mathrm{PE}\ge 1$ within a $20~\si{\nano\second}$ time window. The PMTs must therefore measure the timing of the scintillator photon pulse with a resolution of $\le5~\si{\nano\second}$. The bars will be held in place by a steel frame. A conceptual design of the FORMOSA detector is shown in \cref{fig:formosa}. 

Although omitted for clarity in \cref{fig:formosa}, additional thin scintillator ``panels'' placed on each side of the detector will used in an active veto of cosmic muon shower and beam halo particles. Finally, thin scintillator panels will be placed on the front and back of the detector to aid in the identification of muons resulting from LHC proton collisions. These panels may be segmented (as shown) to allow fine grain tracking of beam muons through the detector. 

In \rfr{milliQan:2021lne}, data from the closely related milliQan prototype was used to predict backgrounds from dark rate pulses and cosmic muon shower particles for a closely related detector design and location. Based on these studies, such backgrounds are expected to be negligible for FORMOSA. Backgrounds from muon after-pulses are considered in \rfr{Foroughi-Abari:2020qar} and can be rejected by vetoing a $10~\si{\micro\second}$ time window in the detector following through-going beam muons.

The signal process is simulated from a range of production modes, as detailed in \rfr{Foroughi-Abari:2020qar}. This provides the expected flux for each mass and charge of the signal and can then be used to determine the expected sensitivity of the experiment. For much of the phase space, the sensitivity is limited by the efficiency of the scintillator bars to detect through-going mcPs. In this regime the mcP flux is very high and so only a small area of higher performance scintillator can allow substantial gains in sensitivity. One possibility is an upgraded design in which an additional $2\times2\times4$ bars are installed using a higher performance scintillator such as LaBr3(Ce). By placing these bars close to the larger plastic scintillator, the active veto background rejection capabilities of the array for sources such as cosmic showers are maintained. Shielding of these would mitigate backgrounds for the larger detector caused by the radioactivity of the LaBr3(Ce). In this scenario, the optimal charge reach could be lowered by as much as a factor of 5.

The FORMOSA collaboration is growing and includes around ten collaborators from multiple institutes. Because of the strong overlap of this group with the proponents of milliQan, it is expected that many milliQan group members and institutes will join this endeavor before the HL-LHC era begins.  

\subsection{HPS}
\label{sec:hps}

\begin{figure}[h]
  \centering
  \includegraphics[width=.57\linewidth]{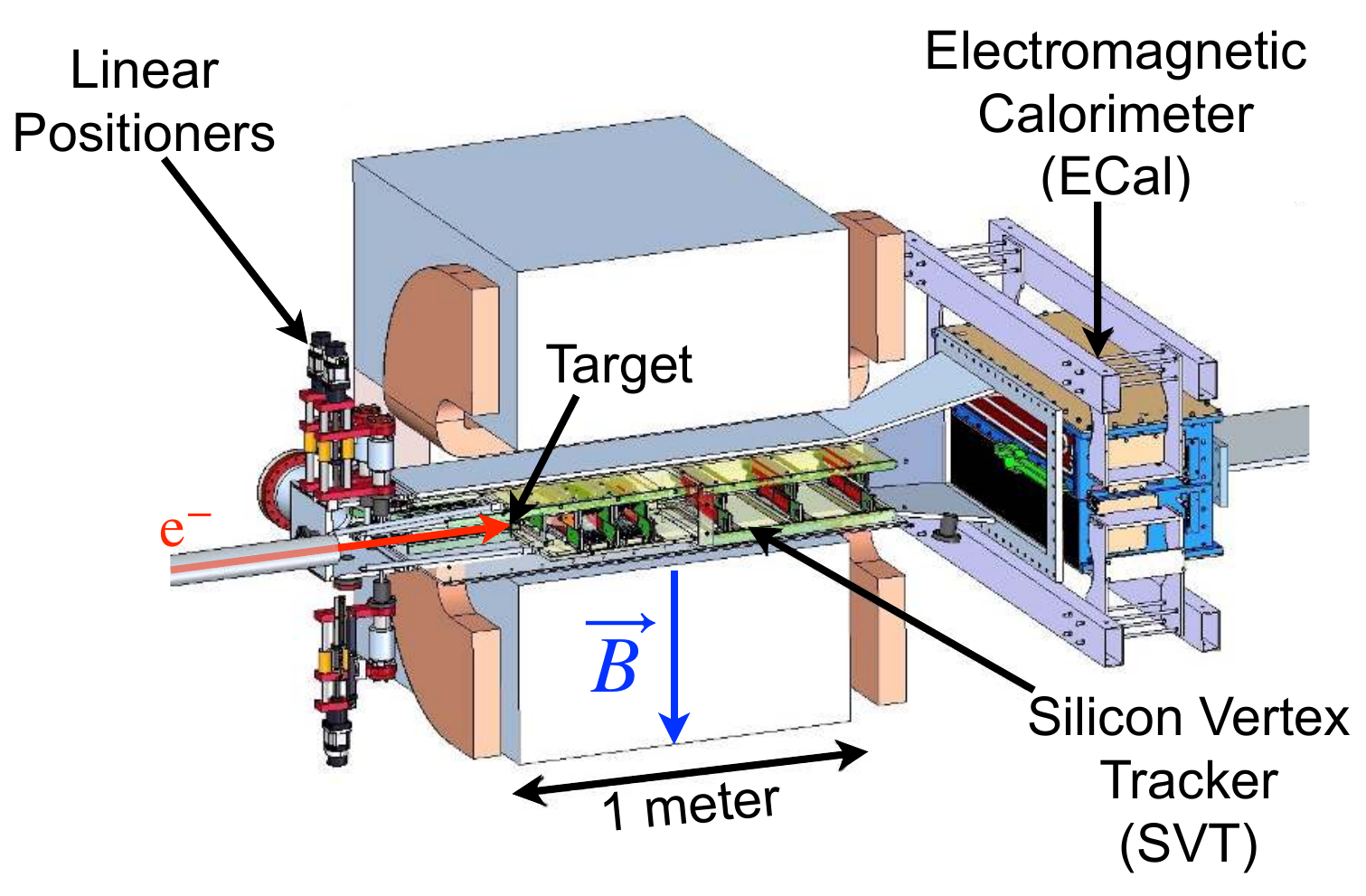}
  \caption{The HPS detector as installed and commissioned in 2015. Not shown are the trigger hodoscope and new inner silicon layers installed in 2019.\label{fig:HPS}}
\end{figure}

The Heavy Photon Search experiment (HPS) is an electron fixed target experiment aimed at searching for dark photons with masses $20-220~\MeV$ and couplings $10^{-10} < \epsilon < 10^{-6}$. HPS operates with a $1 \sim 6~\GeV$ electron beam on $0.125-0.625\%~X_0$ tungsten targets to electroproduce dark photons which would subsequently decay to $e^+ e^-$ pairs in the case that decays to dark matter (DM) are kinematically inaccessible. The HPS detector, shown in \cref{fig:HPS}, utilizes a low-mass, high-rate, silicon vertex tracker (SVT) with excellent forward coverage operating in beam vacuum inside a dipole magnet as a spectrometer and vertex detector to reconstruct the mass and vertex position of $e^+ e^-$ pairs.

Downstream of the SVT, a PbWO$_4$ electromagnetic calorimeter (ECal) and a small scintillator hodoscope are used as a high-rate trigger for the SVT, and offline for particle identification of positrons and electrons. Operating with a high repetition rate ($499~\si{\mega\Hz}$) beam at moderately high intensities ($50-450~\si{\nano\ampere}$), good time resolution in the SVT ($\approx2~\si{\nano\second}$) and ECal ($\approx300~\si{\pico\second}$) are fundamental to associating reconstructed objects to individual events to eliminate hits from the extreme rate of scattered single electrons that dominate the far forward acceptance critical to the dark photon search. The HPS apparatus, completed and commissioned with engineering runs in 2015 and 2016, operates in Hall~B at the JLab CEBAF and has collected physics data during runs in 2019 and 2021 after upgrades to the apparatus to add the hodoscope and an additional SVT layer closer to the target.

HPS uses two techniques to search for dark photons. The first is a simple resonance search for a peak in the $e^+ e^-$ mass distribution atop the continuum background from QED tridents, which is sensitive to dark photons with larger couplings. The second is a search for displaced vertices from long-lived dark photons, which is sensitive at intermediate couplings that are particularly interesting in models where the dark photon interaction is responsible for freeze out at the observed relic abundance of DM, the so-called thermal targets. The displaced vertex search, which has unique sensitivity relative to other experiments, has been the prime motivation in the design of the HPS experiment.

Data taken during the 2015 and 2016 engineering runs have been used to fully develop the analysis techniques and perform preliminary searches for dark photons~\cite{HPS:2018xkw,Moreno:2018tlx}. The results of these analyses have been used to further refine the apparatus and analysis techniques and allow for realistic estimates of reach based on actual results of these searches. Although the experiment is designed with the search for a new vector mediator in mind, it is important to note that HPS is also sensitive to more general models of dark forces with vector, axial-vector, scalar, or pseudo-scalar couplings to matter, since mediators of other spins have production and decay properties similar to that of dark photons. Similarly, HPS has sensitivity to other dark sector models involving long-lived states such as those with strong dynamics in the dark sector (SIMPs) and those with a richer set of dark states with large mass splittings (iDM). Searches for these other scenarios are in development on engineering run data.  In addition to data in hand from 2019 and 2021, the experiment, originally approved for 180 days of operation, has 107 days remaining and expects to operate again in the coming years.

\subsection{JPOS}
\label{sec:jpos}

The JPOS program includes two complementary searches for dark photons ($A'$), a thin-target and active thick-target measurement, to search for light dark matter with positron beam. The two approaches will be introduced followed by a brief discussion of the experimental setup, measurement strategy, data analysis procedure, and foreseen results. JPOS will utilize the proposed $11~\GeV$ positron beam to be developed in the near future at Jefferson Lab~\cite{Marsicano:2018oqf}.

The thin-target measurement exploits the $A'$-strahlung production in electron-positron annihilation. The primary positron beam impinges on a thin target, and a photon-$A'$ pair is produced. By detecting the final-state photon momentum with an electromagnetic calorimeter, the missing mass $M_{\text{miss}}$ per event can be computed. The signal would appear as a peak in the missing mass distribution centered at the $A'$ mass, on top of a smooth background resulting from standard model processes with single photons measured in the calorimeter. The width of the signal peak is mainly determined by the energy and angular resolution of the calorimeter. Several experiments searching for the  $A'$ with this approach have been proposed. The Positron Annihilation into Dark Matter Experiment (PADME) at Laboratori Nazionali di Frascati (LNF)~\cite{Valente:2016hba} is one of the first experiments using the $e^+$ on thin-target approach to search for dark photons. It relies on a $550~\MeV$ positron beam provided by the DA$\Phi$NE Linac at LNF which then impinges on a thin diamond active target. The PADME design can be used as a baseline for future thin target measurements.

\begin{figure}[h]
  \centering
  \includegraphics[width=.56\textwidth]{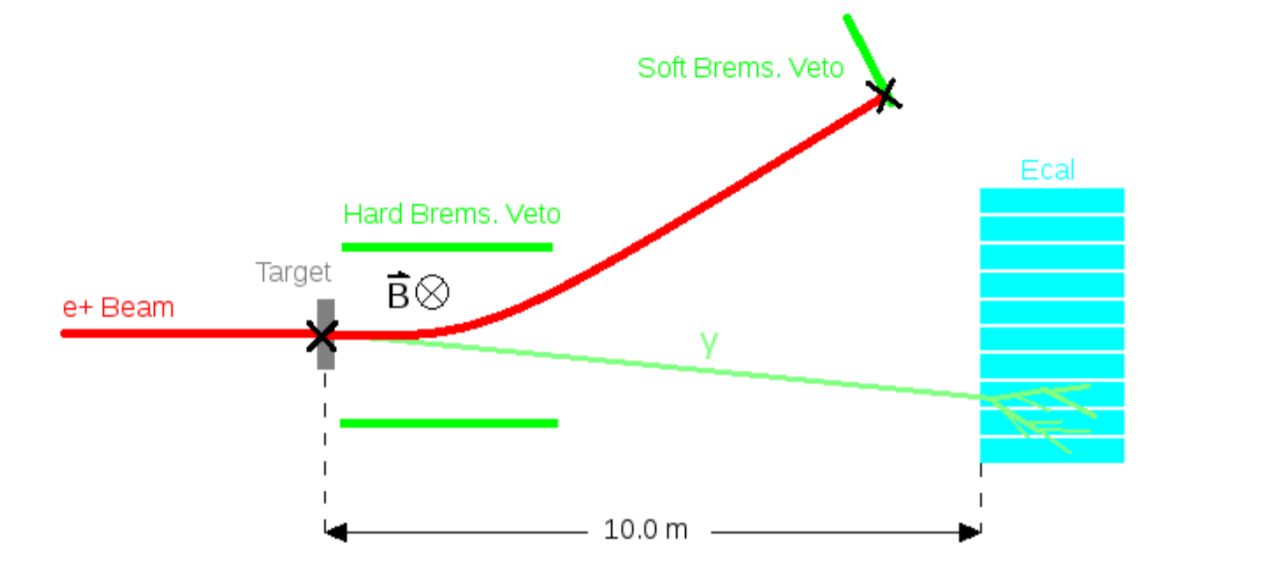}\hfill
  \includegraphics[width=.38\textwidth]{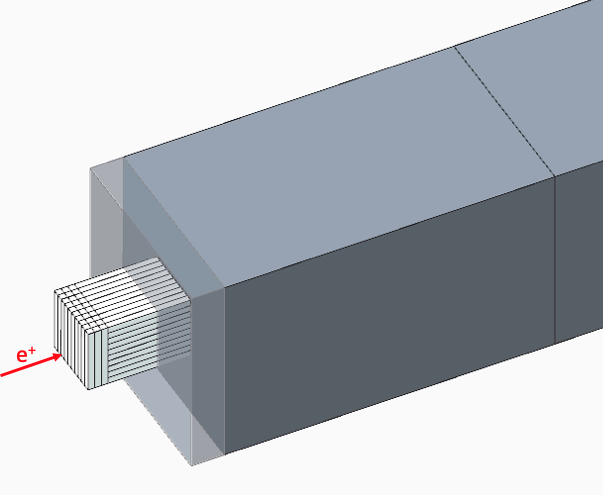}
  \caption{(left) Layout of the proposed thin target setup. (right) Schematic layout of the active thick-target experimental setup, with the (white) ECAL followed and surrounded by the (gray) HCAL. The semi-transparent portion of the HCAL in front is that installed all around the ECAL.\label{fig:jpos}}
\end{figure}

The active thick-target measurement exploits the resonant $A'$ production by positron annihilation on atomic electrons. The primary positron beam impinges on a thick active target, and the missing energy signature of a produced and undetected light dark matter candidate is used to identify the signal~\cite{Izaguirre:2014bca}. The active target measures the energy deposited by individual beam particles; when an energetic $A'$ is produced, its invisible decay products, the $\chi\overline{\chi}$ pair, carry away a significant fraction of the primary beam energy, resulting in a measurable reduction of the expected deposited energy. Signal events are identified when the missing energy $E_{\text{miss}}$, defined as the difference between the beam energy and the detected energy, exceeds a minimum threshold. The signal has a distinct dependence on the missing energy
through the relation $m_{A'}=\sqrt{2 \, m_e \, E_{\text{miss}}}$.

This leads to a specific experimental signature for the signal, appearing as a peak in the missing energy distribution. The peak location depends solely on the dark photon mass. Owing to the emission of soft bremsstrahlung photons, the thick target causes an almost continuous energy loss in impinging positrons. Although the positron energy loss is a quantized process, the finite intrinsic width of the dark photon, much larger than the positron energy differences, and the electron energy and momentum spread induced by atomic motions~\cite{Nardi:2018cxi} will counterbalance this effect. This allows the primary beam energy to be scanned through the full range of dark photon masses from the highest mass (corresponding to the loss of the entire beam energy), to the lowest allowed by the missing energy threshold~\cite{Marsicano:2018krp}. This approach exploits the presence of secondary positrons produced by the developing electromagnetic shower.

\subsection{LDMX}
\label{sec:ldmx}

The Light Dark Matter eXperiment (LDMX) is an electron fixed-target experiment being developed to search for the production of sub-\GeV dark matter via the missing momentum technique at the Linac to End Station A (LESA) facility at SLAC (see \cref{sec:lesa}). LDMX utilizes high-rate ($37.5~\si{\mega\Hz}$), low-current ($\approx 1-5~e^-/\mathrm{bunch}$) $4$ or $8~\GeV$ electron beam impinging on a thin ($10-30\%~X_0$) tungsten or aluminum target to search for events where a massive particle is produced that can escape a thick, sensitive, and hermetic detector without depositing significant energy. In such events, the incoming electron loses the majority of its energy and experiences a large transverse-momentum kick in the hard interaction, providing a clear handle for distinguishing a small number of signal events among a large sample (up to $10^{16}$ incoming electrons) in the presence of finite detector resolutions.  

\begin{figure}[h]
  \centering
  \includegraphics[width=.57\linewidth]{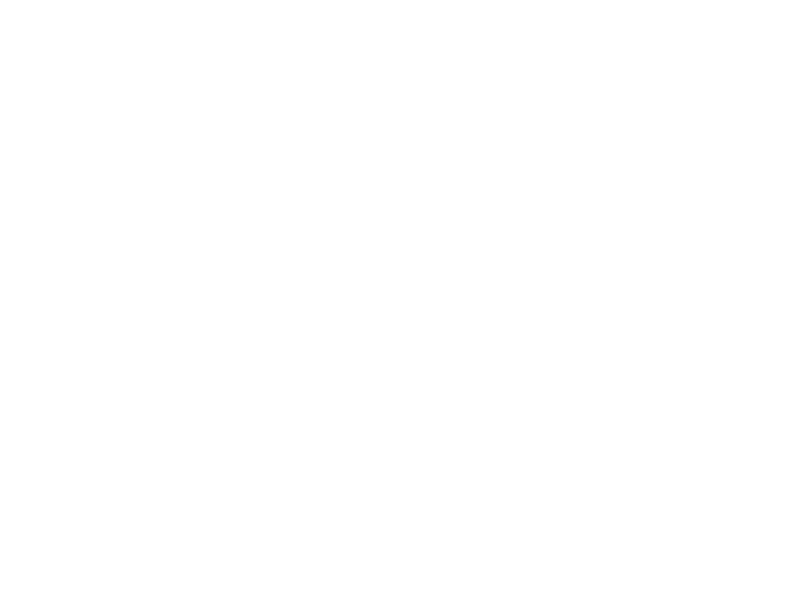}
  \caption{The LDMX detector concept as described in the text.\label{fig:LDMX}}
\end{figure}

The detector, shown in the left panel of \cref{fig:LDMX}, utilizes a pair of trackers upstream and downstream of the target to measure the vector momentum of the incoming and recoiling electron. Downstream of the trackers is an electromagnetic calorimeter (ECal) to measure the energy of the recoiling electron and any bremsstrahlung photons, which can also carry away the majority of the incoming electron energy. Along with a trigger scintillator (TS) to count incoming electrons, the ECal also provides a fast trigger on missing energy. Modest tracker and ECal resolutions are sufficient to veto events where the incoming/outgoing electron and any bremsstrahlung photons are mis-measured. Consequently, the dominant background comes from events where a hard bremsstrahlung photon is produced that undergoes a photo-nuclear reaction in the ECal material resulting in a hadronic final state where only a small fraction of the energy will be deposited in the detector. To veto such events, a very high granularity ECal is employed that allows the imaging of these events and their identification as distinct from electromagnetic showers, and the ECal is surrounded by a large, deep ($15~\lambda$) hadronic calorimeter (HCal) to veto events where significant energy escapes the thick ($40~X_0$) ECal. Studies have shown that this detector is capable of reducing the expected backgrounds to less than one event for a $4\times 10^{14}$ experiment at $4~\GeV$, without using the transverse momentum kick across the target for signal discrimination~\cite{LDMX:2019gvz}. In the case where a signal is observed, the transverse momentum spectrum provides an estimate of the mass of the invisible state produced in the reaction. With early beam and low statistics ($10^{12}$ electrons on target), a missing energy analysis can break new ground without mastering the most difficult backgrounds, and with ab $8~\GeV$ beam, the limiting backgrounds are mitigated, so that a $10^{16}$ electrons on target experiment with little or no background appears feasible.

The LDMX detector employs already available technologies developed for other experiments -- HPS (tracking, see \cref{sec:hps}), CMS (ECal and TS), and Mu2e (HCal) -- to enable the construction of the apparatus without undertaking any detector research and development. A collaboration was formed in 2019 by institutions that are core experts in the required technologies to develop the detector concept. The development of the detector design and project plan has been funded as a dark matter new initiative by the Department of Energy, and approval and construction of the experiment are pending a review of the design and project plan.

\subsection{LHCb}
\label{sec:lhcb}

\begin{figure}[h]
  \centering
  \includegraphics[width=.7\linewidth]{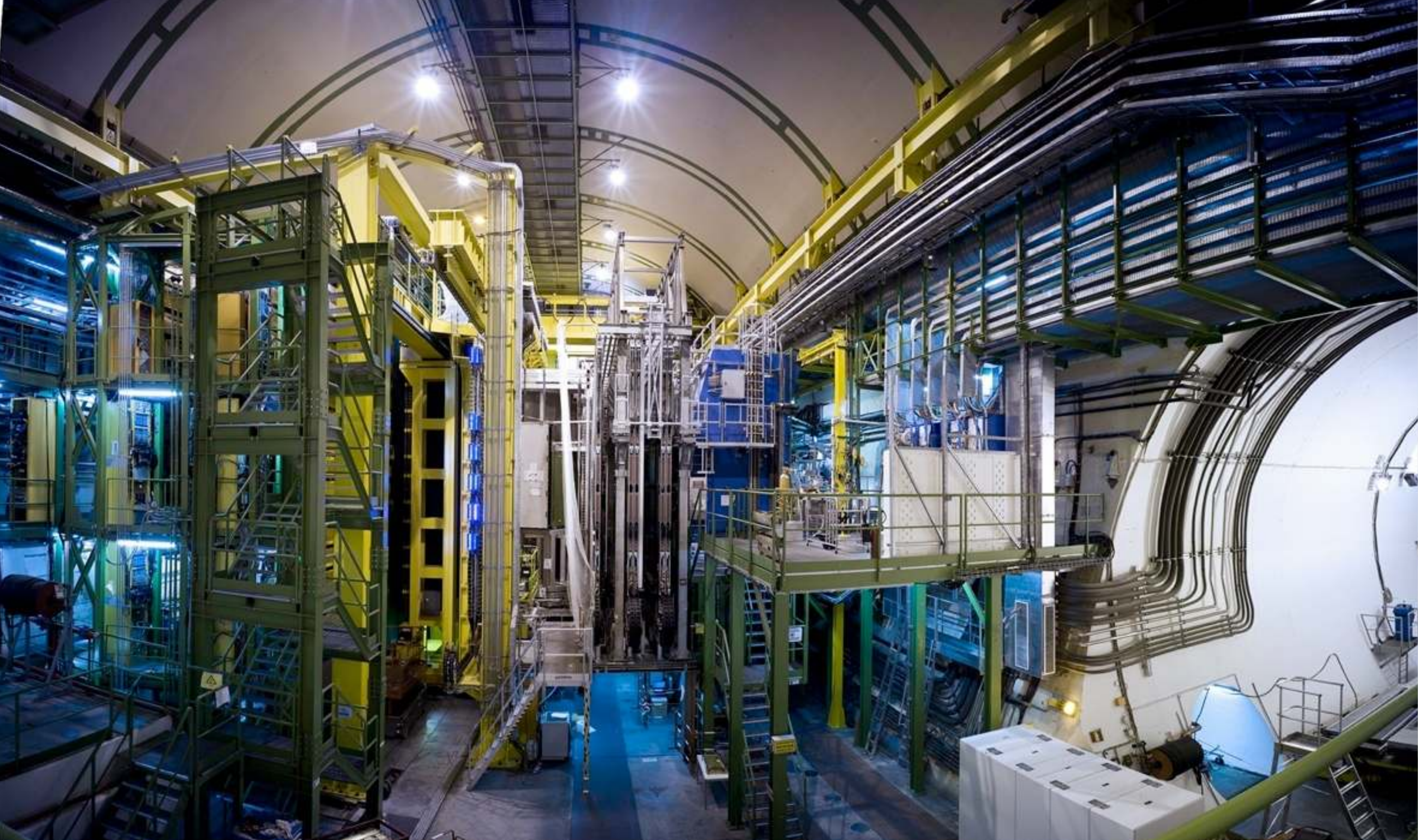}
  \caption{The LHCb detector.\label{fig:LHCb}}
\end{figure}

The LHCb detector (see \cref{fig:LHCb}) is a single-arm spectrometer covering the forward region of $2 < \eta < 5$~\cite{LHCb:2014set}. The detector, built to study the decays of hadrons containing $b$ and $c$ quarks, consists of a high-precision charged-particle tracking system, two ring-imaging Cherenkov (RICH) detectors, an electromagnetic and hadronic calorimeter system, and a system of muon chambers. LHCb collected a total integrated luminosity of $3~\invfb$ at $\sqrt{s}=7$ and $8~\TeV$ in LHC Run~1, and an additional $6~\invfb$ at $\sqrt{s}=13~\TeV$ in Run~2. LHCb has excellent lifetime resolution and a highly flexible trigger system, which are the primary reasons it is one of the premier dark-sector experiments. LHCb has already demonstrated world-leading sensitivity to visible dark-photon decays~\cite{LHCb:2017trq,LHCb:2019vmc} and to \GeV-scale Higgs-portal scalars~\cite{LHCb:2015nkv,LHCb:2016awg} using Run~2 and Run~1 data, respectively.

The LHCb detector is currently undergoing a major upgrade for Run~3.
All of the tracking systems are being replaced to enable increasing the instantaneous luminosity by a factor of 5 -- and to remove the need for a hardware-based trigger, thus greatly increasing the trigger efficiency for most processes of interest, including dark photons. LHCb plans to collect $15~\invfb$ of data in Run~3, then another $25-30~\invfb$ in Run~4. Another major upgrade is planned for Run~5, and by the end of LHC running LHCb plans to collect at least $300~\invfb$ of data. 

\Rfr{Craik:2022riw} provides future LHCb dark-sector sensitivity projections. These include updated projections for dark photons and the Higgs portal, along with new projections for axion-like particles (ALPs) that couple predominantly to gluons. The predicted sensitivity to the ALP-photon coupling is estimated in \rfr{CidVidal:2018blh}. LHCb is projected to be able to cover large regions of unexplored parameter space in all of the high-profile dark-sector models that involve visible decays to SM particles, maintaining its role as a leading dark-sector experiment for the next decade and beyond.

\subsection{milliQan}
\label{sec:milliqan}

The milliQan experiment is composed of two detectors that will begin taking data at the LHC in 2022 to search for beyond the standard model (BSM) particles that have an electrical charge that is a small fraction of that of the electron. Although the value of this fraction can vary over several orders of magnitude, we generically refer to these new states as ``millicharged'' particles (mCPs). General purpose detectors at the LHC are not sensitive to the deposits of such particles and so experimental observation of mCPs requires a dedicated detector. 

\begin{figure}[h]
  \centering
  \includegraphics[width=.4\linewidth]{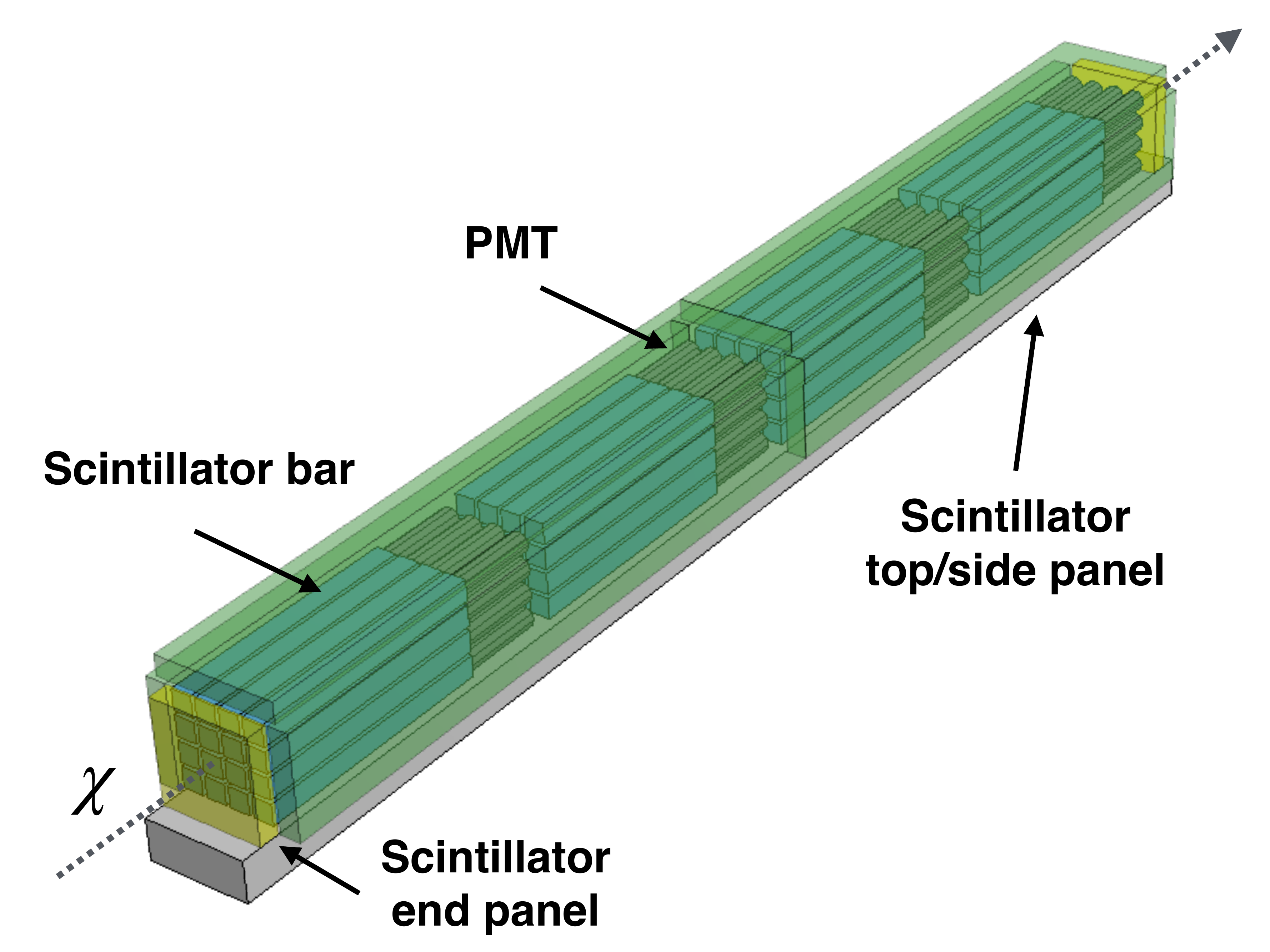}\hfill
  \includegraphics[width=.4\linewidth]{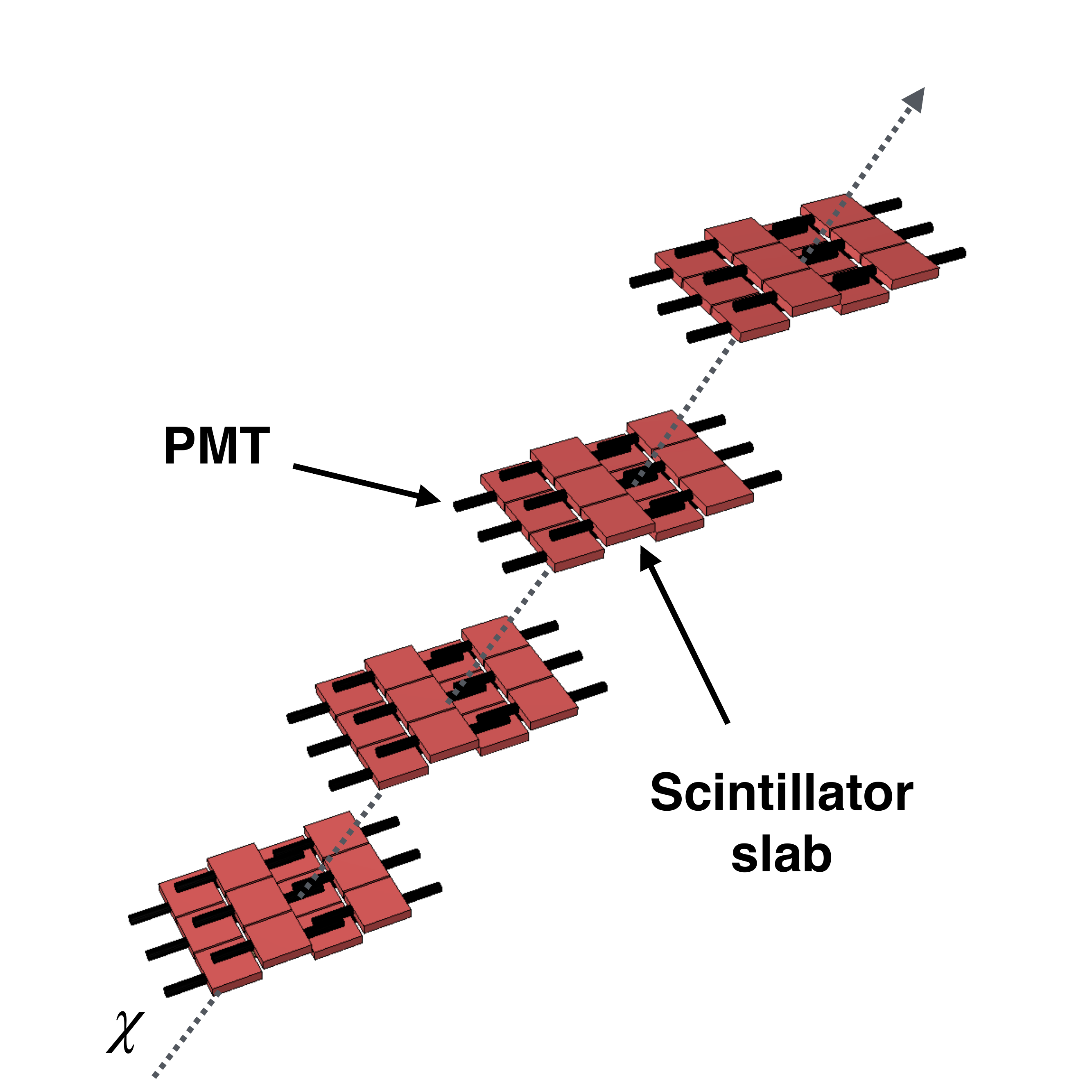}
  \caption{A diagram of the milliQan (left) bar and (right) slab detector components.\label{fig:milliqan}}
\end{figure}

In order to be sensitive to the small $\mathrm{d}E/\mathrm{d}x$ of a particle with $Q \lesssim 0.1e$, an mCP detector must contain a sufficient amount of sensitive material in the longitudinal direction pointing to the IP. The milliQan bar detector is comprised of a $0.2~\mathrm{m}~\times0.2~\mathrm{m}~\times3~\meter^3$ plastic scintillator array. The array will be oriented such that the long axis points at the nominal CMS IP. The array will contain four longitudinal ``layers'', each containing sixteen $5~\mathrm{cm}\times5~\mathrm{cm}\times60~\si{\centi\meter}^3$ scintillator ``bars'' optically coupled to high-gain PMTs in a $4\times4$ array. Surrounding the array is an active muon veto shield composed of six $5~\si{\centi\meter}$ thick scintillator panels that cover the top and sides of the array. Each panel will have two PMTs at opposing ends to increase light collection efficiency and to provide some positional information (using relative pulse sizes and $\sim\si{\nano\second}$ timing resolution). An additional small scintillator panel at each end of the bars will complete the hermeticity of the shield. A diagram of the bar detector may be seen in the left of \cref{fig:milliqan}.

For charges larger than $\sim 0.01e$, the reach is limited by the acceptance rather than the efficiency to reconstruct the millicharged deposits. This motivates the milliQan slab detector. This will be comprised of $40 \times 60 \times 5~\si{\centi\meter}^3$ scintillator ``slabs''. These will be arranged in four layers of $3 \times 4$ slabs. There are therefore a total of 48 slabs in the array. The segmentation of the layers in the slab detector is driven by a compromise between practical considerations, including mechanical constraints and limiting the number of channels, as well as the desire to sharply define pointing paths to the IP to reduce accidental backgrounds. A drawing of the slab detector may be seen in the right of \cref{fig:milliqan}.

Even though the pointing, 4-layered, design will be very effective at reducing background processes, small residual contributions from sources of background that mimic the signal-like quadruple coincidence signature are expected for both detectors. These include overlapping dark rate pulses, cosmic muon shower particles, and beam muon after-pulses. During Run~2 of the LHC, a similar experimental apparatus (the milliQan prototype) was deployed in the PX56 draining gallery at LHC P5 near the CMS IP. This device was used successfully to search for millicharged particles proving the feasibility of such a detector~\cite{Ball:2020dnx}. As detailed in \rfr{milliQan:2021lne}, data from the milliQan prototype has been used to measure the backgrounds and calibrate the expected performance for the milliQan slab and bar detectors. 

Several similar detectors have been proposed that are inspired by the milliQan detector design. These future detectors at a range of facilities including the Forward Physics Facility (see \cref{sec:fpf}) (FORMOSA, see \cref{sec:formosa}), Fermilab (Fermini, see \cref{sec:fermini}) and J-PARC (SUBMET) provide complementary coverage of a wide range of millicharged particle masses and charges.

The milliQan collaboration currently consists of approximately 25 collaborators from 10 institutions in 5 countries. The milliQan prototype was installed and ran successfully from 2017--2019. The milliQan experiment has been constructed and will be installed in time for Run 3 of the LHC. The detectors are fully-funded and will be commissioned and operated following the methods used for the prototype detector. More details on the detector design can be found in \rfr{milliQan:2021lne}.

\subsection{NA64}
\label{sec:na64}

The NA64 experiment pioneered searches for dark sectors in fixed target experiments at CERN. NA64 is running at the CERN Super Proton Synchrotron (SPS) accelerator and was designed as a hermetic general purpose detector to probe dark sectors physics from electron, hadron, and muon scattering off nuclei. In particular, its physics targets are: light dark matter (LDM) models, in which dark matter can be explained as a thermal freeze-out relic; the existence of a light $A'$, $Z'$, or $X$ dark boson as an explanation of the muon ($g-2$) anomaly or the $X17$ boson; and other beyond standard model (BSM) extensions such as axion-like particles (ALPs)  

The dark sector particle (typically $A'$, $Z'$, or a generic $X$ boson) would be produced in electron-nucleus interactions via a dark-bremsstrahlung process, $e^- Z \to e^- Z A'$. After its production, depending on its mass, the $A'$ could promptly decay into a pair of LDM candidate particles, $A' \to \chi\chi$, leaving missing energy as a signature, or could also decay visibly to SM particles. Searching for events with large missing energy allows to probe the mixing strength and the parameter space close to the one predicted by the relic dark matter density. The NA64 experiment took data between 2016 and 2018 using the $100~\GeV$ electron beam from the H4 beam line at the north area of CERN accumulating, $2.84\times10^{11}$ electrons on target (EOT). The latest NA64 results set the most stringent constraints on LDM models in the mass region below $200~\MeV$~\cite{NA64:2017vtt,Banerjee:2019pds}.

NA64 together with the Babar experiment, ruled out the dark photon invisible decay as an explanation of the muon $g-2$ anomaly. An alternative and richer scenario has been proposed, containing two DM species: a lighter one, $\chi_1$, which is stable and constitutes the DM observed in the universe; and a second heavier DM particle $\chi_2$. The $\chi_2$ can then decay into the lighter state and an off-shell dark photon which can then decay visibly to SM particles, leaving a semi-visible signature. This channel recovers the DM thermal freeze-out explanation but additionally can still explain the muon ($g-2$) anomaly and evade the experimental constraints described before, since pure visible and invisible modes are suppressed. NA64 has performed for the first time an analysis of the experiment sensitivity to this decay parameter space~\cite{NA64:2021acr}. 

The results from the combined $2016-2018$ visible data analysis corresponding to $8.4\times10^{10}$ EOT, were summarized in \rfr{NA64:2018lsq}. No signal-like events have been found covering a broad area of the $A' \to e^+e^-$ parameter space and excluding a large fraction of the X17 anomaly coupling parameter space ($\epsilon < 6.8 \times 10^{-4}$) for the vector-like benchmark model. Recently, these searches has been extended also to a pseudo-scalar particle decaying visibly into a lepton pair~\cite{NA64:2021aiq}. Searches for other SM extensions, such as light scalar and pseudo-scalar axion-like particles (ALPs) produced through the Primakoff reaction have also been recently addressed trying to probe the uncovered gap in the parameter space of benchmark axion models between the beam-dump and the LEP searches.

\begin{figure}[h]
  \centering
  \includegraphics[width=0.9\textwidth]{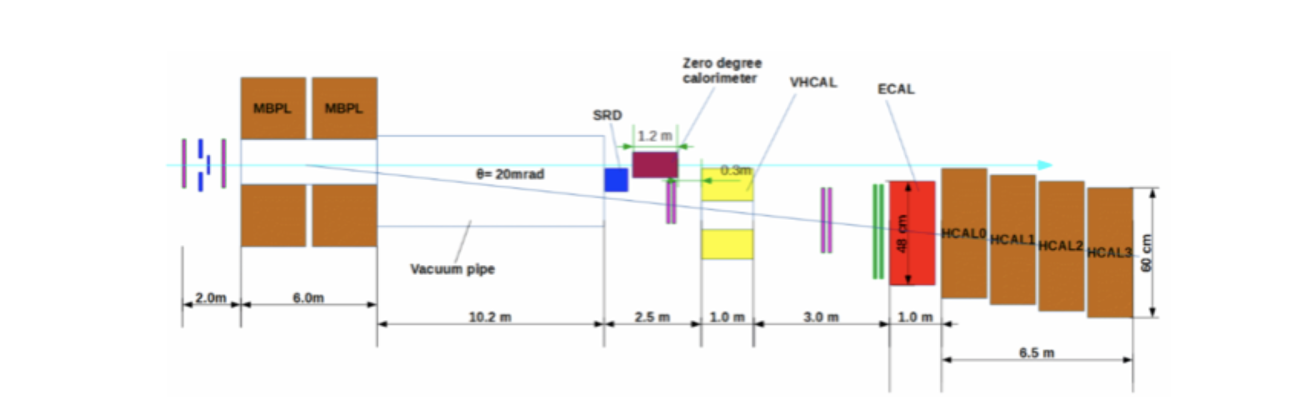}
  \caption{Experimental setup for the 2022 NA64 run at the H4 beam-line at CERN.\label{fig:NA64}}
\end{figure}

Since 2021, NA64 is running in a new dedicated zone at H4 provided by CERN. Data taking restarted August 2021 after the long-shutdown of the LHC acquiring $5\times10^{10}$ EOT, and NA64 foresees to accumulate $10^{12}$ EOT in 2022. The new beam-line was commissioned, and for the first time beam intensities up to $2\times10^7 e^-$/spill were delivered. To fully exploit the beam capabilities, run reliably, and minimize background, a detector and DAQ upgrade was mandatory and was was completed in 2022. An overview of the upgraded setup is illustrated in \cref{fig:NA64}. 

The future projected sensitivity of NA64 will be able to start probing the region suggested by the DM relic. Nevertheless, the signal yield is suppressed at the high energy mass region because the bremsstrahlung $A'$ cross-section emission is proportional to $1/m^2_{A'}$. To enhance the sensitivity in this region after the long shutdown, the collaboration is exploiting two complementary approaches. The first is to explore the resonance $A'$ annihilation channel using the secondary positrons present in the electromagnetic shower in the target induced by an initial $e-/e+$ beam (see \cref{sec:poker}). The second is a complementary approach using a muon beam, the NA64$_{\mu}$ experiment (see \cref{sec:na64Mu}).

\subsection{NA64$_{\mu}$}
\label{sec:na64Mu}

The existence of a new light $Z'$ boson introduced in $L_{\mu}-L_{\tau}$ beyond standard model (BSM) extensions by gauging the difference of the lepton number between the muon and $\tau$ flavor, remains a plausible solution to explain the muon $g-2$ anomaly. This boson can then decay into a pair of neutrinos or dark matter particles. Moreover, the experiment in this configuration will also be sensitive to dark photons in a mass region larger than $0.1~\GeV$. Finally, a dark sector introducing particles predominantly weakly-coupled to the second and possibly third generation of the SM leptons is motivated in many extensions of the SM. A light $Z'$ can be produced in muon-nucleus interactions via the dark-bremsstrahlung process, $\mu + Z \to \mu + Z + Z'$ after a high energy muon beam collides with an active target.

\begin{figure}
  \centering
  \includegraphics[width=0.9\textwidth]{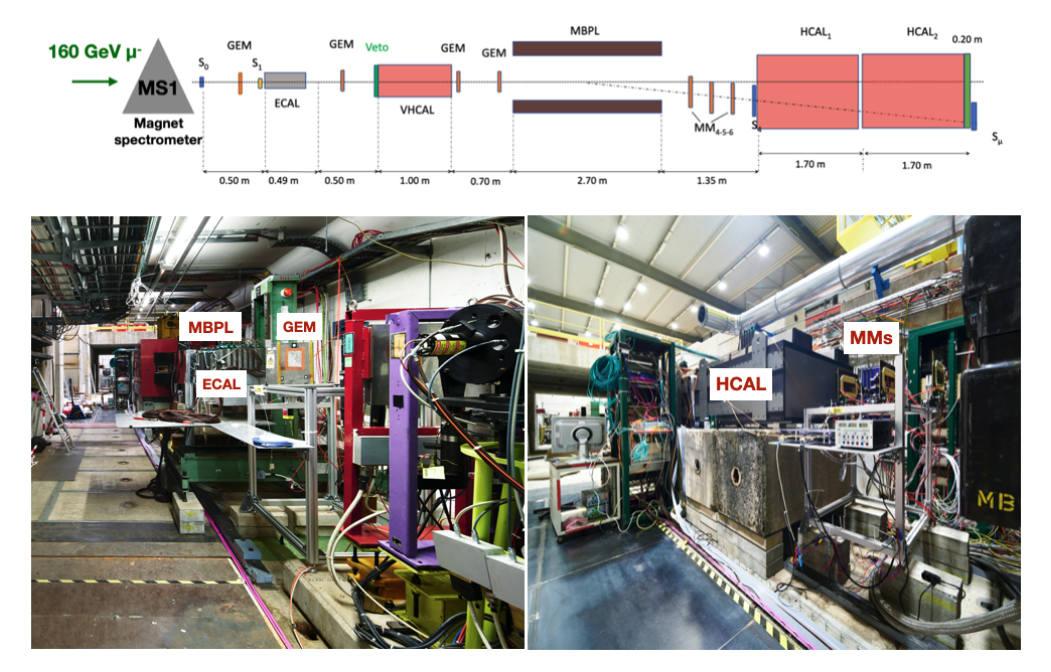}
  \caption{Experimental setup for the 2021 NA64 muon run.\label{fig:NA64mu}}
\end{figure}

NA64$_{\mu}$ uses the unique high intensity and high energy M2 beamline at CERN to search for dark sectors weakly coupled to the second generation of the SM leptons. The experiment is foreseen in two phases. The goal of the first phase, before the next long-shutdown, is to search for invisible decays of the $Z'$ boson as a remaining low mass explanation of the muon $g-2$ anomaly. The second phase, expected after Long Shutdown 3, will be devoted to fully covered LDM parameter space and probe scalar, ALPs, and millicharged particles weakly coupled to muons. A proposal of the experiment was submitted to the SPS committee in 2019. A pilot run of the experiment to study the feasibility of the technique took place during October 2021, collecting $5x10^{9}$ muons on target (MOT). Recently, the SPS committee approved an additional three-week pilot run in April 2022 where the experiment will collect $5x10^{10}$ MOT. The schematics of the setup and the overview of the detectors installed in the M2 area are illustrated in \cref{fig:NA64mu}).

A full simulation and reconstruction package in \textsc{Geant4}~\cite{GEANT4:2002zbu}, including a realistic beam simulation, was developed towards demonstrating the feasibility of the technique to search for such a $Z'$ boson and design and optimize the pilot run setup. The projected sensitivities using the simulation framework show that with 10$^{11}$ MOT the experiment can probe the region of parameter space suggested by the muon $g-2$ anomaly. NA64 electron (see \cref{sec:na64}) and muon experiments can also fully explore the very interesting region of parameter space motivated by light dark matter parameter space.

\subsection{PIONEER}
\label{sec:pioneer}

\begin{figure}[h]
  \centering
  \includegraphics[width=0.44\textwidth]{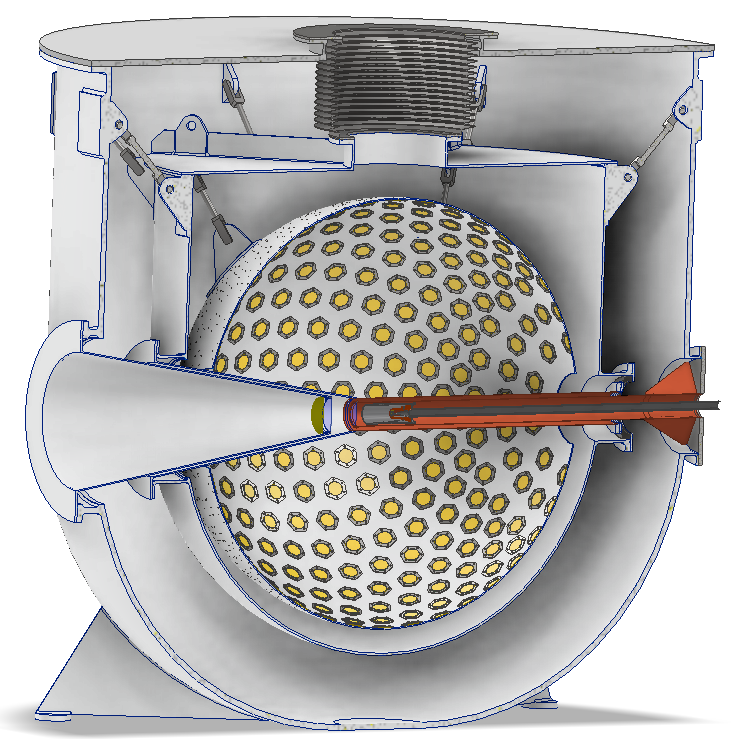} 
  \includegraphics[width=0.55\textwidth]{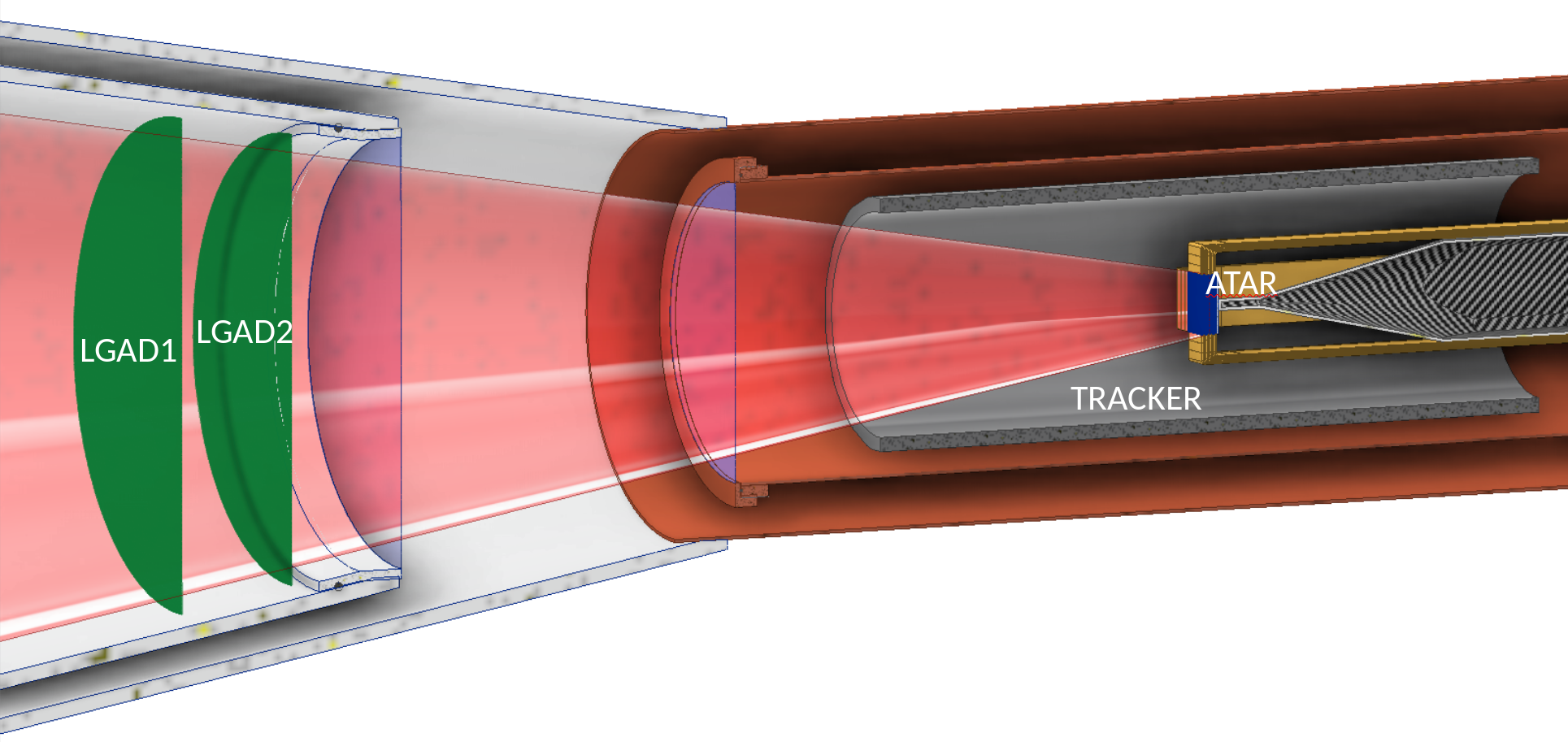}
  \caption{Concept design of the PIONEER experiment. (left) Liquid xenon calorimeter vessel. For scale, the lid has a $3.05~\meter$ diameter. The yellow circles are representative of the photosensors. (right) Beam counters, ATAR, and cylindrical tracker. The ATAR readout is represented by the yellow flexes going to the digitization stage on the outside of the calorimeter vessel.\label{fig:PIONEER}}
\end{figure}

PIONEER~\cite{PIONEER:2022yag} is a next-generation experiment to measure $R_{e/\mu}$ and pion beta decay with order of magnitude improvements in precision from the PIENU, PEN and PIBETA experiments. PIONEER is scheduled to initiate data taking at the $\pi$E5 beamline at PSI within 10 years; schematics of the experiments are shown in \cref{fig:PIONEER}.

The branching ratio $R_{e/\mu} = \frac{\Gamma(\pi^+ \to e^+ \nu (\gamma) )}{\Gamma(\pi^+ \to \mu^+ \nu (\gamma))}$ for pion decays to electrons over muons provides the best test of electron-muon universality in charged-current weak interactions. Furthermore, it is extremely sensitive to new physics at high mass scales. In the standard model (SM), $R_{e/\mu}$ has been calculated with extraordinary precision at the 0.01\% level~\cite{Cirigliano:2007xi,Cirigliano:2007ga,Bryman:2011zz} as perhaps the most precisely calculated weak interaction observable involving quarks. Because the uncertainty of the SM calculation for $R_{e/\mu}$ is very small and the decay $\pi^+ \to e^+ \nu$ is helicity suppressed by the $V-A$ structure of charged currents, a measurement of $R_{e/\mu}$ is extremely sensitive to the presence of pseudo-scalar (and scalar) couplings absent from the SM. A disagreement with the theoretical expectation would unambiguously imply the existence of new physics beyond the SM. Currently, the most accurate $R_{e/\mu}$ measurement was reported by PIENU~\cite{PiENu:2015seu} at the 0.2\%  precision level, still far from the SM expectation precision. With PIONEER's goal of 0.01\% of experimental precision, a direct comparison with the SM calculation can be made, and new physics beyond the SM (BSM) up to the mass scale of $3000~\TeV$ may be revealed if a discrepancy is observed. Such precision would  contribute to  stringent tests of lepton flavor universality (LFU) in a context where several intriguing hints of LFU violation have emerged. In addition, it will allow extended searches for exotic particles such as heavy neutral leptons and dark sector processes. 

Pion beta decay, $\pi^+ \to \pi^0 e^+ \nu (\gamma)$, provides the theoretically cleanest determination of the CKM matrix element $V_{ud}$. The current precision of $0.3\%$ on $V_{ud}$ makes $\pi^+ \to \pi^0 e^+ \nu (\gamma)$ presently irrelevant for the CKM unitarity tests because super-allowed nuclear beta decays provide a nominal precision of $0.03\%$. The final precision of $0.05\%$ that can be achieved after the full phase-III of the PIONEER data taking campaign will provide a higher precision measurement for the $V_{ud}$ element. This is important since recent theoretical developments on radiative corrections and form factors have led to a $3\sigma$ tension with CKM unitarity~\cite{ParticleDataGroup:2020ssz}, which if confirmed would point to new physics in the multi-\TeV scale.

PIONEER was proposed at the Paul Scherrer Institute (PSI) $\pi$E5 beam line, where high-intensity pion beams can be delivered, and was approved in January 2022. To achieve the necessary precision for both $R_{e/\mu}$ and pion beta decay, PIONEER has two main detectors, a high granularity full silicon active target~\cite{Mazza:2021adt} (in short ATAR) and a $28~X_{0}$ liquid xenon segmented calorimeter with high energy resolution. Furthermore, a one layer tracker will be placed around the ATAR to tag exiting positrons. A schematic of the PIONEER LXe vessel and zoomed in active target region can be seen in \cref{fig:PIONEER}.

The ATAR is a $2\times2~\si{\centi\meter}$ active target with a few \si{\milli\meter} of active thickness, sitting at the most probable pion interaction point. It is composed of several layers of $120~\si{\micro\meter}$ thin AC-LGAD strip sensors with a $200~\si{\micro\meter}$ pitch. The ATAR will provide a high precision 4D tracking that can separate the energy deposits of the pion decay products in both position and time. Furthermore it will suppress other significant systematic uncertainties  such as pulse pile-up or decay in flight of slow pions, by up to $<~0.01\%$.

The tracker will detect the positrons exiting from the ATAR and entering the calorimeter.  The LXe calorimeter is a sphere with a $1~\meter$ radius around the ATAR interaction point. It will measure with good position and timing information the exiting positrons and their energy down to the few \% level. The combination of the high precision calorimeter and the highly segmented ATAR will allow the overall uncertainty in the measurements of $R_{e/\mu}$ and pion beta decay to be reduced to $\mathcal{O}(0.01\%)$ and $\mathcal{O}(0.05\%)$, respectively.

The PIONEER experiment will proceed in a phased approach. In Phase I ($\sim$3-year), it will bridge the gap of a factor 15 between theoretical and experimental precision for $R_{e/\mu}$, reaching the $0.01\%$ level in precision. New physics up to the \si{\peta\eV} scale~\cite{Bryman:2011zz} may be revealed. This first phase of PIONEER ($R_{e/\mu}$) will employ a beam with a pion stopping rate in the ATAR of approximately $3\times 10^5/\second$ at a momentum of $55-70~\MeV/c$ with $\frac{\Delta p}{p}\leq 2\%$ in a spot size $\leq2\si{\centi\meter}$ in diameter. Muon and positron contamination will be reduced to $<10\%$ with the use of a separator. These requirements are compatible with the beams available at the $\pi$E5 beamline. The total number of $\pi^+\to e^+ \nu$ events for a 3-year run will be $2\times 10^8$, satisfying the statistics goal. 

In later Phases II ($\sim$4-year) and III (no timeline yet), PIONEER will also study pion beta decay $\pi^+\to \pi^0 e^+ \nu (\gamma)$, ultimately aiming at an order of magnitude improvement in precision to determine $V_{ud}$ in a theoretically pristine manner and test CKM unitarity. For the $\pi^+ \to \pi^0 e^+ \nu$ experiment, the positive pion stop rate must be higher, $\geq 10^7/\second$, possibly with a larger momentum bite $\frac{\Delta p}{p}\approx 3\%$, and likely using higher pion momentum. These beam conditions are compatible with the $\pi$E5 beam line. This would result in $7\times 10^5$ $\pi^+ \to \pi^0 e^+ \nu$ events collected over 4 years. This would be sufficient to achieve the required statistical precision to improve the pion beta decay branching ratio measurement precision by a factor of 3 in Phase II.

Systematic effects are expected to be reduced to the$ 0.06\%$ level ($10\times$ lower than for the previous PIBETA experiment) due to the combined improvements to the calorimetry (the time and energy resolutions) and the ATAR which may facilitate the observation of the positron in $\pi^+ \to \pi^0 e^+ \nu$ decay in coincidence with the $\pi^0$ detection. Running at higher rates may be possible, leading to a further precision improvement of Phase III and will depend on the ability of the spectrometer to deal with higher rates of pile-up events.

\subsection{PIP2-BD}
\label{sec:pip2bd}

Using the PIP-II proton fixed-target facility (see \cref{sec:pip2}), PIP2-BD envisions a $100$~ton scale liquid argon single-phase scintillation-only detector located $18~\meter$ from a carbon target~\cite{Toups:2022yxs}. An $\mathcal{O}(1~\GeV)$ proton beam colliding with the fixed target produces charged mesons. Dark matter models predict dark matter production via neutral mesons such as pions and $\eta$ mesons, also produced by the proton collisions with the fixed target. A custom \textsc{Geant4}-based (v4.10.6p01) simulation~\cite{Allison:2006ve,Allison:2016lfl} was built to study the accumulator ring scenarios described in \cref{sec:pip2} while implementing a simplified GDML-based target geometry to understand the production rate of charged mesons for each proton on target. Another \textsc{Geant4}-based (v4.10.6p01) simulation models the detector response at nuclear recoil energies relevant to light dark matter searches. In the model, the liquid argon scintillation light is detected by TPB-coated Hamamatsu R5912-02MOD photomultiplier tubes (PMTs) that surround the cylindrical active volume bounded by a TPB-coated Teflon reflector with holes for placement of the PMTs.

PIP2-BD is sensitive to a wide class of dark matter models. Initial sensitivity to a vector portal kinetic mixing model~\cite{deNiverville:2011it,deNiverville:2015mwa,deNiverville:2016rqh} includes the simulated detector response for both the dark matter signal and the neutrino backgrounds coming from CEvNS and the electron recoil background from $^{39}$Ar, greatly reduced by pulse shape discrimination and the proton beam duty factor. These studies assume that beam-related neutron backgrounds can be suppressed to negligible levels at this facility. The dark matter production simulations are performed using the BdNMC code~\cite{deNiverville:2016rqh} and the detector response for signal and backgrounds using the \textsc{Geant4}-based model. Another model is a leptophobic model where the dark matter couples to quarks instead of leptons, providing a similar phenomenology to the vector portal model with identical background estimates.

PIP2-BD has the capability to be a world-leading probe of both the vector portal kinetic mixing and leptophobic models assuming a 5-year run of the C-PAR scenario from \cref{sec:pip2}. A different phenomenology arises if there are two new particles $\chi_1$ and $\chi_2$, where $\Delta=(m_{\chi_2} - m_{\chi_1})/m_{\chi_1}>0$. The $\chi_2$ travels to the detector and decays into a $\chi_1$ and an $e^+e^-$ pair if the $\chi_2$ is sufficiently long lived. If the $\chi_2$ decay is not kinematically allowed, the dark matter signal is detectable through up- or down-scattering off of electrons. A recent study of the detection of inelastic dark matter at JSNS$^2$~\cite{Jordan:2018gcd} showed the possibility of signal/background separation in a scintillation-only detector. 

The protons colliding with the carbon target not only produce hadrons, but also photons, electrons, and positrons which makes these facilities natural places to search for axion-like particles (ALPs)~\cite{Dent:2019ueq,AristizabalSierra:2020rom,Brdar:2020dpr}. ALPs are produced in a proton fixed-target facility through the Primakoff process with a $Z^2$ enhancement and then subsequently decay into two photons within the detector. The photon flux above $100~\keV$ is produced using the \textsc{Geant4}-based proton beam and target model. The Primakoff production cross section and the probability for the subsequent ALP to decay within the detector volume is computed following \rfr{Brdar:2020dpr}. In 5 years of data-taking, PIP2-BD will cover the ``cosmological triangle'' region where no lab-based constraints exist to rule out axion-like particles.

\subsection{POKER}
\label{sec:poker}

The POsitron resonant annihilation into darK mattER (POKER/NA64$_{e^{+}}$, see \cref{sec:na64} for details on NA64) is a positron beam, missing energy experiment at CERN. It focuses on exploring the thermal relic light dark matter (LDM) parameter space in the mass range of a few hundred \MeV. The main production channel exploited in the experiment is the resonant annihilation of beam positrons with atomic electrons, mediated by the $s-$channel exchange of a massive mediator. The experimental signature associated to the signal production is a peak in the missing energy distribution, whose position depends only on the mass of the mediator $m_\mathrm{MED}$ via the resonant relation $E_r=\frac{m^2_\mathrm{MED}}{2m_e}$, and whose line-shape depends on the properties of the mediator-SM and mediator-LDM interaction. The use of a thick target, providing an almost continuous energy loss for the impinging positrons, allows the primary beam to ``scan'' the full range of dark photon masses from the maximum value corresponding to almost the full beam energy, to the minimum value fixed by the missing energy threshold. Values out of this range are also accessible thanks to the finite mediator width, although with a less-prominent kinematic signature.

\begin{figure}
  \centering
  \includegraphics[width=.5\textwidth]{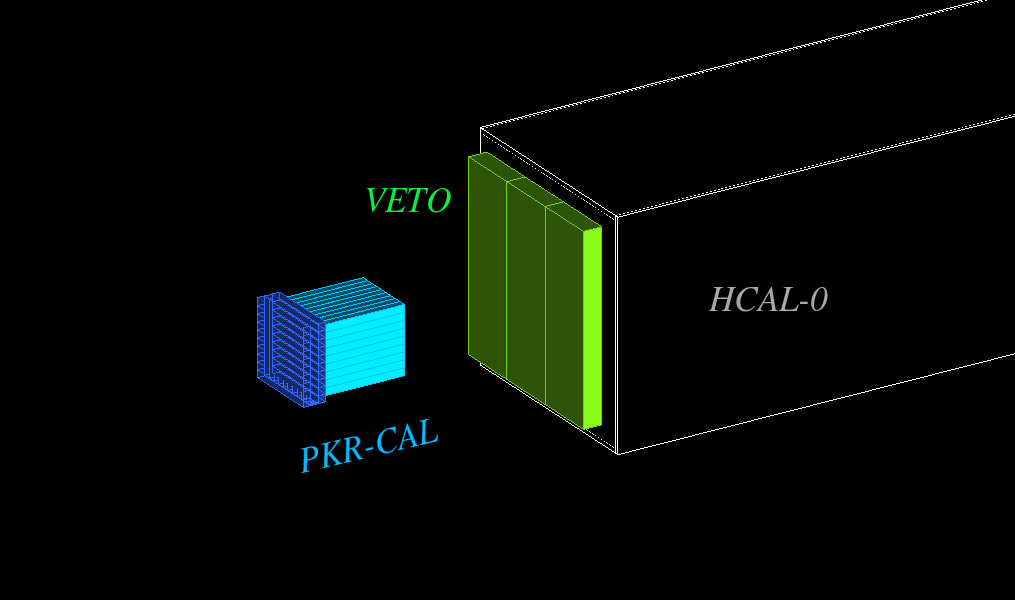}
  \caption{Conceptual design of the POKER PKR-CAL detector within the NA64 geometry (see \cref{sec:na64}). For simplicity, only a subset of the NA64 detectors is shown, the VETO and the first HCAL module. The different PKR-CAL sections, the pre-shower, and the main detector are shown respectively in dark blue and cyan.\label{fig:POKER}}
\end{figure}

The POKER active thick target (PKR-CAL) used to measure the energy released by each primary positron and thus to compute the missing energy, will be a high-resolution homogeneous PbWO$_4$ EM calorimeter, with a length of about $35~X_0$. The energy resolution target is $\sigma_E/E \simeq \frac{2\div3\%}{\sqrt{E}}\oplus(0.5\div1)\%$, in order to properly measure and reconstruct the annihilation line-shape, and also for LDM model variations, foreseeing a narrow mediator, \eg a vector mediator decaying to scalar LDM). The PKR-CAL is planned to be installed and integrated inside the existing NA64 setup at CERN~\cite{Banerjee:2019pds}.

The preliminary detector configuration is based on a $9\times9$ matrix of $2\times2\times22~\si{\centi\meter}^3$ PbWO$_4$ crystals, with the crystal axis oriented along the nominal beam direction. The matrix is preceded by a dual-layer pre-shower section, each layer being made by ten 2x2x20 cm$^3$ crystals, with the crystal axis oriented perpendicularly to the beam axis. The two layers are rotated by 90$^\circ$ around the $z$ axis to better determine impact point of particles on the calorimeter front face, as well as to minimize the bleed-through probability. The other existing NA64 detector elements will be exploited during the measurement, including the tracking system, the SRD detector, the beam-defining plastic scintillator counters, the VETO, and the HCAL.

The potential of the $e^+e^-$ annihilation channel in missing energy experiments has been recently demonstrated by an experimental re-analysis of the existing NA64 dataset, based on an accumulated statistics of $2.84\times10^{11}$ $100~\GeV$ electrons on target~\cite{Andreev:2021fzd}. By including explicitly this production mechanism, due to the contribution of secondary positrons, the existing LDM limits in the $m_\mathrm{MED}$ range between 200 and 300 MeV region are extended by almost an order of magnitude, touching for the first time the dark matter relic density constrained parameter combinations. 

The first goal of POKER will be to demonstrate the feasibility of the positron-beam missing energy approach in a pilot run, with accumulated statistics of about $10^{11}$ positrons on target. Assuming a conservative missing energy threshold of $50~\GeV$, the pilot run will explore the LDM parameter space up to the thermal targets for $m_\mathrm{MED}$ in the range $200-300~\MeV$, for a mediator-to-LDM mass ratio of three.  The pilot run is expected to take place in 2024 or 2025, depending on the H4 beam-time allocation. During the summer 2021 run, a first, low-statistics exploratory test with a $100~\GeV$ $e^+$ beam was performed exploiting the current NA64 detector. A preliminary analysis showed that no unforeseen backgrounds were present; applying the standard NA64 analysis procedure, no events were observed in the NA64 signal window. POKER is supported by a 2020 ERC Starting Grant (grant agreement 947715) from $2021-2025$. The POKER initiative is officially part of the NA64 experiment.

\subsection{REDTOP}
\label{sec:redtop}

The Rare $\eta$/$\eta'$ Decays To Probe New Physics experiment (REDTOP) is a fixed-target meson factory~\cite{REDTOP:2022slw} searching for new physics in flavor-conserving rare decays of the $\eta$ and $\eta'$ mesons. Such particles are almost unique in the particle universe, carrying the same quantum number as the Higgs boson except for parity, and no standard model (SM) charges, the dynamics of their decays is highly constrained. Conservation of their quantum numbers impose that all electromagnetic and strong decays are forbidden at the tree-level. Rare decays, are, therefore, enhanced compared to the remaining, flavor-carrying mesons. REDTOP aims at producing $10^{14}$($10^{12}$) $\eta(\eta')$ mesons. Such a sample can probe many recent theoretical models, providing enough sensitivity to explore all four portals connecting the dark sector with the SM while also probing conservation laws.

The $\eta$/$\eta'$ hadro-production mechanism is based on a proton beam with energy equal to $1.8~\GeV$ ($3.6~\GeV$ for the $\eta'$). Several intra-nuclear baryonic resonances ($\Delta$s, N(1440), N(1535), \etc) are created in nucleon-nucleon collisions, whose decay produce an $\eta$ or $\eta'$ meson. The inclusive $\eta/\eta'$ production cross-section has been calculated for $p$-Li scattering, using several theoretical models. The average value is $1.2\times10^{-23}~\si{centi\meter}^{-2}$ for the $\eta$ meson, and $6.8\times10^{-20}~\si{centi\meter}^{-2}$ for the $\eta$' meson. Assuming $10^{18}$ protons on target (POT) per year, the desired yield can be achieved in a few years of running. A CW proton beam delivering $1\times10{}^{11}$ POT$/\second$ would generate a rate of inelastic interactions of about $0.7~\GHz$ and an $\eta$-meson yield of $4\times10{}^{6}~\eta/\second$, corresponding to $4\times10{}^{13}~\eta/\mathrm{year}$. The beam power corresponding to the above parameters is approximately $30~\watt$ ($60~\watt$ for the $\eta$'), of which less than 1\% (or $300~\si{\milli\watt}$) is absorbed in the target systems. 

\begin{figure}[h]
  \centering
  \includegraphics[width=\textwidth]{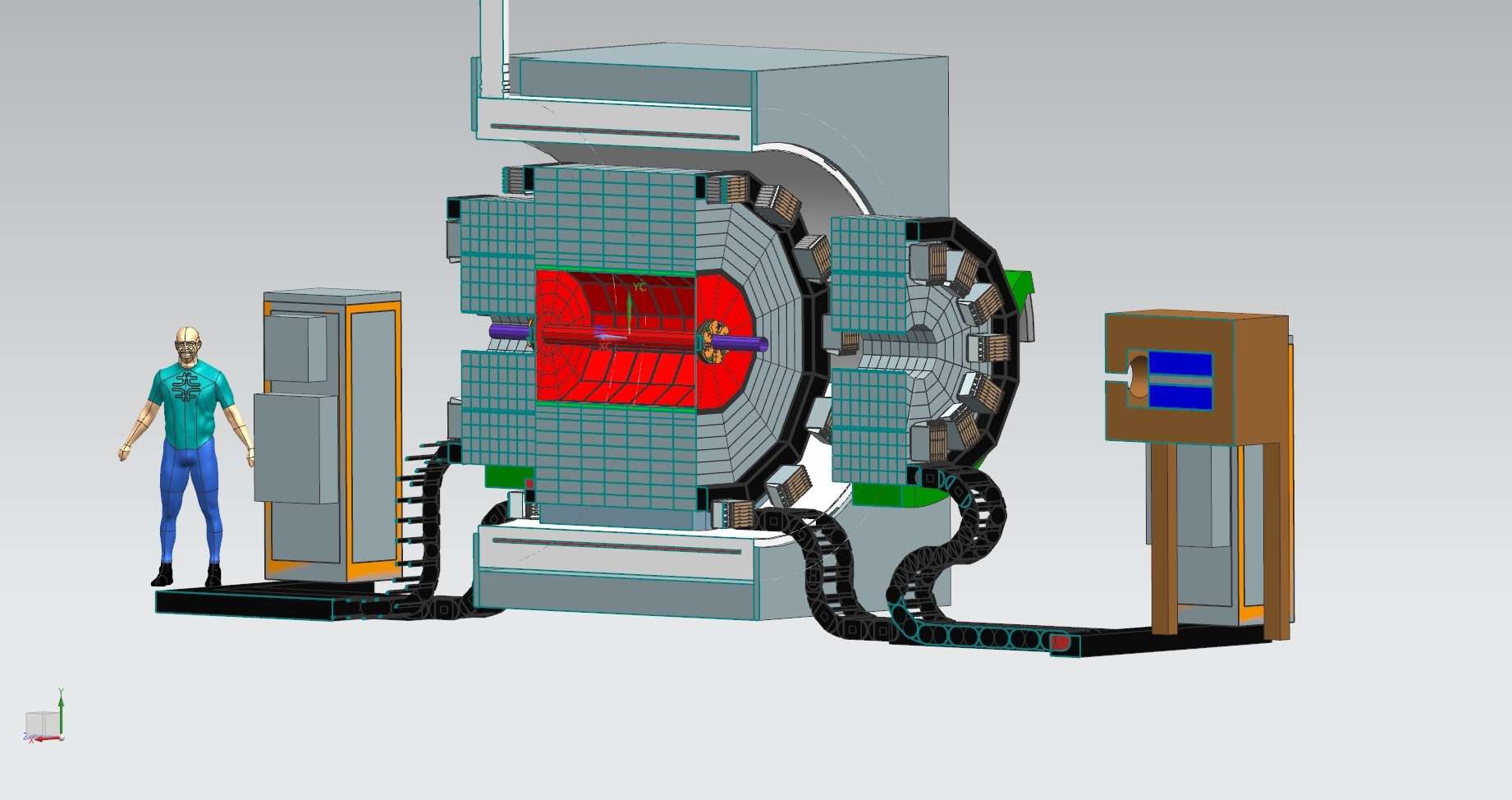}
  \caption{The REDTOP detector.\label{fig:REDTOP}}
\end{figure}

At the beam energies considered above, the $\eta$ and $\eta$' mesons are produced almost at rest in the lab frame, receiving only a small boost in the direction of the incoming beam. Consequently, a hermetic, collider-style, detector covering most of the solid angle, is one of the requirements for REDTOP. A schematics of REDTOP detector is shown in \cref{fig:REDTOP}.

The REDTOP target system is split in 10 foils, each $\sim0.7~\si{milli\meter}$ thick, of a low-$Z$ material (Li or Be), spaced $10~\si{\centi\meter}$ apart. This improves the measurement of the $z$ coordinate of the $\eta$ production vertex, while minimizing the multiple scattering affecting the $\eta$ decay products when they escape from the target. It also allows any eventual BSM, long-lived particle, and its decay products, to leave the target undisturbed. The vertex detector is the innermost active component of REDTOP. It has four main tasks: identifying events with a detached vertex; participating in the reconstruction of charged tracks originating in the target region; rejecting photon converting in the target materials; and reconstructing tracks with very low transverse momentum. An event with a detached vertex found not belonging to the very few produced $K_{short}$ will be a clear indication that new physics is being observed.

The REDTOP sensitivity studies performed probe the conservation of discrete symmetries of the universe. $CP$ symmetry can be explored at REDTOP in a number of ways, thanks to the large statistics available for certain channels and a well known background. In all cases, $CP$-violation is observed via asymmetries of the $\eta$ decays. An almost unique technique proposed by REDTOP is based on the measurement of the polarization of the muon, copiously produced in some decays of the $\eta$ mesons and tagged by fully reconstructing the decay of the latter. This could be achieved either in the high-granularity, polarization-conserving, calorimeter or in a dedicated polarimeter.

These sensitivity studies performed for REDPTOP also include the search for new particles produced in rare decays of the $\eta$ and $\eta'$ mesons. Most of these particles could be visibly observed in the detector, assuming that they decay within about $50~\si{\centi\meter}$ from the $\eta$ or $\eta'$ meson production point. With the proposed statistics, REDTOP has excellent sensitivity to all four portals connecting the standard model with the dark sector.

REDTOP is especially well suited to appreciably improve the precision of lepton universality measurements, thanks to the very large number of semi-leptonic decays of the $\eta$ meson which can be fully reconstructed, as well as a well known background. Several sensitivity studies on lepton universality have been performed, and indicate that the experiment is able to decisively probe such symmetry with unprecedented precision. Despite being more challenging at an $\eta$/$\eta'$ factory, the violation of leptonic flavor could also be probed. REDTOP could not compete on this measurement with dedicated experiments based on intense muon beams. Still, the different mechanism of production of the muons would make this exploration interesting and worth pursuing. 

The beam requirements for REDTOP are modest, and several existing laboratories worldwide have the capability to host the experiment. The REDTOP collaboration, consisting of more than one hundred scientists from fifty-three institutions, has engaged in a broad exploration, and identified BNL, Fermilab, GSI, and HIAF as possible candidates to host REDTOP.

\subsection{SHADOWS}
\label{sec:shadows}

\begin{figure}[h]
  \centering
  \includegraphics[width=0.6\textwidth]{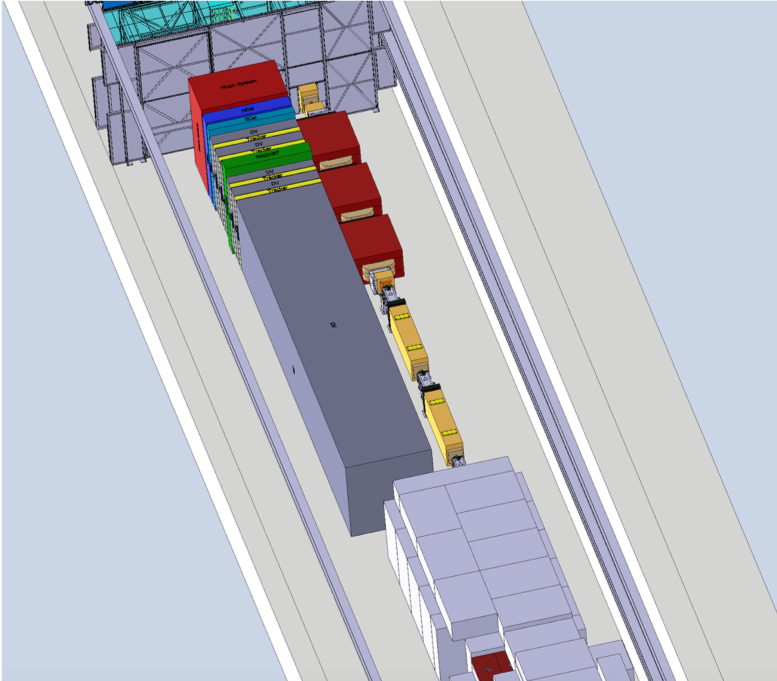}
  \caption{3D view of the SHADOWS detector located adjacent to the beam line and close to the dump zone (gray blocks).\label{fig:SHADOWS}}
\end{figure}

The Search for Hidden And Dark Object With the SPS (SHADOWS) is a new proton beam dump experiment able to search for a large variety of feebly-interacting particles (FIPs) possibly produced in the interactions of a $400~\GeV$ proton beam with a high-$Z$ material dump. SHADOWS will use the $400~\GeV$ primary proton beam extracted from the CERN SPS currently serving the NA62 experiment in the CERN north area and will take data off-axis  when the P42 beam line is operated in beam-dump mode. SHADOWS can accumulate up to $\sim 1.2 \times 10^{19}$ protons on target (POT) per year and expand the exploration for a large variety of FIPs well beyond the state-of-the-art in the mass range of \MeV to \GeV in a parameter space that is allowed by cosmological and astrophysical observations.

About $2 \times 10^{18}$ POT are delivered by the P42 line to NA62 every year. This assumes roughly 200 days of SPS uptime per year, a beam intensity of $3.3 \times 10^{12}$ protons per $4.8~\second$ long spill, and 3000 spills delivered per day. With some upgrade of the target and dump complex, the intensity of the P42 primary proton beam can be increased by up to a factor $6-7$ and still remain compatible with the operation of all the other existing experiments in the CERN north area.

At the SPS energy, FIPs with masses above the kaon mass are mostly produced in the decays of charmed and beauty hadrons from the interactions of the protons with the dump. This means the FIPs emerge with a non-negligible polar angle, opening up the possibility of placing a spectrometer off axis and very close to the dump. This has the double advantage of maximizing the acceptance for a given transverse dimension, while being less affected by backgrounds that are mostly produced in the forward direction. The residual beam-induced muon background will be swept away by a magnetic system. A conceptual layout of the spectrometer is shown in \cref{fig:SHADOWS}. The spectrometer is made of a $20~\meter$ long in-vacuum decay volume, followed by a tracker, a timing detector, an electro-magnetic calorimeter, a hadron calorimeter, and a muon detector. SHADOWS aims to take data after the Long Shutdown 3 (2029) and accumulate about $10^{19}$ POT per year during five nominal years of data taking. The project is under the review of the SPS committee at CERN and is expected to produce a proposal by 2023.

\subsection{SND@LHC}
\label{sec:snd}

The Scattering and Neutrino Detector at the LHC (SND@LHC) is a compact and stand-alone experiment designed to perform measurements with neutrinos produced at the LHC in a previously unexplored pseudorapidity region of $7.2 < \eta < 8.4$, complementary to all the other experiments at the LHC, including FASER (see \cref{sec:faser}). The SND@LHC collaboration submitted a letter of intent in August 2020~\cite{Collaboration:2729015}. Following investigations that confirmed the possibility of preparing the experimental area and installing the detector during 2021, with the LHC in cold operating conditions, the LHCC recommended the collaboration to proceed with the preparation of a technical proposal, submitted in January 2021~\cite{Ahdida:2750060}. Based on this document, the experiment was approved in March 2021 by the research board. 

\begin{figure}[h]
  \centering
  \includegraphics[width=0.8\textwidth]{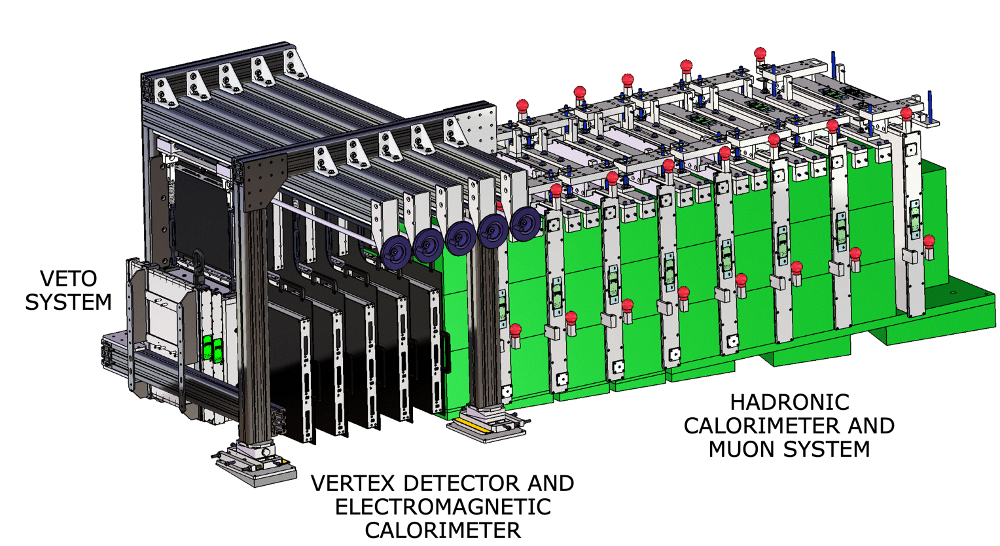}
  \caption{Detector layout of SND@LHC, the veto system is located upstream of the tungsten target, interleaved with emulsion and SciFi planes. Downstream of the target, 8 iron walls are followed by scintillating stations with the function of a hadronic calorimeter and a muon identification system.\label{fig:sndDetector}}
\end{figure}

The experiment is located $480~\meter$ downstream of IP1 in the unused TI18 tunnel. The detector is composed of a hybrid system based on an $800~\si{\kilo\gram}$ target mass of tungsten plates, interleaved with emulsion and electronic trackers, followed downstream by an hadronic calorimeter and a muon identification system, as shown in \cref{fig:sndDetector}. The configuration allows efficiently distinguishing between all three neutrino flavors, opening a unique opportunity to probe physics of heavy flavor production at the LHC in the region that is not accessible to ATLAS, CMS, and LHCb (see \cref{sec:lhcb}). The detector concept is also well suited to searching for feebly interacting particles via signatures of scattering in the detector target~\cite{Boyarsky:2021moj}. The first phase aims at operating the detector throughout Run~3 to collect more than $250~\invfb$ of data. 

\begin{figure}[h]
  \centering
  \includegraphics[width=0.9\textwidth]{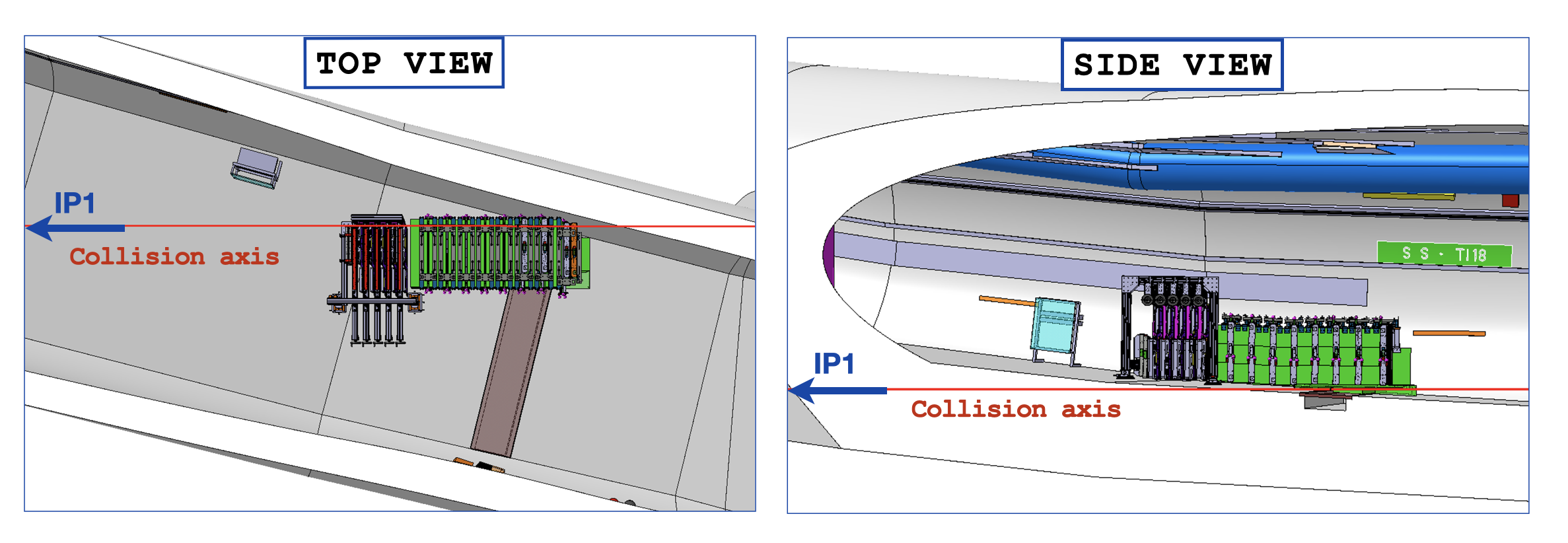}
  \caption{Views of the SND@LHC detector from the (left) top and (right) side in the TI18 tunnel~\cite{Ahdida:2750060}.\label{fig:sndSide}}
\end{figure}

The SND@LHC detector takes full advantage of the space available in the TI18 tunnel to cover the desired range in pseudorapidity. \Cref{fig:sndSide} shows the side and top views of the detector positioned inside the tunnel. It is worth noting that the tunnel floor is sloped, as can be seen from the side view, with the floor sloping down along the length of the detector. As shown in the top view, the nominal collision axis from IP1 comes out of the floor very close to the wall of the tunnel. The location is ideal to explore the off-axis region. Since no civil engineering work could be done in time for operation during Run 3, the tunnel geometry imposed several constraints. The following guidelines were adopted for the optimization of the detector design: a good calorimetry measurement of the energy requires $\sim10~\lambda_{\mathrm{int}}$; a good muon identification efficiency requires enough material to absorb hadrons; and for a given transverse size of the target region, the azimuthal angular acceptance decreases with distance from the beam axis. The energy measurement and the muon identification set a constraint on the minimum length of the detector. With the constraints from the tunnel, this requirement competes with the azimuthal angular acceptance that determines the overall intercepted flux, and therefore the total number of observed interactions. The combination of position and size of the proposed detector is an optimal compromise between these competing requirements. The geometrical constraints also restrict the detector to the first quadrant only around the nominal collision axis, as shown in \cref{fig:sndSide}.

The result is a compact detector, $2.6~\meter$ in length. The energy measurement and the muon identification limit the target region to a length of about $80~\si{\centi\meter}$. The transverse size downstream of about $80(\textrm{H})\times60(\textrm{W})~\si{\centi\meter}^2$ is limited by the constraint of the tunnel side wall. The transverse size of the target region is proportionally smaller in order to match the acceptance of the energy measurement and the muon identification for the vertices identified in the target volume. In order to maximize the number of neutrino interactions, tungsten has been selected as the passive material. The emulsion target will be replaced a few times per year during technical stops of the LHC.

With data from Run~3, SND@LHC will be able to study about two thousand high-energy neutrino interactions. \Cref{tab:sndFlux} reports the expected number of charged-current and neutral-current neutrino interactions in the detector target, assuming $150~\invfb$ and an equal weight of upward and downward crossing-angle configurations. The average energies of the interactions are also reported.  Performance studies show that the charmed-hadron production in the SND@LHC pseudorapidity range can be determined with a statistical and systematic accuracy of $5\%$ and $35\%$, respectively. The result may be further used to constrain the gluon PDF~in the very-small-$x$ region~\cite{Ahdida:2750060}. Unique tests of lepton flavor universality with neutrino interactions can reach 10\% for both statistical and systematic uncertainties for $\nu_e$ and $\nu_{\mu}$ at high energy, and about 30\% for $\nu_e$ and $\nu_{\tau}$~\cite{Ahdida:2750060}. 

\begin{table}[hbtp]
  \centering
  \begin{tabular}{c | c r | c r}
    \toprule 
    &  \multicolumn{2}{c|}{CC neutrino interactions} & \multicolumn{2}{c}{NC neutrino interactions} \\
    \midrule
    Flavor     &  $\langle\text{E}\rangle$ [\GeV]  & Yield &  $\langle\text{E}\rangle$ [\GeV]  & Yield \\
    \midrule
    $\nu_\mu$       &  452  & 606 & 480 & 182 \\
    $\bar{\nu}_\mu$ & 485  & 248 & 480 & 93\\
    $\nu_e$         & 760  & 182 & 720 & 54\\
    $\bar{\nu}_e$   & 680  &  97 & 720 & 35\\
    $\nu_\tau$      & 740  &  14 & 740 & 4\\
    $\bar{\nu}_\tau$& 740  &   6 & 740 & 2\\
    \midrule
    total &   & 1153  & & 370\\
    \bottomrule
  \end{tabular}
  \caption{Expected number of charged-current and neutral-current neutrino interactions in the SND@LHC acceptance.\label{tab:sndFlux}}
\end{table}

\begin{figure}
  \centering
  \includegraphics[width=0.7\linewidth]{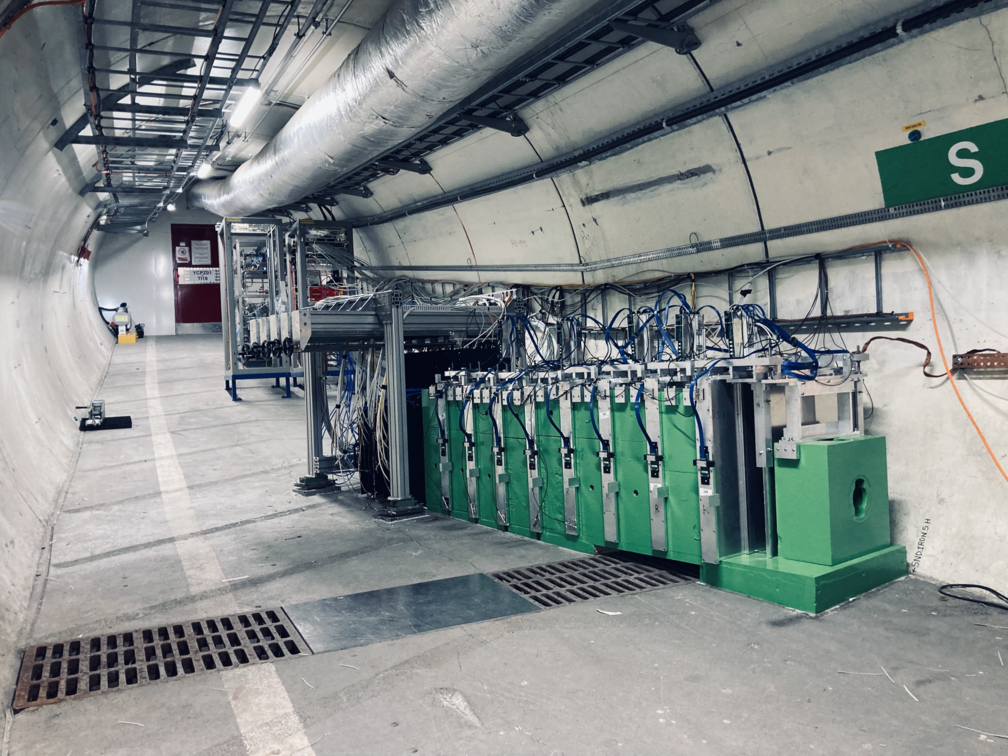}
  \caption{Picture of the installed SND@LHC detector in the TI18 tunnel.\label{fig:sndTunnel}}
\end{figure}

All the detector systems were constructed in the labs by summer 2021 and were assembled and tested at CERN. In October 2021, a test beam was performed at the SPS with protons of different energies in order to calibrate the response of the hadronic calorimeter, 7 scintillating bar stations interleaved by real-size iron blocks were used for the measurement. Moreover, the full detector was commissioned on the surface at CERN with penetrating muons in the H6 experimental hall. In the beginning of November 2021, the installation underground started. \Cref{fig:sndTunnel} shows the full detector installed in the middle of December 2021. A borated polyethylene shielding box has been added to surround the target and absorb low-energy neutrons originated from beam-gas interactions: its installation was completed on March $8^{\mathrm{th}}$, 2022. The detector installation was completed in March, by adding the target walls, so as to be ready to take data as soon as the LHC operation resumes with the Run~3.

\clearpage
\section{Facilities}
\label{sec:facilities}
\subsection{NM4 Experimental Hall at Fermilab 120 GeV Main Injector}
\label{sec:nm4}

The Fermilab Main Injector ramps up $8~\GeV$ protons from the Recycler Ring to $120~\GeV$. The beam is extracted in a slow spill lasting just under four seconds. Typically, the time between the beginning of spills is just over one minute. Beam is extracted using a resonant process and the extracted beam retains the $53.1~\MHz$ structure of the Main Injector RF, dividing the beam into ``RF buckets'' that are less than $2~\si{\nano\second}$ long and occur every $18.8~\si{\nano\second}$. The beam is delivered along the neutrino-muon beamline to the NM4 Experimental Hall in the fixed-target area. The beamline is currently operational and delivers beam to the SpinQuest experiment. The beam intensity is approximately $10^4-10^5$~protons per second with an intensity of roughly $10^{12}$~protons per second. This leads to a total integrated luminosity of approximately $10^{18}$~protons-on-target (POT) per year.  

The NM4 Experimental Hall houses the SpinQuest experiment which is a $25~\meter$ long spectrometer and has space for additional detectors up to roughly $40~\meter$ in length. It operates as a high intensity proton fixed target experiment to precisely study the Drell-Yan process. The facility can house a variety of proton fixed-target dark sector experiments (see \cref{sec:darkQuest} and \cref{sec:fermini}). With modifications to the beamline, the beamline could potentially deliver muons for a muon fixed-target experimental program and in the past was home to the KTEV experiment which studied rare neutral kaon decays.

\subsection{Forward Physics Facility}
\label{sec:fpf}

\begin{figure}[h]
  \centering
  \includegraphics[width=0.90\textwidth]{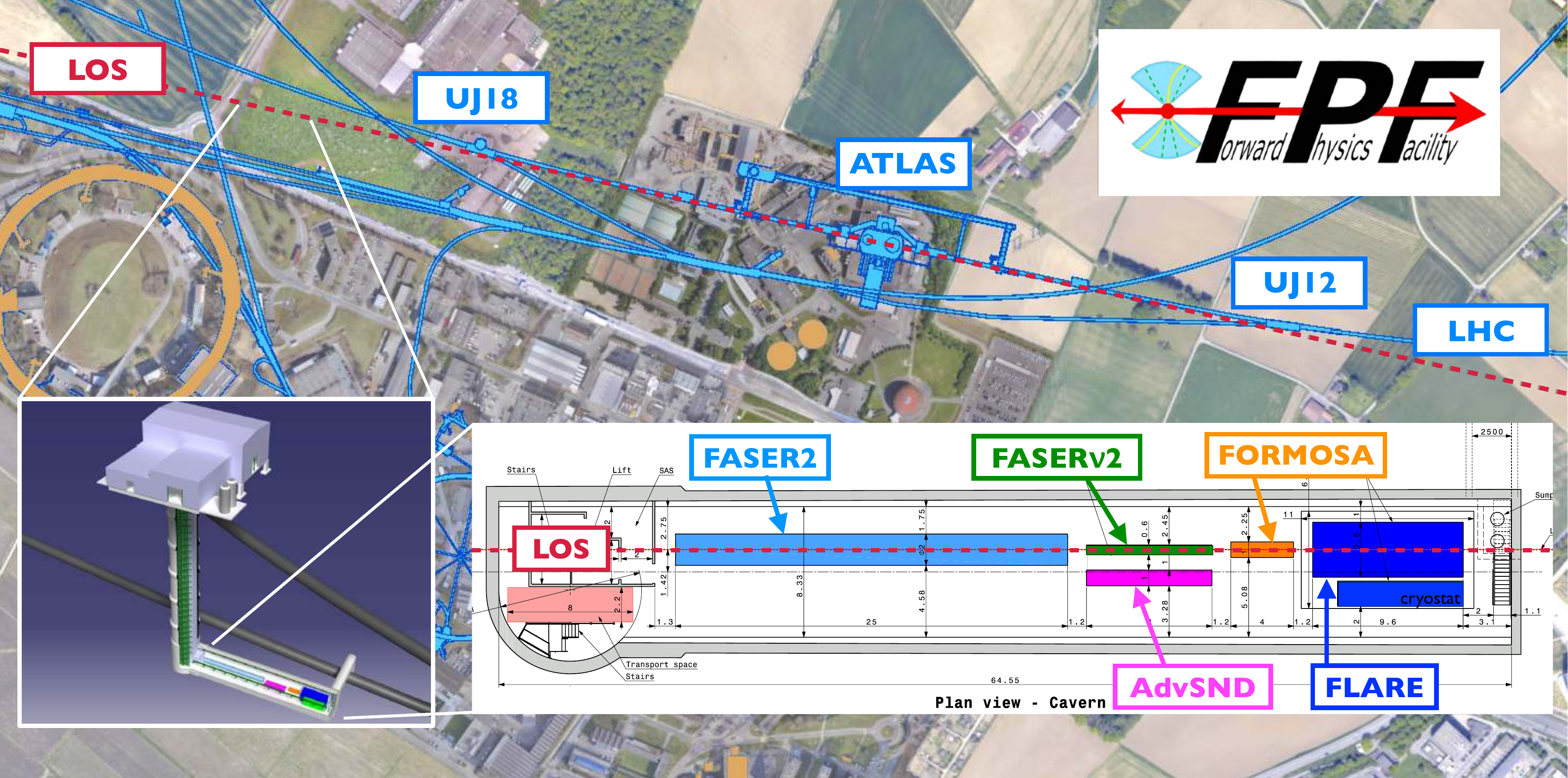}
  \caption{The preferred location for the Forward Physics Facility, a proposed new cavern for the High-Luminosity LHC era. The FPF will be $65~\meter$-long and $8.5~\meter$-wide, and will house a diverse set of experiments to explore the many physics opportunities in the far-forward region.\label{fig:FPF}}
\end{figure}

The Forward Physics Facility (FPF)~\cite{MammenAbraham:2020hex, Anchordoqui:2021ghd, Feng:2022inv} is a proposal to build a new underground cavern at the Large Hadron Collider (LHC) to host a suite of far-forward experiments during the High-Luminosity LHC era. The existing large LHC detectors have holes along the beam line, and so miss the physics opportunities provided by the enormous flux of particles produced in the far-forward direction. The FPF will realize this physics potential. A preferred site for the FPF is along the beam collision axis, $617-682~\meter$ west of the ATLAS interaction point (IP), see \cref{fig:FPF}. This location is shielded from the ATLAS IP by over $200~\meter$ of concrete and rock, providing an ideal location to search for rare processes and very weakly-interacting particles. FPF experiments will detect $\sim 10^6$ neutrino interactions at the highest human-made energies ever recorded, expand our understanding of proton and nuclear structure and the strong interactions to new regimes, and carry out world-leading searches for a wide range of new phenomena, enhancing the LHC's physics program through to its conclusion in 2040.

The FPF is uniquely suited to exploit physics opportunities in the far-forward region because it will house a diverse set of experiments, each optimized for particular physics goals. FASER2 (see \cref{sec:faser2}), a magnetic spectrometer and tracker, will search for light and weakly-interacting states, including long-lived particles, new force carriers, axion-like particles, and dark sector particles. FASER$\nu$2 (see \cref{sec:faserNu2}) and Advanced SND (see \cref{sec:snd}), proposed emulsion and electronic detectors, respectively, will detect $\sim 10^6$ neutrinos and anti-neutrinos at \TeV energies, including $\sim 10^3$ $\tau$ neutrinos, the least well-understood of all known particles.  FLArE (see \cref{sec:flare}), a proposed 10-ton-scale noble liquid detector, will detect neutrinos and also search for light dark matter.  Finally FORMOSA (see \cref{sec:formosa}), a detector composed of scintillating bars, will provide world-leading sensitivity to millicharged particles and other very weakly-interacting particles across a large range of masses.

The FPF is well aligned with the 2020 European Strategy Update's first recommendation that ``the full physics potential of the LHC and the HL-LHC \ldots should be exploited.''  To fully exploit the far-forward physics opportunities, many of which will disappear for several decades if not explored now, the FPF should be available for as much of the HL-LHC era as possible. The FPF requires no modifications to the LHC, and all of the planned experiments are relatively small, inexpensive, and fast to construct. A very preliminary costing for the FPF has yielded estimates of 25~MCHF for the construction of the new shaft and cavern and 15~MCHF for all necessary services. To this must be added the cost of the individual experiments. A possible timeline is for the FPF to be built during Long Shutdown 3 from $2026-2028$, the support services and experiments to be installed starting in 2029, and the experiments to begin taking data not long after the beginning of Run~4. Such a timeline is guaranteed to produce exciting physics results through studies of very high energy neutrinos, QCD, and other SM topics, and will additionally enhance the LHC's potential for groundbreaking discoveries that will clarify the path forward for decades to come.

\subsection{SLAC Linac to End Station A (LESA)}
\label{sec:lesa}

The SLAC Linac to End Station A (LESA) will extract low-current near-CW $4-8~\GeV$ beams from the SLAC LCLS-II superconducting linac~\cite{Raubenheimer:2018mwt} and transport this beam to end station A, enabling low-current electron beam experiments in the end station.

LESA is optimized to deliver currents ranging from \si{\pico\ampere} to \si{\nano\ampere}, and does so parasitically to the LCLS-II Free Electron Laser (FEL) program. To achieve this, LESA uses a high-rate kicker to extract low-current beam during the $1.1~\si{\micro\second}$ gaps between FEL pulses. The low-current LESA beam can be either the dark current from the gun or sourced by a gun laser at $46~\MHz$ or subharmonics thereof (with potential upgrades up to $186~\MHz$). This provides a near-CW electron beam at the \si{\nano\ampere} current scale to the A-Line. The beam can be attenuated to lower currents at precise energies using a spoiler and collimators in the A-Line, as was done for the End Station Test Beam (ESTB) program \cite{Sheppard:2017kjh}. 

Construction of the LESA beamline is ongoing. Operation with a $4~\GeV$ dark-current-sourced electron beam is expected in 2023, with laser installation in 2024 and operation with an $8~\GeV$ beam expected in 2027. The scientific impact of multi-\GeV CW electron beams is underscored by the high demand for beam time at Jefferson Lab's CEBAF (the only facility in the world delivering such a beam) from both nuclear and high-energy physics experiments.  While CEBAF is optimized for the needs of the new physics community, including precisely tunable energy and beam polarization, LESA will enable the longer run-times required by high-energy physics experiments as well as readily offering the low-current CW beam configurations required by missing momentum experiments. The planned LESA program includes electron-nuclear scattering for neutrino physics, test beam, and dark sectors (see \cref{sec:ldmx}) in the near term and potentially, with upgrades, to support $\sim 1~\si{\micro\ampere}$ currents for complementary searches for new forces.

\subsection{MESA}
\label{sec:mesa}

The Mainz Energy-Recovering Superconducting Accelerator (MESA) is a low-energy continuous-wave recirculating electron accelerator for particle and nuclear physics experiments, currently under construction at the Johannes Gutenberg University Mainz, Germany~\cite{MESA:1923127}. It is housed at the Institute of Nuclear Physics that has been operating electron accelerators for more than 50 years, most notably the MAinz MIkrotron (MAMI). The superconducting cavities, together with the ERL option, are novel features in Mainz. MESA will offer an up to $105~\MeV$ at an $1~\si{\milli\ampere}$ (upgradable to $10~\si{\milli\ampere}$) unpolarized electron beam for fixed internal target experiments in energy recovery operation (ERL mode) and $150-200~\MeV$ at $150~\si{\micro\ampere}$ polarized electron beam for extracted beam operation (EB mode). 

Three experimental areas are foreseen of which two have a focus on light dark matter searches.
\begin{enumerate}
\item The MAinz Gas Injection target eXperiment (MAGIX) is employing the ERL mode in combination with a gas jet target and two high-resolution magnetic spectrometers. The very clean and high-intensity beam allows for dark photon visible decay searches at low masses and small couplings. For invisible decay searches, missing mass experiments with the spectrometer setup are planned that include a silicon detector within the scattering chamber for the detection of the recoil target protons in coincidence with the scattered beam electron.
\item The DarkMESA experiment makes use of the beam dump for the EB mode and will search for dark matter production and rescattering events. Several detectors will be placed $23~\meter$ behind the beam dump and will be shielded by the walls of the accelerator hall from neutrons or other standard model particles that escape the beam dump.
\end{enumerate}

Civil construction for the new experimental hall is expected to be completed in late 2022, when installation of the accelerator components and the experimental setup will start. The commissioning of the accelerator and of the experiments is expected to start in 2023. A total of about 4,000\,h per year of beam-time for experiments is planned.

\subsection{PIP2}
\label{sec:pip2}

The completion of the PIP-II superconducting linac at the Fermilab accelerator complex in the late 2020s provides an opportunity to design an HEP-focused dark sector search facility at Fermilab. This facility is motivated by recent theoretical advancements in sub-\GeV dark matter models explaining the cosmological dark matter abundance and that accelerator-based, fixed-target experiments complement existing direct dark matter detection experiments~\cite{Alexander:2016aln,Battaglieri:2017aum,BRN}. The recent observations of coherent elastic neutrino-nucleus scattering (CEvNS)~\cite{Freedman:1973yd,Kopeliovich:1974mv} by the COHERENT experiment~\cite{COHERENT:2017ipa,COHERENT:2020iec} provides a novel experimental tool and guide for searching for light dark matter using an accelerator-based fixed target facility. The novelty of the PIP2-BD facility (see \cref{sec:pip2bd}) is that the target is envisioned as a lighter nucleus to suppress neutron production through spallation on the target material as opposed to other currently existing facilities using heavy targets. The facility also envisions the ability to host multiple 100-ton scale detectors with the possibility of different distances and angles with respect to the target allowing for sensitive dark sector searches. 

The proposed PIP-II facility enables a sensitive dark sector search program by coupling the PIP-II superconducting linac with an accumulator ring in order to reduce backgrounds to low enough levels through beam pulse timing. The facility has the capability to support $\mathcal{O}(1~\GeV$) proton beams that produce secondary hadrons such as pions and $\eta$ mesons from the proton collisions with the fixed target. The addition of an accumulator ring is envisioned as an upgrade to the Fermilab accelerator complex. Reduction in the beam-related neutron and neutrino backgrounds is also possible if the beam pulse width is reduced to the $\mathcal{O}(10)~\si{\nano\second}$ level. A baseline accumulator ring with a beam power of $\sim100~\si{\kilo\watt}$ and a beam pulse width of $\sim1~\si{\micro\second}$, called ``PAR'', is possible within the decade and enables an initial dark sector program using the PIP-II fixed-target facility. Two other scenarios extend the physics capabilities of the facility by either reducing the beam pulse width to $\mathcal{O}(10)~\si{\nano\second}$ (``C-PAR'') or increase the beam power to $\sim1~\si{\mega\watt}$ (``RCS-SR''). The three accumulator ring scenarios will produce between $5\times10^{22}$ protons on target (C-PAR) and $5\times10^{23}$ protons on target (RCS-SR) over a five-year run period. The three scenarios are summarized in \cref{tab:pip2}. 

\begin{table}[h]
  \centering
  \begin{tabular}{p{2.0cm}|p{2.0cm}|p{2.0cm}|p{2.0cm}|p{2.0cm}|p{2.0cm}}
    \toprule
    Facility & Beam energy [\GeV] & Repetition rate [\Hz] & Pulse length [\second] & Beam power [\si{\mega\watt}] & POT/5~years\\
    \midrule
    PAR & 0.8 & 100 & $2 \times 10^{-6}$ & 0.1 & $1.2\times10^{23}$ \\
    C-PAR & 1.2 & 100 & $2\times 10^{-8}$ & 0.09 & $5.7\times10^{22}$\\
    RCS-SR & 2 & 120 & $2 \times 10^{-6}$ & 1.3 & $5.0\times10^{23}$ \\
    \bottomrule
  \end{tabular}
  \caption{The parameters of three possible accumulator ring scenarios considered to bunch the beam current from the PIP-II linac. The PAR is considered as the baseline scenario and the C-PAR and RCS-SR are upgrade paths coinciding with further upgrades to the Fermilab accelerator complex to support $2.4~\si{\mega\watt}$ operation for DUNE.}
  \label{tab:pip2}
\end{table}

\clearpage
\bibliography{references}
\end{document}